\documentstyle[sprocl]{article}
\def\Journal#1#2#3#4{{#1} {\bf #2}, #3 (#4)}

\def\NPB{{\em Nucl. Phys.} B}

\def\PRL{\em Phys. Rev. Lett.}
\def\PRD{{\em Phys. Rev.} D}
\def\PRC{{\em Phys. Rev.} C}

\def\lag{{\cal L}}    % script L for Lagrangian
\def\cutoff{{(\Lambda)}}
\def\Jv{{\bf J}}

\def\quarter{\mbox{$\frac{1}{4}$}}

\newcommand{\be}{\begin{equation}}
\newcommand{\ee}{\end{equation}}
\newcommand{\eq}[1]{Eq.~(\ref{#1})}
\newcommand{\order}{{\cal O}}
\newcommand{\nl}{\nonumber \\}
\newcommand{\bearray}{\begin{eqnarray}}
\newcommand{\eearray}{\end{eqnarray}}

\newcommand{\nablav}{\mbox{\boldmath $\nabla$}}
\newcommand{\sigmav}{\mbox{\boldmath $\sigma$}}
\newcommand{\tauv}{\mbox{\boldmath $\tau$}}
\newcommand{\e}{{\rm e}}
\newcommand{\eff}{{\rm eff}}
\newcommand{\true}{{\rm true}}
\newcommand{\erf}{{\rm erf}}
\newcommand{\erfc}{{\rm erfc}}
\newcommand{\appr}{{\rm app}}
\newcommand{\pv}{{\bf p}}
\newcommand{\qv}{{\bf q}}
\newcommand{\kv}{{\bf k}}
\newcommand{\pfour}{\pv^4}
\newcommand{\rv}{{\bf r}}
\newcommand{\Lv}{{\bf L}}
\newcommand{\ebar}{\overline{e}}
\newcommand{\bbar}{\overline{b}}

\newcommand{\omittext}[1]{}

\newlength{\minuslength}
\newcommand{\msp}{\rule{\minuslength}{0pt}}

\newlength{\digitlength}
\newcommand{\dsp}{\rule{\digitlength}{0pt}}

\newcommand{\tstrut}{\rule{0pt}{3ex}}

\begin{document}
\title{HOW TO RENORMALIZE THE SCHR\"ODINGER EQUATION\\
\small Lectures at the VIII Jorge Andr\'e Swieca Summer School
(Brazil, Feb.\ 1997)}
\author{ G. P. LEPAGE}
\address{Newman Laboratory of Nuclear Studies, Cornell University\\
Ithaca, NY 14853\\E-mail: gpl@mail.lns.cornell.edu}
\maketitle\abstracts{These lectures illustrate the key ideas of modern
renormalization theory and effective field theories in the context of
simple nonrelativistic quantum mechanics and the Schr\"odinger
equation. They also discuss problems in QED, QCD and nuclear physics
for which rigorous potential models can be derived using 
renormalization techniques. They end with an analysis of
nucleon-nucleon scattering based effective theory.}

\section{Renormalization Revisited}
These lectures are about effective field theories\,---\,low-energy
approximations to arbitrary high-energy physics\,---\,and therefore
they are about modern renormalization theory.\cite{tasitalk}

Despite the complexity of most textbook accounts,
renormalization is based upon a very familiar and simple idea: a
probe of wavelength $\lambda$ is insensitive to details of structure
at distances much smaller than $\lambda$. This means that we can mimic
the {\em real\/} short-distance structure of the target and probe by
{\em simple\/} short-distance structure. For example, a complicated current
source~$\Jv(\rv,t)$ of size~$d$ that generates radiation with wavelengths
$\lambda\gg d$ is accurately mimicked by a sum of point-like multipole
currents ($E1$, $M1$, etc). In thinking about the long-wavelength
radiation it is generally much easier to treat the source as a sum of
multipoles than to deal with the true current directly. This is
particularly true since usually only one or two multipoles are needed
for sufficient accuracy. The multipole expansion is a simple example
of a renormalization analysis.

In a quantum field theory, QED for example, 
the quantum fluctuations probe arbitrarily short distances.
This is evident when one computes radiative corrections in
perturbation theory. Ultraviolet divergences, coming from loop momenta
$k\to\infty$ (or wavelengths $\lambda\to 0$), result in infinite
contributions\,---\,radiative corrections seem infinitely sensitive to short
distance behavior. Even ignoring the infinities, this poses a serious
conceptual problem since we don't really know what happens as
$k\to\infty$. For example, there might be new supersymmetric
interactions, or superstring properties might become important, or
electrons and muons might have internal structure.
The situation is saved by renormalization theory which tells us that 
we don't really need to know what
happens at very large momenta in order to understand low-momentum
experiments. As in the multipole expansion, we can mimic the complex
high-momentum, short-distance structure of the real theory,
whatever it is, by a generic set of simple point-like interactions.

The transformation from the real theory to a simpler effective theory,
valid for low-momentum processes, is achieved in two steps. First we
introduce a momentum cutoff~$\Lambda$ that is of order the momentum at
which new as yet unknown physics becomes important. Only momenta $k<\Lambda$
are retained when calculating radiative corrections.\footnote{For
clarity's sake we
adopt a simple cutoff as our regulator here. In
actual calculations one generally tailors the regulator to optimize
the calculation.}
This means that our radiative corrections include only physics that we
understand, and that there are no longer infinities. Of course we
don't really know the scale $\Lambda$ at which new physics will be
discovered, but, as we shall see, results are almost independent of
$\Lambda$ provided it is much larger than the momenta in the 
range being probed experimentally.

The second step is to add local interactions to the lagrangian (or
hamiltonian). These mimic the effects of the true short-distance
physics. Any radiative correction that involves momenta above the
cutoff is necessarily highly virtual, and, by the uncertainty
principle, must occur over distances of order $1/\Lambda$ or less.
Such corrections will appear to be local to low-momentum probes
whose wavelengths $\lambda\approx1/p$ are large compared with
$1/\Lambda$. Thus the correct lagrangian for cutoff QED consists of the normal
lagrangian together with a series of correction terms:
\bearray
\lag^\cutoff &=& \overline{\Psi}\left( i\partial\cdot\gamma - e\cutoff
A\cdot\gamma - m\cutoff\right)\Psi - \quarter F_{\mu\nu}F^{\mu\nu} \nl
&+& \frac{e\cutoff c_1\cutoff}{\Lambda}\, 
\overline{\Psi}\sigma_{\mu\nu} F^{\mu\nu} \Psi \nl
&+& \frac{e\cutoff c_2\cutoff}{2\Lambda^2}\, \overline{\Psi}
i\partial_\mu F^{\mu\nu}\gamma_\nu \Psi
+ \frac{d_2\cutoff}{\Lambda^2} \, \left( \overline{\Psi}\gamma_\mu\Psi\right)^2
+ \cdots, \label{cutoff-qed}
\eearray
where couplings $e\cutoff$, $c_1\cutoff$, $c_2\cutoff$ and $d_2\cutoff$
are dimensionless. The correction terms are nonrenormalizable, but
that does not lead to problems because we keep the cutoff finite.
These new interactions are far simpler to work with than the
supersymmetric/superstring/\ldots interactions that they simulate.

The correction terms in cutoff QED modify the predictions of the
theory. The modifications, however, are small if $\Lambda$ is large. 
Contributions from
the $\sigma_{\mu\nu} F^{\mu\nu}$~term, for example, are suppressed by
$p/\Lambda$, where $p$ is the typical momentum in the process under
study. The next two terms are suppressed by $(p/\Lambda)^2$, and so
on. In principle there are infinitely many correction terms in
$\lag^\cutoff$, forming a series in $1/\Lambda$; but, when working
to a given precision (ie, to a given order in $p/\Lambda$), only a
finite number of these terms is important. Indeed none of these
correction terms seems important in any high-precision test of QED.
This indicates that the scale for new physics, $\Lambda$, is quite
large\,---\,probably of order a few TeV or larger.

Expansion~(\ref{cutoff-qed}) provides a useful 
parameterization for the effects of new physics on low-momentum
processes. The form of the cutoff lagrangian is independent of the new
physics. It is only the numerical values of the couplings $c\cutoff$,
$d\cutoff$\ldots that contain
information about the new physics. (The couplings are analogous to the
multipole moments of a current in our example above.) Thus the
implications of a high-precision test of QED can be expressed in a
model-independent way as limits on or values for these couplings.

In these lectures I illustrate the powerful techniques of modern
renormalization theory in a series of fully worked-out examples. 
These examples are all based upon the standard
Schr\"odinger equation; they require nothing more
than elementary quantum mechanics and some simple numerical
analysis. And yet, as I discuss, several important problems in QED,
QCD and nuclear physics can be rigorously formulated as potential
models using renormalization techniques.

I begin, in Section~2, with an illustration of both nonperturbative
and perturbative renormalization in the context of the Schr\"odinger
equation. We will explore many aspects of renormalization 
familiar from applications in quantum field theory.
In particular we will see in detail how to design an effective theory
to model a particular set of low-energy data. In Section~3 I discuss
the physical conditions that lead to potential models, and, briefly, 
how they are used 
in QED and QCD. Finally, in Section~4, I describe how to use
our effective potential theory in a systematic analysis of low-energy
nucleon-nucleon interactions.

\section{Renormalizing the Schr\"odinger Equation}\label{ren-schrod}

In this section I illustrate the construction and use of an effective theory
by designing one that reproduces a given collection of low-energy data. I
begin by discussing the data we will use. I then describe 
an obvious, well-known
procedure for modelling such data, and its 
limitations. Next I construct a somewhat less obvious effective
theory that overcomes these limitations. This modern approach allows us to
model the data with arbitrary precision at low energies. My goal
throughout is
to demonstrate how to use effective theories; I am less interested in
proving the formalism mathematically correct. Consequently
most examples are solved numerically. Nevertheless the insight
gained from analytic calculations is ultimately indispensable. Thus
I end this section with a very simple analytic calculation, 
using one-loop perturbation
theory, that illustrates several of the key ideas.

\subsection{Synthetic Data}
To begin we need a collection of low-energy data that 
describe a particular physical system. While we could use real experimental
data from a physically interesting system, it is better here to avoid the
complexities associated with experimental error by
generating ``synthetic data.'' I illustrate the application to real
problems later in the lectures, once we have worked through the formalism. 

I generated synthetic data for use in these lectures by inventing a simple
physical system, and then solving the Schr\"odinger equation that
describes it. 
I obtained binding energies, low-energy phase shifts, and matrix elements. For
the system, I chose the familiar one-particle Coulombic atom, but with a
short-range potential~$V_s(\rv)$ in addition to the Coulomb potential:
\be \label{trueH}
H = \frac{\pv^2}{2m} + V(\rv)
\ee
where
\be \label{trueV}
V(r) = -\frac{\alpha}{r} + V_s(\rv).
\ee
I arbitrarily set $m\!=\!1$ and $\alpha\!=\!1$.

For our purposes, the short-range potential may be
anything; I made one up. The whole point of effective theories is that we can
systematically design them directly from low-energy data, with no knowledge
of the short-distance dynamics. Thus the form of $V_s(\rv)$ is irrelevant to
our analysis. To underscore this point I will not reveal the functional form
of the $V_s(\rv)$ I used. In what follows we need only know
that it has finite range. 

Such short-range interactions are common in real Coulombic atoms.  The
effect of the proton's finite size on the hydrogen atom's spectrum is
an example. Another is the weak interaction between the electron and
proton, which generates very short-range potentials.

Having chosen a particular $V_s(\rv)$, I wrote a simple computer code to
numerically solve for the
radial wavefunctions of energy eigenstates. I used this to compute a variety
of binding energies, phase shifts and matrix elements. Some of the $S$-state
binding energies are listed in Table~\ref{s-state-energies-table}. Note that
the energies would have been given by $-\alpha^2m/2n^2 = -1/2n^2$, with
$n=1,2\ldots$, had there been no
$V_s$. Thus the 1$S$~energy is more than doubled by the short-range
potential; $V_s$ is not a small perturbation. Also the short-range nature of
$V_s$ is evident in this data since the energies approach the Coulombic
energies for the very low-energy, large-$n$ states. 

Sample $S$-wave phase shifts are given
in Table~\ref{phase-shift-table}. These
phase shifts depend upon the radius at which they are measured because of
the long Coulomb tail in the potential~$V(\rv)$; I
arbitrarily chose $r\!=\!50$ for my phase-shift measurements, this being much
larger than the Bohr radius, $1/\alpha\,m$, of my atom, and also much larger
than the range of~$V_s(\rv)$. (Alternatively, one can 
compute the phase shift with and without $V_s(\rv)$, and
take the difference; the Coulomb divergence cancels in the difference.)
I also computed $\langle \pfour \rangle$ for several
$S$-states, as well as the wavefunctions at the origin; these are
tabulated in later sections.

\begin{table}
\caption{Binding energies for various $S$-wave eigenstates of
the synthetic hamiltonian.}
\begin{center}
\begin{tabular}{|cclcc|} \hline
% & & & & \\
level &  binding energy & &level & binding energy \tstrut \\ \hline
1$S$ &  1.28711542\dsp\dsp     & & 6$S$ &   0.0155492598\dsp\tstrut   \\
2$S$ &  0.183325753\dsp        & & ... &  \\
3$S$ &  0.0703755485             && 10$S$ & 0.00534541931\\
4$S$ &  0.0371495726             && 20$S$ & 0.00129205010 \\
5$S$ &  0.0229268241     &&  & \\ \hline
\end{tabular}
\end{center}
\label{s-state-energies-table}
\end{table}

\begin{table}
\caption{$S$-wave phase shifts for the synthetic hamiltonian.
Phase shifts are computed for $r\!=\!50$.}
\begin{center}
\begin{tabular}{|cclcc|} \hline
energy &  phase shift & &energy & phase shift \tstrut \\ \hline
% & & & & \\
$10^{-10}$ &   $-$0.000421343353    && .03  & \msp 1.232867297 \tstrut \\
$10^{-5}$  &   $-$0.133227246       && .07  &    $-$0.579619620      \\
 .001      &   $-$1.319383451        && .1   &   $-$1.156444634       \\
 .003      &   \msp 0.900186195        && .3   &   $-$0.106466466      \\
 .007      &   $-$0.146570028       && .7   &    $-$1.457426179       \\
 .01       &   $-$0.654835316       && 1    &    \msp 1.160634967   \\
\hline
\end{tabular}
\end{center}
\label{phase-shift-table}
\end{table}

The numerical analysis required to generate such data is minimal.
The computation can easily be done using standard numerical analysis
packages or symbolic manipulation programs on a personal computer. 
I strongly suggest that you invent your own $V_s(\rv)$ and
subject your own synthetic data to the following analysis.

Our challenge is to design a simple theory that
reproduces our low-energy data with arbitrarily high precision. We must do
this using only the data and knowledge of the long-range structure of the
theory (that is, we are given $m$ and~$\alpha$).

\subsection{A Naive Approximation}\label{PTh-Sec}\label{naive-Sec}

A standard textbook approach to modelling our data is to approximate the
unknown short-range potential by a delta function, whose effect is computed
using first-order perturbation theory. The approximate hamiltonian is
\be
H_\appr = \frac{\pv^2}{2m} -\frac{\alpha}{r} + c\,\delta^3(\rv),
\ee
where $c$ is a parameter. Using first-order perturbation
theory, the energy levels in our approximate theory are
\bearray 
E_n^\appr &=& E_n^{\rm coul} + c \left| \psi^{\rm coul}_n(0) \right|^2 \nl
&=& -\frac{1}{2n^2} + c\,\frac{\delta_{l,0}}{\sqrt{\pi}\,n^3}. \label{approxE}
\eearray
Our approximation has only a single parameter, $c$. This we must determine
from the data. We do this by fitting formula~(\ref{approxE}) to 
our lowest-energy data, 
the 20$S$~binding energy; this implies $c\!=\!-.5963$.
We use the lowest-energy state because that is the state for which the
replacement $V_s(\rv) \to c\,\delta^3(\rv)$ is most accurate. 

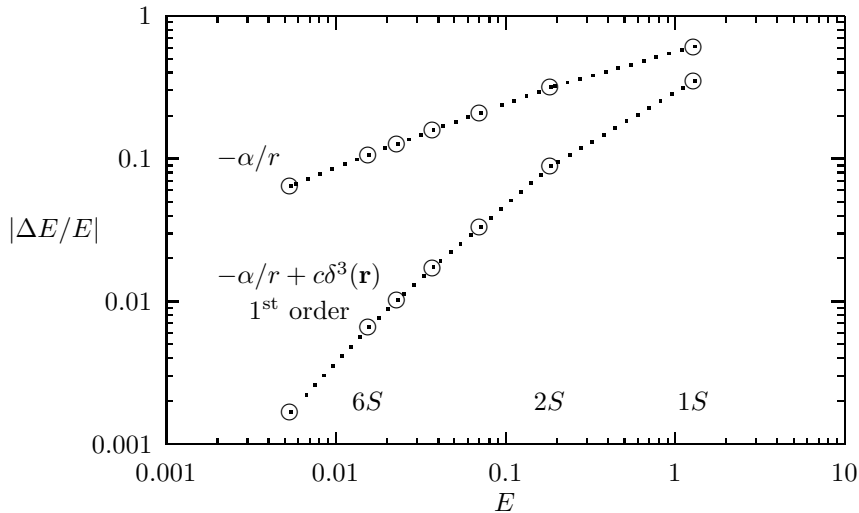
\begin{figure}
\begin{center}
% GNUPLOT: LaTeX picture
\setlength{\unitlength}{0.240900pt}
\ifx\plotpoint\undefined\newsavebox{\plotpoint}\fi
\sbox{\plotpoint}{\rule[-0.200pt]{0.400pt}{0.400pt}}%
\begin{picture}(1349,809)(0,0)
\font\gnuplot=cmr10 at 10pt
\gnuplot
\sbox{\plotpoint}{\rule[-0.200pt]{0.400pt}{0.400pt}}%
\put(220.0,113.0){\rule[-0.200pt]{4.818pt}{0.400pt}}
\put(198,113){\makebox(0,0)[r]{0.001}}
\put(1265.0,113.0){\rule[-0.200pt]{4.818pt}{0.400pt}}
\put(220.0,181.0){\rule[-0.200pt]{2.409pt}{0.400pt}}
\put(1275.0,181.0){\rule[-0.200pt]{2.409pt}{0.400pt}}
\put(220.0,220.0){\rule[-0.200pt]{2.409pt}{0.400pt}}
\put(1275.0,220.0){\rule[-0.200pt]{2.409pt}{0.400pt}}
\put(220.0,248.0){\rule[-0.200pt]{2.409pt}{0.400pt}}
\put(1275.0,248.0){\rule[-0.200pt]{2.409pt}{0.400pt}}
\put(220.0,270.0){\rule[-0.200pt]{2.409pt}{0.400pt}}
\put(1275.0,270.0){\rule[-0.200pt]{2.409pt}{0.400pt}}
\put(220.0,288.0){\rule[-0.200pt]{2.409pt}{0.400pt}}
\put(1275.0,288.0){\rule[-0.200pt]{2.409pt}{0.400pt}}
\put(220.0,303.0){\rule[-0.200pt]{2.409pt}{0.400pt}}
\put(1275.0,303.0){\rule[-0.200pt]{2.409pt}{0.400pt}}
\put(220.0,316.0){\rule[-0.200pt]{2.409pt}{0.400pt}}
\put(1275.0,316.0){\rule[-0.200pt]{2.409pt}{0.400pt}}
\put(220.0,327.0){\rule[-0.200pt]{2.409pt}{0.400pt}}
\put(1275.0,327.0){\rule[-0.200pt]{2.409pt}{0.400pt}}
\put(220.0,337.0){\rule[-0.200pt]{4.818pt}{0.400pt}}
\put(198,337){\makebox(0,0)[r]{0.01}}
\put(1265.0,337.0){\rule[-0.200pt]{4.818pt}{0.400pt}}
\put(220.0,405.0){\rule[-0.200pt]{2.409pt}{0.400pt}}
\put(1275.0,405.0){\rule[-0.200pt]{2.409pt}{0.400pt}}
\put(220.0,444.0){\rule[-0.200pt]{2.409pt}{0.400pt}}
\put(1275.0,444.0){\rule[-0.200pt]{2.409pt}{0.400pt}}
\put(220.0,472.0){\rule[-0.200pt]{2.409pt}{0.400pt}}
\put(1275.0,472.0){\rule[-0.200pt]{2.409pt}{0.400pt}}
\put(220.0,494.0){\rule[-0.200pt]{2.409pt}{0.400pt}}
\put(1275.0,494.0){\rule[-0.200pt]{2.409pt}{0.400pt}}
\put(220.0,512.0){\rule[-0.200pt]{2.409pt}{0.400pt}}
\put(1275.0,512.0){\rule[-0.200pt]{2.409pt}{0.400pt}}
\put(220.0,527.0){\rule[-0.200pt]{2.409pt}{0.400pt}}
\put(1275.0,527.0){\rule[-0.200pt]{2.409pt}{0.400pt}}
\put(220.0,540.0){\rule[-0.200pt]{2.409pt}{0.400pt}}
\put(1275.0,540.0){\rule[-0.200pt]{2.409pt}{0.400pt}}
\put(220.0,551.0){\rule[-0.200pt]{2.409pt}{0.400pt}}
\put(1275.0,551.0){\rule[-0.200pt]{2.409pt}{0.400pt}}
\put(220.0,562.0){\rule[-0.200pt]{4.818pt}{0.400pt}}
\put(198,562){\makebox(0,0)[r]{0.1}}
\put(1265.0,562.0){\rule[-0.200pt]{4.818pt}{0.400pt}}
\put(220.0,629.0){\rule[-0.200pt]{2.409pt}{0.400pt}}
\put(1275.0,629.0){\rule[-0.200pt]{2.409pt}{0.400pt}}
\put(220.0,669.0){\rule[-0.200pt]{2.409pt}{0.400pt}}
\put(1275.0,669.0){\rule[-0.200pt]{2.409pt}{0.400pt}}
\put(220.0,697.0){\rule[-0.200pt]{2.409pt}{0.400pt}}
\put(1275.0,697.0){\rule[-0.200pt]{2.409pt}{0.400pt}}
\put(220.0,718.0){\rule[-0.200pt]{2.409pt}{0.400pt}}
\put(1275.0,718.0){\rule[-0.200pt]{2.409pt}{0.400pt}}
\put(220.0,736.0){\rule[-0.200pt]{2.409pt}{0.400pt}}
\put(1275.0,736.0){\rule[-0.200pt]{2.409pt}{0.400pt}}
\put(220.0,751.0){\rule[-0.200pt]{2.409pt}{0.400pt}}
\put(1275.0,751.0){\rule[-0.200pt]{2.409pt}{0.400pt}}
\put(220.0,764.0){\rule[-0.200pt]{2.409pt}{0.400pt}}
\put(1275.0,764.0){\rule[-0.200pt]{2.409pt}{0.400pt}}
\put(220.0,776.0){\rule[-0.200pt]{2.409pt}{0.400pt}}
\put(1275.0,776.0){\rule[-0.200pt]{2.409pt}{0.400pt}}
\put(220.0,786.0){\rule[-0.200pt]{4.818pt}{0.400pt}}
\put(198,786){\makebox(0,0)[r]{1}}
\put(1265.0,786.0){\rule[-0.200pt]{4.818pt}{0.400pt}}
\put(220.0,113.0){\rule[-0.200pt]{0.400pt}{4.818pt}}
\put(220,68){\makebox(0,0){0.001}}
\put(220.0,766.0){\rule[-0.200pt]{0.400pt}{4.818pt}}
\put(300.0,113.0){\rule[-0.200pt]{0.400pt}{2.409pt}}
\put(300.0,776.0){\rule[-0.200pt]{0.400pt}{2.409pt}}
\put(347.0,113.0){\rule[-0.200pt]{0.400pt}{2.409pt}}
\put(347.0,776.0){\rule[-0.200pt]{0.400pt}{2.409pt}}
\put(380.0,113.0){\rule[-0.200pt]{0.400pt}{2.409pt}}
\put(380.0,776.0){\rule[-0.200pt]{0.400pt}{2.409pt}}
\put(406.0,113.0){\rule[-0.200pt]{0.400pt}{2.409pt}}
\put(406.0,776.0){\rule[-0.200pt]{0.400pt}{2.409pt}}
\put(427.0,113.0){\rule[-0.200pt]{0.400pt}{2.409pt}}
\put(427.0,776.0){\rule[-0.200pt]{0.400pt}{2.409pt}}
\put(445.0,113.0){\rule[-0.200pt]{0.400pt}{2.409pt}}
\put(445.0,776.0){\rule[-0.200pt]{0.400pt}{2.409pt}}
\put(460.0,113.0){\rule[-0.200pt]{0.400pt}{2.409pt}}
\put(460.0,776.0){\rule[-0.200pt]{0.400pt}{2.409pt}}
\put(474.0,113.0){\rule[-0.200pt]{0.400pt}{2.409pt}}
\put(474.0,776.0){\rule[-0.200pt]{0.400pt}{2.409pt}}
\put(486.0,113.0){\rule[-0.200pt]{0.400pt}{4.818pt}}
\put(486,68){\makebox(0,0){0.01}}
\put(486.0,766.0){\rule[-0.200pt]{0.400pt}{4.818pt}}
\put(566.0,113.0){\rule[-0.200pt]{0.400pt}{2.409pt}}
\put(566.0,776.0){\rule[-0.200pt]{0.400pt}{2.409pt}}
\put(613.0,113.0){\rule[-0.200pt]{0.400pt}{2.409pt}}
\put(613.0,776.0){\rule[-0.200pt]{0.400pt}{2.409pt}}
\put(647.0,113.0){\rule[-0.200pt]{0.400pt}{2.409pt}}
\put(647.0,776.0){\rule[-0.200pt]{0.400pt}{2.409pt}}
\put(672.0,113.0){\rule[-0.200pt]{0.400pt}{2.409pt}}
\put(672.0,776.0){\rule[-0.200pt]{0.400pt}{2.409pt}}
\put(693.0,113.0){\rule[-0.200pt]{0.400pt}{2.409pt}}
\put(693.0,776.0){\rule[-0.200pt]{0.400pt}{2.409pt}}
\put(711.0,113.0){\rule[-0.200pt]{0.400pt}{2.409pt}}
\put(711.0,776.0){\rule[-0.200pt]{0.400pt}{2.409pt}}
\put(727.0,113.0){\rule[-0.200pt]{0.400pt}{2.409pt}}
\put(727.0,776.0){\rule[-0.200pt]{0.400pt}{2.409pt}}
\put(740.0,113.0){\rule[-0.200pt]{0.400pt}{2.409pt}}
\put(740.0,776.0){\rule[-0.200pt]{0.400pt}{2.409pt}}
\put(753.0,113.0){\rule[-0.200pt]{0.400pt}{4.818pt}}
\put(753,68){\makebox(0,0){0.1}}
\put(753.0,766.0){\rule[-0.200pt]{0.400pt}{4.818pt}}
\put(833.0,113.0){\rule[-0.200pt]{0.400pt}{2.409pt}}
\put(833.0,776.0){\rule[-0.200pt]{0.400pt}{2.409pt}}
\put(880.0,113.0){\rule[-0.200pt]{0.400pt}{2.409pt}}
\put(880.0,776.0){\rule[-0.200pt]{0.400pt}{2.409pt}}
\put(913.0,113.0){\rule[-0.200pt]{0.400pt}{2.409pt}}
\put(913.0,776.0){\rule[-0.200pt]{0.400pt}{2.409pt}}
\put(939.0,113.0){\rule[-0.200pt]{0.400pt}{2.409pt}}
\put(939.0,776.0){\rule[-0.200pt]{0.400pt}{2.409pt}}
\put(960.0,113.0){\rule[-0.200pt]{0.400pt}{2.409pt}}
\put(960.0,776.0){\rule[-0.200pt]{0.400pt}{2.409pt}}
\put(978.0,113.0){\rule[-0.200pt]{0.400pt}{2.409pt}}
\put(978.0,776.0){\rule[-0.200pt]{0.400pt}{2.409pt}}
\put(993.0,113.0){\rule[-0.200pt]{0.400pt}{2.409pt}}
\put(993.0,776.0){\rule[-0.200pt]{0.400pt}{2.409pt}}
\put(1007.0,113.0){\rule[-0.200pt]{0.400pt}{2.409pt}}
\put(1007.0,776.0){\rule[-0.200pt]{0.400pt}{2.409pt}}
\put(1019.0,113.0){\rule[-0.200pt]{0.400pt}{4.818pt}}
\put(1019,68){\makebox(0,0){1}}
\put(1019.0,766.0){\rule[-0.200pt]{0.400pt}{4.818pt}}
\put(1099.0,113.0){\rule[-0.200pt]{0.400pt}{2.409pt}}
\put(1099.0,776.0){\rule[-0.200pt]{0.400pt}{2.409pt}}
\put(1146.0,113.0){\rule[-0.200pt]{0.400pt}{2.409pt}}
\put(1146.0,776.0){\rule[-0.200pt]{0.400pt}{2.409pt}}
\put(1179.0,113.0){\rule[-0.200pt]{0.400pt}{2.409pt}}
\put(1179.0,776.0){\rule[-0.200pt]{0.400pt}{2.409pt}}
\put(1205.0,113.0){\rule[-0.200pt]{0.400pt}{2.409pt}}
\put(1205.0,776.0){\rule[-0.200pt]{0.400pt}{2.409pt}}
\put(1226.0,113.0){\rule[-0.200pt]{0.400pt}{2.409pt}}
\put(1226.0,776.0){\rule[-0.200pt]{0.400pt}{2.409pt}}
\put(1244.0,113.0){\rule[-0.200pt]{0.400pt}{2.409pt}}
\put(1244.0,776.0){\rule[-0.200pt]{0.400pt}{2.409pt}}
\put(1259.0,113.0){\rule[-0.200pt]{0.400pt}{2.409pt}}
\put(1259.0,776.0){\rule[-0.200pt]{0.400pt}{2.409pt}}
\put(1273.0,113.0){\rule[-0.200pt]{0.400pt}{2.409pt}}
\put(1273.0,776.0){\rule[-0.200pt]{0.400pt}{2.409pt}}
\put(1285.0,113.0){\rule[-0.200pt]{0.400pt}{4.818pt}}
\put(1285,68){\makebox(0,0){10}}
\put(1285.0,766.0){\rule[-0.200pt]{0.400pt}{4.818pt}}
\put(220.0,113.0){\rule[-0.200pt]{256.558pt}{0.400pt}}
\put(1285.0,113.0){\rule[-0.200pt]{0.400pt}{162.126pt}}
\put(220.0,786.0){\rule[-0.200pt]{256.558pt}{0.400pt}}
\put(45,449){\makebox(0,0){$|\Delta E/E|$}}
\put(752,23){\makebox(0,0){$E$}}
\put(300,562){\makebox(0,0)[l]{$-\alpha/r$}}
\put(300,355){\makebox(0,0)[l]{\shortstack{$-\alpha/r+c\delta^3(\rv)$ \\ $1^{\rm st}$ order}}}
\put(1048,181){\makebox(0,0){$1S$}}
\put(822,181){\makebox(0,0){$2S$}}
\put(537,181){\makebox(0,0){$6S$}}
\put(220.0,113.0){\rule[-0.200pt]{0.400pt}{162.126pt}}
\sbox{\plotpoint}{\rule[-0.500pt]{1.000pt}{1.000pt}}%
\put(1048,738){\usebox{\plotpoint}}
\multiput(1048,738)(-19.964,-5.679){12}{\usebox{\plotpoint}}
\multiput(823,674)(-19.526,-7.037){5}{\usebox{\plotpoint}}
\multiput(712,634)(-19.498,-7.114){4}{\usebox{\plotpoint}}
\multiput(638,607)(-19.318,-7.589){3}{\usebox{\plotpoint}}
\multiput(582,585)(-19.416,-7.335){2}{\usebox{\plotpoint}}
\multiput(537,568)(-19.282,-7.681){7}{\usebox{\plotpoint}}
\put(414,519){\usebox{\plotpoint}}
\put(1048,738){\circle{24}}
\put(823,674){\circle{24}}
\put(712,634){\circle{24}}
\put(638,607){\circle{24}}
\put(582,585){\circle{24}}
\put(537,568){\circle{24}}
\put(414,519){\circle{24}}
\put(1048,684){\usebox{\plotpoint}}
\multiput(1048,684)(-17.833,-10.620){13}{\usebox{\plotpoint}}
\multiput(823,550)(-15.769,-13.496){7}{\usebox{\plotpoint}}
\multiput(712,455)(-15.699,-13.577){5}{\usebox{\plotpoint}}
\multiput(638,391)(-15.345,-13.975){4}{\usebox{\plotpoint}}
\multiput(582,340)(-15.006,-14.339){3}{\usebox{\plotpoint}}
\multiput(537,297)(-14.092,-15.238){8}{\usebox{\plotpoint}}
\put(414,164){\usebox{\plotpoint}}
\put(1048,684){\circle{24}}
\put(823,550){\circle{24}}
\put(712,455){\circle{24}}
\put(638,391){\circle{24}}
\put(582,340){\circle{24}}
\put(537,297){\circle{24}}
\put(414,164){\circle{24}}
\end{picture}
\end{center}
\caption{Relative errors in the $S$-wave 
binding energies are plotted versus binding energy for the
Coulomb theory, and for the Coulomb theory augmented with a delta function
in first-order perturbation theory.}
\label{bindingE-fig}
\end{figure}

In Figure~\ref{bindingE-fig} I show the relative errors in several
$S$-wave
binding energies, plotted versus binding energy, obtained 
using just a Coulomb
potential ($c\!=\!0$) and using our approximate formula, \eq{approxE}, with
$c\!=\!-.5963$. Both approximations become more accurate as the binding energy
decreases, but adding the delta function in first order gives substantially
better results. Our approximation to $V_s$ is quite successful.

The limitations of this approximation become evident if we seek greater
accuracy. We have made two approximations. First we used only
first-order perturbation theory. 
There will be large contributions from second and higher orders in
perturbation theory if the short-range potential is strong. 
So we might wish to compute the second-order
contribution to $E_n^\appr$:
\be
\sum_{m\ne n} \frac{\langle n | c\,\delta^3(\rv) | m\rangle\,\,
\langle m | c\,\delta^3(\rv) | n\rangle}{E_n-E_m}. 
\ee
Unfortunately this expression gives an infinite shift; the sum over
scattering eigenstates diverges as scattering 
momentum $\pv\!\to\!\infty$. The delta
function is too singular to be meaningful beyond first-order perturbation
theory.

The second approximation is replacing $V_s$ by a delta function.
The nature of this approximation can be appreciated by 
Fourier transforming~$V_s$. Since $V_s$ has a very short range,
its transform~$v_s(q^2)$ 
depends only weakly on the momentum transfer $q$. Thus
we might try approximating $v_s$ by the first
few terms in its Taylor expansion when calculating 
low-energy matrix elements:
\be
v_s(q^2) =  v_s(0)  + q^2\, v_s^\prime(0) +\cdots.
\ee
This is an economical parameterization of the short-distance dynamics 
since it replaces a function, $v_s(q^2)$, by a small set of numbers:
$v_s(0)$, $v_s^\prime(0)\ldots$. Keeping just the first term is equivalent
to approximating $V_s(r)$ by a delta function. The additional terms
correspond to derivatives of a delta function, and correct for the fact that
the range of $V_s$ is not infinitely small\,---\,the expansion is 
in powers of~$q$ times the range. This suggests that we might
improve our approximation by adding a second parameter, again to be tuned to
reproduce data:
\be \label{Vs-expansion}
V_s(\rv) \to c\,\delta^3(\rv) + d\,\nabla^2\delta^3(\rv).
\ee
Unfortunately the additional first-order shift in $E_n^\appr$,
$\langle n|d\,\nabla^2\delta^3(\rv)|n\rangle$, is infinite.
Again the delta function is too singular at short
distances for such matrix elements to make sense.

Our first attempt to refine the approximate model has failed. We 
replaced the true, but unknown short-distance behavior of the potential by
behavior that is pathologically singular as $r\!\to\!0$. Conventional wisdom,
often to be seen in books and even recent conference proceedings, is that
our approximations must be abandoned once infinities
appear, and that only knowledge of the true
potential can get us beyond such difficulties. This wisdom is incorrect, as
I now show.

\subsection{Effective Theory}\label{eff-theory-sec}
The infinities discussed in the previous section are exactly analogous
to those found in relativistic quantum field theories. Among other
things, they indicate that even low-energy processes are sensitive to
physics at short distances. 
Modern renormalization theory, however, 
tells us that the low-energy (infrared)
behavior of a theory is independent of the {\em details\/} 
of the short-distance
(ultraviolet) dynamics. Insensitivity
to the short-distance details 
means that there are infinitely many theories that have the
same low-energy behavior; all are identical at large distances but each
is quite different from the others at short distances. 
Thus we can generally
replace the short-distance dynamics of a theory by something different, and
perhaps simpler, without changing the low-energy behavior.

The freedom to redesign at short-distances allows us to create
effective theories that model arbitrary low-energy data sets with
arbitrary precision. There are three steps:
\begin{enumerate}
\item Incorporate the correct long-range behavior: The long-range behavior of
the underlying theory must be known, and it must be built into the effective
theory.

\item Introduce an ultraviolet cutoff to exclude high-momentum states, 
or, equivalently, to soften the short-distance behavior: The cutoff
has two effects. First it
excludes high-momentum states, which are sensitive to the unknown
short-distance dynamics; only states that we understand are retained. 
Second, it makes
all interactions regular at $r\!=\!0$, thereby avoiding the infinities
that plague the naive approach of the previous section.

\item Add local correction terms to the effective hamiltonian: These 
mimic the effects of the high-momentum states excluded by the cutoff in
step~2. Each correction term consists of a theory-specific coupling constant,
a number, multiplied by a  theory-independent local
operator. The correction terms systematically remove dependence on the
cutoff.
Their locality implies that only a finite number of corrections is
needed to achieve any given level of precision.
\end{enumerate}
We now apply this algorithm to design an effective hamiltonian that
describes our data.

To begin, our effective theory is specified by a hamiltonian,
\be
H_\eff = \frac{\pv^2}{2m} + V_\eff(\rv),
\ee
where the effective potential, $V_\eff$, must become Coulombic at large~$r$:
$V_\eff(r)\!\to\!-\alpha/r$, with $\alpha\!=\!1$, for  large~$r$. We also
need an ultraviolet cutoff. I chose to introduce a cutoff into the Coulomb
potential through its Fourier transform:
\bearray
\frac{1}{r} &\stackrel{{\rm F.T.}}{\longrightarrow}& \frac{4\pi}{q^2} \nl
&\stackrel{{\rm cutoff}}{\longrightarrow} & 
  \frac{4\pi}{q^2}\,\e^{-q^2a^2/2} \nl
&\stackrel{{\rm F.T.}}{\longrightarrow}& \frac{\erf(r/\sqrt{2}a)}{r},
\eearray
where 
\be
\erf(x) = \frac{2}{\sqrt{\pi}}\int_{0}^x \e^{-t^2} dt
\ee 
is the standard error function. The new, regulated
potential is finite at $r\!=\!0$, but goes to $1/r$ for $r\!\gg\!a$.
It inhibits momentum transfers of order 
$\Lambda\!\equiv\!1/a$ or larger. The exact form of the cutoff is irrelevant;
there are infinitely many choices all of which give similar
results.

It is not really necessary to regulate the Coulomb
potential since it is only mildly singular at the
origin. Nevertheless, 
in many applications, it is still a good idea. For example 
the potential and the wavefunctions are analytic at
$r\!=\!0$ if we use the 
regulator described above. 
Numerical techniques are often much more accurate or convergent
for analytic functions.

Short-distance dynamics is explicitly excluded from the effective
theory by the cutoff. We mimic the effects of the true 
short-distance structure by adding local correction terms. As discussed in the
previous section, the low-momentum behavior of any short-range potential is
efficiently described in terms of the Taylor expansion in
momentum space.  Transforming back to
coordinate space gives a series  that is a polynomial in the momentum
operator~$\pv\!\equiv\!\nablav/{\rm i}$ multiplied by 
a delta function. We need an
ultraviolet cutoff to avoid infinities, 
and therefore we smear the delta function over a
volume whose radius is approximately the cutoff distance~$a$. I chose a
smeared delta function defined by
\be
\delta^3_a(\rv) \equiv \frac{\e^{-r^2/2a^2}}{(2\pi)^{3/2}\,a^3},
\ee
but, again, the detailed 
structure of this function is irrelevant; other choices
work just as well.

Remarkably, the structure of the
correction terms is now completely determined, even though we have yet
to examine the data. The effective potential must have the form
\bearray 
V_\eff(\rv) &=& -\frac{\alpha}{r}\,\erf(r/\sqrt{2}a) \nl
&& + c\,a^2\,\delta^3_a(\rv) \nl
&& + d_1\,a^4\,\nabla^2\delta_a^3(\rv) 
   + d_2\,a^4\,\nablav\!\cdot\!\delta_a^3(\rv)\nablav \nl
&& + \cdots \nl
&& + g\,a^{n+2}\,\nabla^n \delta_a^3(\rv) \nl
&& + \cdots,
\label{Veff}
\eearray
where coupling constants $c$, $d_1$, $d_2$\ldots are
dimensionless. It consists of a long-range part together with a series
of local ``contact'' potentials. The contact terms are
indistinguishable, to a low-momentum particle, from the the true
short-distance potential, provided the coupling constants are properly tuned.
The potential is nonrenormalizable in the traditional sense, but that
is not a problem here since the cutoff prevents infinities.

Generally the effective potential reflects the symmetries of the true
theory. Here our effective potential is rotationally invariant because
our data are. When the data are not rotationally invariant, 
additional terms like
$a^3\,{\bf g}\cdot\nablav\,\delta_a^3(\rv)$ must be included in
$V_\eff$, where now
the coupling constants include vectors, tensors\,\ldots\,that characterize the
rotational asymmetries of the true theory. 

One might worry that our Taylor expansion of the short-distance
dynamics would fail when we computed physical quantities like the
scattering amplitude. This is because high-momentum states affect even 
low-energy processes through quantum fluctuations. 
For example, in the true theory  
the scattering amplitude is
\be
\langle f|T(E)|i\rangle = \langle f|V|i\rangle + \sum_n
\frac{\langle f|V|n\rangle\,\langle n|V|i\rangle}{E-E_n}+\cdots.
\ee
The sums over intermediate states in second-order and beyond
include states with arbitrarily large momentum. A momentum-space Taylor
expansion of the potential would not converge in matrix elements involving
such states. Furthermore when we replace the exact potential~$V$ by our
effective potential~$V_\eff$, our ultraviolet cutoff in effect limits the
sums to states with momenta less than the cutoff~$1/a$; contributions
from high-momentum intermediate states apparently are completely absent from
our effective theory. Our effective theory is saved by the fact that
high-momentum intermediate states are necessarily
highly virtual if, as we assume, the initial and final states have a low
energy.  By the uncertainty principle, such states cannot propagate
for long times or over large distances. Thus any contribution from these
states is very local and can be incorporated into the effective theory using 
the same set of (smeared) delta-function potentials already included
in~$V_\eff$. In this way,
the high-momentum states that are
explicitly excluded, or badly distorted, by the cutoff are included
implicitly through the coupling constants. Note that this means the
coupling constants in our effective theory depend nonlinearly on the true
potential~$V(r)$.

While the true theory is obviously independent of the value of
the cutoff~$a$, results computed in the 
effective theory are only approximately $a$~independent.
The residual $a$~dependence in such a result
is typically a power series in $qa$, where $q$ is the
characteristic momentum associated with the process under study (for example,
the initial momentum or the momentum transfer in a scattering
amplitude).
The contact terms remove these $a$-dependent errors order-by-order
in~$a$. Thus, for example, 
the $\order(a^2)$~term in~$V_\eff$ removes errors of
order~$(qa)^2$ and is the most important correction. Errors of
order~$a^4$ are removed by the $\order(a^4)$ terms, of which there may
be only two since there are only two independent ways of combining two
momentum operators $\nablav/{\rm i}$ with the smeared delta function.
Obviously only a finite number of contact terms is needed to remove
errors through any finite order in~$a$.

The coupling constants  vary with~$a$
since  they must account for quantum fluctuations excluded by the
cutoff. More or less is excluded as $a$~is increased or decreased, 
and therefore the coupling constants
must be adjusted to compensate. They are said to be ``running coupling
constants.'' When the short-distance potential is
weak, the coupling to high-momentum intermediate states is weak and
the coupling constants change only slowly
with~$a$.\footnote{This behavior is modified if the cutoff
distance~$a$ is very large. When~$a$ is larger than the range
of the {\em long-range\/} potential, or when there is no long-range
potential, the coefficients of the contact operators tend to go
to a constant. Thus for example, $c$ in $V_\eff$ will decrease like
$1/a^2$ as $a$~increases 
so that the coefficient  of the delta function operator
becomes constant.}
On the other hand, the coupling constants
are often very cutoff dependent when the
short-distance interactions are strong. The dependence of the
effective theory on its coupling constants becomes highly nonlinear in
this case. In fact this behavior can be so
strong that the relative importance of different contact terms can
change: an $a^4$ operator can act more like an $a^2$~operator, and
vice versa. The physical systems I discuss here do not suffer from
this complication, although it is easy to study the phenomenon 
by modifying the synthetic data appropriately. 

Highly nonlinear behavior also results when the cutoff distance~$a$ is made
too small. From the previous section we know that taking $a\!\to\!0$
is a bad idea. In general it makes little sense to reduce~$a$ below
the range~$r_s$ of the true potential. 
By assumption, our data involves energies that are too low\,---\,wavelengths
that are too long\,---\,to probe the true structure of the theory at distances
as small as $r_s$. When $a\!<\!r_s$, high-momentum states are included
that are sensitive to structure at distances smaller than~$r_s$.
But the structure they see there is almost certainly wrong. Thus
taking~$a$ smaller than~$r_s$ cannot improve results obtained from the
effective theory. In fact, as the nonlinearities develop for
small~$a$'s, results often degrade, or, in more extreme cases, the
theory may become unstable or untunable.

Finally I should comment on the physical significance of the cutoff.
A common practice in applications of potential models, for example to
nuclear or atomic physics, is to use a cutoff like our's 
to account for some physical effect, for example finite nuclear
size. Then one must worry about whether the functional form of the cutoff is
physically correct. 
We are {\em not\/} doing this here. Our cutoff is
only a cutoff; physical effects like finite nuclear size are incorporated
systematically through the  contact terms. We need
never worry, for example, about what the true charge distribution is inside a
nucleus. Indeed this is the great strength of the effective theory.
The effective theory gives us a simple, {\em universal\/} parameterization for
the effects of short-distance structure. The form of the contact terms is
theory independent. Only the numerical values of the
coupling constants $c$, $d_1$\ldots are theory specific; for our
(low-energy) purposes, everything that we need to know about the
short-distance dynamics of the true theory is contained in these coupling
constants. As discussed earlier, this situation is conceptually 
similar to a multipole
analysis of a classical field. The couplings here are the analogues of the
multipole moments, while the delta-function potentials are analogous to the
multipoles that generate the field.

The exact form of the cutoff, and the exact definition of the
smeared delta function are irrelevant. I encourage you to make up
your own versions of these and repeat the analysis given here. You will get
similar results for binding energies and phase shifts, although all of
your coupling constants will probably be different, and possibly quite
different.

\subsection{Tuning the Effective Theory; Results}

We now tune the parameters of 
our effective theory so that it reproduces our low-energy data through
order~$a^4$.
For simplicity, we examine only $S$-wave properties. Thus we need only
the $c$~and $d_1$~terms in $V_\eff$, \eq{Veff}; we can drop
the $d_2$~term in $V_\eff(\rv)$ since it is important only for $P$-wave
states.\footnote{It is obvious that the $d_2$~term
couples only to $P$-waves in the limit $a\!\to\!0$. When $a$ is
nonzero, however,
this term has a small residual coupling to $S$-waves. This results
in an interaction of order $a^6$, which may be ignored to the order we are
working here. One can easily design contact terms that contribute only to
a single channel in orbital angular momentum. For example, our smeared delta
function could be replaced by a local potential that is separable.}

To generate the results discussed here I first chose a reasonable value for
the cutoff distance: to begin with, 
$a\!=\!1$, the Bohr radius for the Coulombic part of
the interaction. Then I varied the coupling constants~$c$
and~$d_1$ until the $S$-wave phase shifts from the effective hamiltonian
agreed with the data at energies
$E=10^{-5}$ and~$10^{-10}$. I found $c/4\pi = -3.18$ and
$d/4\pi = -0.199$. Having found
the coupling constants,  I then generated binding energies,
phase shifts, and matrix elements using the
effective theory. I also generated results with only the $a^2$ correction
($d_1\!=\!0$); the effective theory 
requires a different value of $c$ in this case,
since the two contact terms affect each other.

I tuned the couplings using phase shifts at very low
energies. This was 
to minimize the effect of the $a^6$ errors, which arise because we
truncated our effective potential at order~$a^4$. 
These ``truncation'' errors
are smallest for the lowest-energy data, since they are generally
proportional to $qa$ raised to some positive power. In general
one should use the most infrared data available when tuning. Alternatively
one might attempt a global fit to all of the data; but then it is crucial to
give greater weight in the fit to low-energy data than to high-energy data:
some estimate of the $(qa)^n$ error due to truncation must be added to the
experimental errors used to weight the data in the fit.\cite{angelos}

So long as $a\!>\!0$, $V_\eff(\rv)$ is a simple, nonsingular
function of~$r$. Consequently the effective theory is simple to solve
numerically. To
compute the results here, I reused the  code that I wrote to generate
the synthetic data, but with the true potential~$V(\rv)$ replaced by the
effective potential~$V_\eff(\rv)$. The effective theory is no harder to solve
than the other; 
in particular, because of the cutoff, there are no infinities.

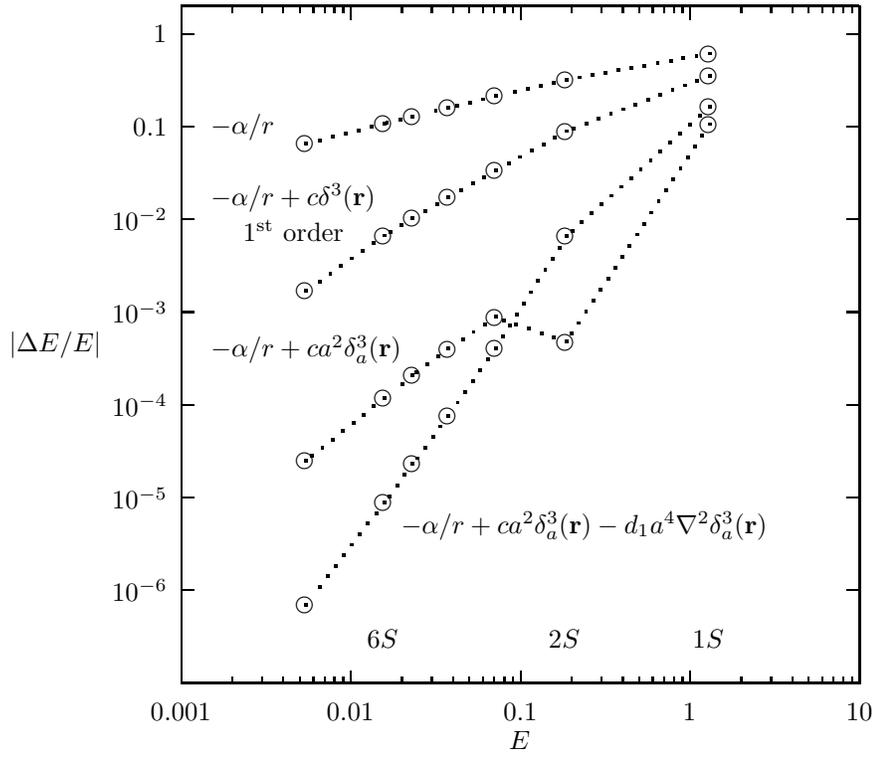
\begin{figure}
\begin{center}
% GNUPLOT: LaTeX picture
\setlength{\unitlength}{0.240900pt}
\ifx\plotpoint\undefined\newsavebox{\plotpoint}\fi
\sbox{\plotpoint}{\rule[-0.200pt]{0.400pt}{0.400pt}}%
\begin{picture}(1349,1200)(0,0)
\font\gnuplot=cmr10 at 10pt
\gnuplot
\sbox{\plotpoint}{\rule[-0.200pt]{0.400pt}{0.400pt}}%
\put(220.0,259.0){\rule[-0.200pt]{4.818pt}{0.400pt}}
\put(198,259){\makebox(0,0)[r]{$10^{-6}$}}
\put(1265.0,259.0){\rule[-0.200pt]{4.818pt}{0.400pt}}
\put(220.0,404.0){\rule[-0.200pt]{4.818pt}{0.400pt}}
\put(198,404){\makebox(0,0)[r]{$10^{-5}$}}
\put(1265.0,404.0){\rule[-0.200pt]{4.818pt}{0.400pt}}
\put(220.0,550.0){\rule[-0.200pt]{4.818pt}{0.400pt}}
\put(198,550){\makebox(0,0)[r]{$10^{-4}$}}
\put(1265.0,550.0){\rule[-0.200pt]{4.818pt}{0.400pt}}
\put(220.0,696.0){\rule[-0.200pt]{4.818pt}{0.400pt}}
\put(198,696){\makebox(0,0)[r]{$10^{-3}$}}
\put(1265.0,696.0){\rule[-0.200pt]{4.818pt}{0.400pt}}
\put(220.0,842.0){\rule[-0.200pt]{4.818pt}{0.400pt}}
\put(198,842){\makebox(0,0)[r]{$10^{-2}$}}
\put(1265.0,842.0){\rule[-0.200pt]{4.818pt}{0.400pt}}
\put(220.0,987.0){\rule[-0.200pt]{4.818pt}{0.400pt}}
\put(198,987){\makebox(0,0)[r]{$0.1$}}
\put(1265.0,987.0){\rule[-0.200pt]{4.818pt}{0.400pt}}
\put(220.0,1133.0){\rule[-0.200pt]{4.818pt}{0.400pt}}
\put(198,1133){\makebox(0,0)[r]{$1$}}
\put(1265.0,1133.0){\rule[-0.200pt]{4.818pt}{0.400pt}}
\put(220.0,113.0){\rule[-0.200pt]{0.400pt}{4.818pt}}
\put(220,68){\makebox(0,0){0.001}}
\put(220.0,1157.0){\rule[-0.200pt]{0.400pt}{4.818pt}}
\put(300.0,113.0){\rule[-0.200pt]{0.400pt}{2.409pt}}
\put(300.0,1167.0){\rule[-0.200pt]{0.400pt}{2.409pt}}
\put(347.0,113.0){\rule[-0.200pt]{0.400pt}{2.409pt}}
\put(347.0,1167.0){\rule[-0.200pt]{0.400pt}{2.409pt}}
\put(380.0,113.0){\rule[-0.200pt]{0.400pt}{2.409pt}}
\put(380.0,1167.0){\rule[-0.200pt]{0.400pt}{2.409pt}}
\put(406.0,113.0){\rule[-0.200pt]{0.400pt}{2.409pt}}
\put(406.0,1167.0){\rule[-0.200pt]{0.400pt}{2.409pt}}
\put(427.0,113.0){\rule[-0.200pt]{0.400pt}{2.409pt}}
\put(427.0,1167.0){\rule[-0.200pt]{0.400pt}{2.409pt}}
\put(445.0,113.0){\rule[-0.200pt]{0.400pt}{2.409pt}}
\put(445.0,1167.0){\rule[-0.200pt]{0.400pt}{2.409pt}}
\put(460.0,113.0){\rule[-0.200pt]{0.400pt}{2.409pt}}
\put(460.0,1167.0){\rule[-0.200pt]{0.400pt}{2.409pt}}
\put(474.0,113.0){\rule[-0.200pt]{0.400pt}{2.409pt}}
\put(474.0,1167.0){\rule[-0.200pt]{0.400pt}{2.409pt}}
\put(486.0,113.0){\rule[-0.200pt]{0.400pt}{4.818pt}}
\put(486,68){\makebox(0,0){0.01}}
\put(486.0,1157.0){\rule[-0.200pt]{0.400pt}{4.818pt}}
\put(566.0,113.0){\rule[-0.200pt]{0.400pt}{2.409pt}}
\put(566.0,1167.0){\rule[-0.200pt]{0.400pt}{2.409pt}}
\put(613.0,113.0){\rule[-0.200pt]{0.400pt}{2.409pt}}
\put(613.0,1167.0){\rule[-0.200pt]{0.400pt}{2.409pt}}
\put(647.0,113.0){\rule[-0.200pt]{0.400pt}{2.409pt}}
\put(647.0,1167.0){\rule[-0.200pt]{0.400pt}{2.409pt}}
\put(672.0,113.0){\rule[-0.200pt]{0.400pt}{2.409pt}}
\put(672.0,1167.0){\rule[-0.200pt]{0.400pt}{2.409pt}}
\put(693.0,113.0){\rule[-0.200pt]{0.400pt}{2.409pt}}
\put(693.0,1167.0){\rule[-0.200pt]{0.400pt}{2.409pt}}
\put(711.0,113.0){\rule[-0.200pt]{0.400pt}{2.409pt}}
\put(711.0,1167.0){\rule[-0.200pt]{0.400pt}{2.409pt}}
\put(727.0,113.0){\rule[-0.200pt]{0.400pt}{2.409pt}}
\put(727.0,1167.0){\rule[-0.200pt]{0.400pt}{2.409pt}}
\put(740.0,113.0){\rule[-0.200pt]{0.400pt}{2.409pt}}
\put(740.0,1167.0){\rule[-0.200pt]{0.400pt}{2.409pt}}
\put(753.0,113.0){\rule[-0.200pt]{0.400pt}{4.818pt}}
\put(753,68){\makebox(0,0){0.1}}
\put(753.0,1157.0){\rule[-0.200pt]{0.400pt}{4.818pt}}
\put(833.0,113.0){\rule[-0.200pt]{0.400pt}{2.409pt}}
\put(833.0,1167.0){\rule[-0.200pt]{0.400pt}{2.409pt}}
\put(880.0,113.0){\rule[-0.200pt]{0.400pt}{2.409pt}}
\put(880.0,1167.0){\rule[-0.200pt]{0.400pt}{2.409pt}}
\put(913.0,113.0){\rule[-0.200pt]{0.400pt}{2.409pt}}
\put(913.0,1167.0){\rule[-0.200pt]{0.400pt}{2.409pt}}
\put(939.0,113.0){\rule[-0.200pt]{0.400pt}{2.409pt}}
\put(939.0,1167.0){\rule[-0.200pt]{0.400pt}{2.409pt}}
\put(960.0,113.0){\rule[-0.200pt]{0.400pt}{2.409pt}}
\put(960.0,1167.0){\rule[-0.200pt]{0.400pt}{2.409pt}}
\put(978.0,113.0){\rule[-0.200pt]{0.400pt}{2.409pt}}
\put(978.0,1167.0){\rule[-0.200pt]{0.400pt}{2.409pt}}
\put(993.0,113.0){\rule[-0.200pt]{0.400pt}{2.409pt}}
\put(993.0,1167.0){\rule[-0.200pt]{0.400pt}{2.409pt}}
\put(1007.0,113.0){\rule[-0.200pt]{0.400pt}{2.409pt}}
\put(1007.0,1167.0){\rule[-0.200pt]{0.400pt}{2.409pt}}
\put(1019.0,113.0){\rule[-0.200pt]{0.400pt}{4.818pt}}
\put(1019,68){\makebox(0,0){1}}
\put(1019.0,1157.0){\rule[-0.200pt]{0.400pt}{4.818pt}}
\put(1099.0,113.0){\rule[-0.200pt]{0.400pt}{2.409pt}}
\put(1099.0,1167.0){\rule[-0.200pt]{0.400pt}{2.409pt}}
\put(1146.0,113.0){\rule[-0.200pt]{0.400pt}{2.409pt}}
\put(1146.0,1167.0){\rule[-0.200pt]{0.400pt}{2.409pt}}
\put(1179.0,113.0){\rule[-0.200pt]{0.400pt}{2.409pt}}
\put(1179.0,1167.0){\rule[-0.200pt]{0.400pt}{2.409pt}}
\put(1205.0,113.0){\rule[-0.200pt]{0.400pt}{2.409pt}}
\put(1205.0,1167.0){\rule[-0.200pt]{0.400pt}{2.409pt}}
\put(1226.0,113.0){\rule[-0.200pt]{0.400pt}{2.409pt}}
\put(1226.0,1167.0){\rule[-0.200pt]{0.400pt}{2.409pt}}
\put(1244.0,113.0){\rule[-0.200pt]{0.400pt}{2.409pt}}
\put(1244.0,1167.0){\rule[-0.200pt]{0.400pt}{2.409pt}}
\put(1259.0,113.0){\rule[-0.200pt]{0.400pt}{2.409pt}}
\put(1259.0,1167.0){\rule[-0.200pt]{0.400pt}{2.409pt}}
\put(1273.0,113.0){\rule[-0.200pt]{0.400pt}{2.409pt}}
\put(1273.0,1167.0){\rule[-0.200pt]{0.400pt}{2.409pt}}
\put(1285.0,113.0){\rule[-0.200pt]{0.400pt}{4.818pt}}
\put(1285,68){\makebox(0,0){10}}
\put(1285.0,1157.0){\rule[-0.200pt]{0.400pt}{4.818pt}}
\put(220.0,113.0){\rule[-0.200pt]{256.558pt}{0.400pt}}
\put(1285.0,113.0){\rule[-0.200pt]{0.400pt}{256.318pt}}
\put(220.0,1177.0){\rule[-0.200pt]{256.558pt}{0.400pt}}
\put(23,645){\makebox(0,0){$|\Delta E/E|$}}
\put(752,23){\makebox(0,0){$E$}}
\put(267,987){\makebox(0,0)[l]{$-\alpha/r$}}
\put(267,853){\makebox(0,0)[l]{\shortstack{$-\alpha/r+c\delta^3(\rv)$ \\ $1^{\rm st}$ order}}}
\put(267,638){\makebox(0,0)[l]{$-\alpha/r+ca^2\delta_a^3(\rv)$}}
\put(566,361){\makebox(0,0)[l]{$-\alpha/r+ca^2\delta_a^3(\rv)-d_1a^4\nabla^2\delta_a^3(\rv)$}}
\put(1048,183){\makebox(0,0){$1S$}}
\put(822,183){\makebox(0,0){$2S$}}
\put(537,183){\makebox(0,0){$6S$}}
\put(220.0,113.0){\rule[-0.200pt]{0.400pt}{256.318pt}}
\sbox{\plotpoint}{\rule[-0.500pt]{1.000pt}{1.000pt}}%
\put(1048,1102){\usebox{\plotpoint}}
\multiput(1048,1102)(-20.419,-3.721){12}{\usebox{\plotpoint}}
\multiput(823,1061)(-20.209,-4.734){5}{\usebox{\plotpoint}}
\multiput(712,1035)(-20.167,-4.906){4}{\usebox{\plotpoint}}
\multiput(638,1017)(-20.136,-5.034){2}{\usebox{\plotpoint}}
\multiput(582,1003)(-20.162,-4.928){3}{\usebox{\plotpoint}}
\multiput(537,992)(-20.087,-5.226){6}{\usebox{\plotpoint}}
\put(414,960){\usebox{\plotpoint}}
\put(1048,1102){\circle{24}}
\put(823,1061){\circle{24}}
\put(712,1035){\circle{24}}
\put(638,1017){\circle{24}}
\put(582,1003){\circle{24}}
\put(537,992){\circle{24}}
\put(414,960){\circle{24}}
\put(1048,1067){\usebox{\plotpoint}}
\multiput(1048,1067)(-19.359,-7.485){12}{\usebox{\plotpoint}}
\multiput(823,980)(-18.120,-10.121){6}{\usebox{\plotpoint}}
\multiput(712,918)(-18.051,-10.245){4}{\usebox{\plotpoint}}
\multiput(638,876)(-17.882,-10.537){3}{\usebox{\plotpoint}}
\multiput(582,843)(-17.798,-10.679){3}{\usebox{\plotpoint}}
\multiput(537,816)(-16.945,-11.986){7}{\usebox{\plotpoint}}
\put(414,729){\usebox{\plotpoint}}
\put(1048,1067){\circle{24}}
\put(823,980){\circle{24}}
\put(712,918){\circle{24}}
\put(638,876){\circle{24}}
\put(582,843){\circle{24}}
\put(537,816){\circle{24}}
\put(414,729){\circle{24}}
\put(1048,991){\usebox{\plotpoint}}
\multiput(1048,991)(-11.408,-17.340){20}{\usebox{\plotpoint}}
\multiput(823,649)(-19.582,6.880){6}{\usebox{\plotpoint}}
\multiput(712,688)(-17.198,-11.620){4}{\usebox{\plotpoint}}
\multiput(638,638)(-16.604,-12.453){4}{\usebox{\plotpoint}}
\multiput(582,596)(-16.383,-12.743){2}{\usebox{\plotpoint}}
\multiput(537,561)(-16.169,-13.014){8}{\usebox{\plotpoint}}
\put(414,462){\usebox{\plotpoint}}
\put(1048,991){\circle{24}}
\put(823,649){\circle{24}}
\put(712,688){\circle{24}}
\put(638,638){\circle{24}}
\put(582,596){\circle{24}}
\put(537,561){\circle{24}}
\put(414,462){\circle{24}}
\put(1048,1018){\usebox{\plotpoint}}
\multiput(1048,1018)(-15.410,-13.904){15}{\usebox{\plotpoint}}
\multiput(823,815)(-11.072,-17.556){10}{\usebox{\plotpoint}}
\multiput(712,639)(-11.881,-17.019){6}{\usebox{\plotpoint}}
\multiput(638,533)(-12.312,-16.709){5}{\usebox{\plotpoint}}
\multiput(582,457)(-12.321,-16.702){4}{\usebox{\plotpoint}}
\multiput(537,396)(-12.650,-16.455){9}{\usebox{\plotpoint}}
\put(414,236){\usebox{\plotpoint}}
\put(1048,1018){\circle{24}}
\put(823,815){\circle{24}}
\put(712,639){\circle{24}}
\put(638,533){\circle{24}}
\put(582,457){\circle{24}}
\put(537,396){\circle{24}}
\put(414,236){\circle{24}}
\end{picture}
\end{center}
\caption{Relative errors in the $S$-wave 
binding energies are plotted versus binding energy for the
Coulomb theory,  the Coulomb theory augmented with a delta function
in first-order perturbation theory, the nonperturbative effective
theory through~$a^2$, and the effective theory through~$a^4$.}
\label{bindingE2-fig}
\end{figure}

In Figure~\ref{bindingE2-fig} I compare the errors in the binding energies
obtained from the tuned effective theory with those obtained in
Section~\ref{PTh-Sec} from first-order perturbation theory. Even with only
the $a^2$ correction there is a sizable reduction in errors, due to
contributions from second and higher orders in the correction\,---\,our
numerical solution of the effective theory is nonperturbative, and so
includes all orders. The accuracy at low energies is further
enhanced by the $a^4$ correction, which begins to account for the
finite range of the short-range interaction. The crossing of the
curves for the $a^2$ and $a^4$ theories around~$E\!=\!0.1$ has no
significance; the error in the $a^2$ theory changes sign there, passing
through zero.

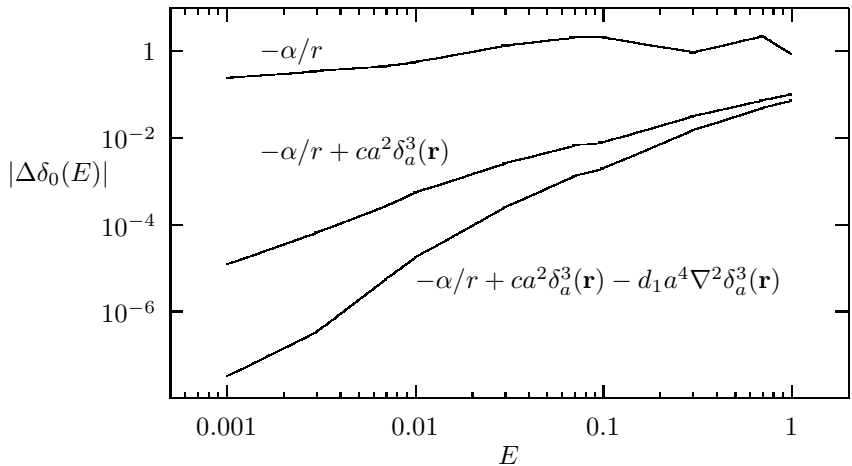
\begin{figure}
\begin{center}
% GNUPLOT: LaTeX picture
\setlength{\unitlength}{0.240900pt}
\ifx\plotpoint\undefined\newsavebox{\plotpoint}\fi
\sbox{\plotpoint}{\rule[-0.200pt]{0.400pt}{0.400pt}}%
\begin{picture}(1349,749)(0,0)
\font\gnuplot=cmr10 at 10pt
\gnuplot
\sbox{\plotpoint}{\rule[-0.200pt]{0.400pt}{0.400pt}}%
\put(220.0,249.0){\rule[-0.200pt]{4.818pt}{0.400pt}}
\put(198,249){\makebox(0,0)[r]{$10^{-6}$}}
\put(1265.0,249.0){\rule[-0.200pt]{4.818pt}{0.400pt}}
\put(220.0,385.0){\rule[-0.200pt]{4.818pt}{0.400pt}}
\put(198,385){\makebox(0,0)[r]{$10^{-4}$}}
\put(1265.0,385.0){\rule[-0.200pt]{4.818pt}{0.400pt}}
\put(220.0,522.0){\rule[-0.200pt]{4.818pt}{0.400pt}}
\put(198,522){\makebox(0,0)[r]{$10^{-2}$}}
\put(1265.0,522.0){\rule[-0.200pt]{4.818pt}{0.400pt}}
\put(220.0,658.0){\rule[-0.200pt]{4.818pt}{0.400pt}}
\put(198,658){\makebox(0,0)[r]{$1$}}
\put(1265.0,658.0){\rule[-0.200pt]{4.818pt}{0.400pt}}
\put(220.0,113.0){\rule[-0.200pt]{0.400pt}{2.409pt}}
\put(220.0,716.0){\rule[-0.200pt]{0.400pt}{2.409pt}}
\put(243.0,113.0){\rule[-0.200pt]{0.400pt}{2.409pt}}
\put(243.0,716.0){\rule[-0.200pt]{0.400pt}{2.409pt}}
\put(263.0,113.0){\rule[-0.200pt]{0.400pt}{2.409pt}}
\put(263.0,716.0){\rule[-0.200pt]{0.400pt}{2.409pt}}
\put(280.0,113.0){\rule[-0.200pt]{0.400pt}{2.409pt}}
\put(280.0,716.0){\rule[-0.200pt]{0.400pt}{2.409pt}}
\put(295.0,113.0){\rule[-0.200pt]{0.400pt}{2.409pt}}
\put(295.0,716.0){\rule[-0.200pt]{0.400pt}{2.409pt}}
\put(309.0,113.0){\rule[-0.200pt]{0.400pt}{4.818pt}}
\put(309,68){\makebox(0,0){0.001}}
\put(309.0,706.0){\rule[-0.200pt]{0.400pt}{4.818pt}}
\put(398.0,113.0){\rule[-0.200pt]{0.400pt}{2.409pt}}
\put(398.0,716.0){\rule[-0.200pt]{0.400pt}{2.409pt}}
\put(450.0,113.0){\rule[-0.200pt]{0.400pt}{2.409pt}}
\put(450.0,716.0){\rule[-0.200pt]{0.400pt}{2.409pt}}
\put(487.0,113.0){\rule[-0.200pt]{0.400pt}{2.409pt}}
\put(487.0,716.0){\rule[-0.200pt]{0.400pt}{2.409pt}}
\put(516.0,113.0){\rule[-0.200pt]{0.400pt}{2.409pt}}
\put(516.0,716.0){\rule[-0.200pt]{0.400pt}{2.409pt}}
\put(539.0,113.0){\rule[-0.200pt]{0.400pt}{2.409pt}}
\put(539.0,716.0){\rule[-0.200pt]{0.400pt}{2.409pt}}
\put(559.0,113.0){\rule[-0.200pt]{0.400pt}{2.409pt}}
\put(559.0,716.0){\rule[-0.200pt]{0.400pt}{2.409pt}}
\put(576.0,113.0){\rule[-0.200pt]{0.400pt}{2.409pt}}
\put(576.0,716.0){\rule[-0.200pt]{0.400pt}{2.409pt}}
\put(591.0,113.0){\rule[-0.200pt]{0.400pt}{2.409pt}}
\put(591.0,716.0){\rule[-0.200pt]{0.400pt}{2.409pt}}
\put(605.0,113.0){\rule[-0.200pt]{0.400pt}{4.818pt}}
\put(605,68){\makebox(0,0){0.01}}
\put(605.0,706.0){\rule[-0.200pt]{0.400pt}{4.818pt}}
\put(694.0,113.0){\rule[-0.200pt]{0.400pt}{2.409pt}}
\put(694.0,716.0){\rule[-0.200pt]{0.400pt}{2.409pt}}
\put(746.0,113.0){\rule[-0.200pt]{0.400pt}{2.409pt}}
\put(746.0,716.0){\rule[-0.200pt]{0.400pt}{2.409pt}}
\put(783.0,113.0){\rule[-0.200pt]{0.400pt}{2.409pt}}
\put(783.0,716.0){\rule[-0.200pt]{0.400pt}{2.409pt}}
\put(811.0,113.0){\rule[-0.200pt]{0.400pt}{2.409pt}}
\put(811.0,716.0){\rule[-0.200pt]{0.400pt}{2.409pt}}
\put(835.0,113.0){\rule[-0.200pt]{0.400pt}{2.409pt}}
\put(835.0,716.0){\rule[-0.200pt]{0.400pt}{2.409pt}}
\put(855.0,113.0){\rule[-0.200pt]{0.400pt}{2.409pt}}
\put(855.0,716.0){\rule[-0.200pt]{0.400pt}{2.409pt}}
\put(872.0,113.0){\rule[-0.200pt]{0.400pt}{2.409pt}}
\put(872.0,716.0){\rule[-0.200pt]{0.400pt}{2.409pt}}
\put(887.0,113.0){\rule[-0.200pt]{0.400pt}{2.409pt}}
\put(887.0,716.0){\rule[-0.200pt]{0.400pt}{2.409pt}}
\put(900.0,113.0){\rule[-0.200pt]{0.400pt}{4.818pt}}
\put(900,68){\makebox(0,0){0.1}}
\put(900.0,706.0){\rule[-0.200pt]{0.400pt}{4.818pt}}
\put(989.0,113.0){\rule[-0.200pt]{0.400pt}{2.409pt}}
\put(989.0,716.0){\rule[-0.200pt]{0.400pt}{2.409pt}}
\put(1041.0,113.0){\rule[-0.200pt]{0.400pt}{2.409pt}}
\put(1041.0,716.0){\rule[-0.200pt]{0.400pt}{2.409pt}}
\put(1078.0,113.0){\rule[-0.200pt]{0.400pt}{2.409pt}}
\put(1078.0,716.0){\rule[-0.200pt]{0.400pt}{2.409pt}}
\put(1107.0,113.0){\rule[-0.200pt]{0.400pt}{2.409pt}}
\put(1107.0,716.0){\rule[-0.200pt]{0.400pt}{2.409pt}}
\put(1130.0,113.0){\rule[-0.200pt]{0.400pt}{2.409pt}}
\put(1130.0,716.0){\rule[-0.200pt]{0.400pt}{2.409pt}}
\put(1150.0,113.0){\rule[-0.200pt]{0.400pt}{2.409pt}}
\put(1150.0,716.0){\rule[-0.200pt]{0.400pt}{2.409pt}}
\put(1167.0,113.0){\rule[-0.200pt]{0.400pt}{2.409pt}}
\put(1167.0,716.0){\rule[-0.200pt]{0.400pt}{2.409pt}}
\put(1182.0,113.0){\rule[-0.200pt]{0.400pt}{2.409pt}}
\put(1182.0,716.0){\rule[-0.200pt]{0.400pt}{2.409pt}}
\put(1196.0,113.0){\rule[-0.200pt]{0.400pt}{4.818pt}}
\put(1196,68){\makebox(0,0){1}}
\put(1196.0,706.0){\rule[-0.200pt]{0.400pt}{4.818pt}}
\put(1285.0,113.0){\rule[-0.200pt]{0.400pt}{2.409pt}}
\put(1285.0,716.0){\rule[-0.200pt]{0.400pt}{2.409pt}}
\put(220.0,113.0){\rule[-0.200pt]{256.558pt}{0.400pt}}
\put(1285.0,113.0){\rule[-0.200pt]{0.400pt}{147.672pt}}
\put(220.0,726.0){\rule[-0.200pt]{256.558pt}{0.400pt}}
\put(45,464){\makebox(0,0){$|\Delta \delta_0(E)|$}}
\put(752,23){\makebox(0,0){$E$}}
\put(361,658){\makebox(0,0)[l]{$-\alpha/r$}}
\put(361,501){\makebox(0,0)[l]{$-\alpha/r+ca^2\delta_a^3(\rv)$}}
\put(605,297){\makebox(0,0)[l]{$-\alpha/r+ca^2\delta_a^3(\rv)-d_1a^4\nabla^2\delta_a^3(\rv)$}}
\put(220.0,113.0){\rule[-0.200pt]{0.400pt}{147.672pt}}
\put(309,323){\usebox{\plotpoint}}
\multiput(309.00,323.58)(1.416,0.498){97}{\rule{1.228pt}{0.120pt}}
\multiput(309.00,322.17)(138.451,50.000){2}{\rule{0.614pt}{0.400pt}}
\multiput(450.00,373.58)(1.303,0.498){81}{\rule{1.138pt}{0.120pt}}
\multiput(450.00,372.17)(106.638,42.000){2}{\rule{0.569pt}{0.400pt}}
\multiput(559.00,415.58)(1.104,0.496){39}{\rule{0.976pt}{0.119pt}}
\multiput(559.00,414.17)(43.974,21.000){2}{\rule{0.488pt}{0.400pt}}
\multiput(605.00,436.58)(1.540,0.498){89}{\rule{1.326pt}{0.120pt}}
\multiput(605.00,435.17)(138.248,46.000){2}{\rule{0.663pt}{0.400pt}}
\multiput(746.00,482.58)(1.964,0.497){53}{\rule{1.657pt}{0.120pt}}
\multiput(746.00,481.17)(105.561,28.000){2}{\rule{0.829pt}{0.400pt}}
\multiput(855.00,510.59)(4.940,0.477){7}{\rule{3.700pt}{0.115pt}}
\multiput(855.00,509.17)(37.320,5.000){2}{\rule{1.850pt}{0.400pt}}
\multiput(900.00,515.58)(1.729,0.498){79}{\rule{1.476pt}{0.120pt}}
\multiput(900.00,514.17)(137.937,41.000){2}{\rule{0.738pt}{0.400pt}}
\multiput(1041.00,556.58)(2.203,0.497){47}{\rule{1.844pt}{0.120pt}}
\multiput(1041.00,555.17)(105.173,25.000){2}{\rule{0.922pt}{0.400pt}}
\multiput(1150.00,581.58)(2.372,0.491){17}{\rule{1.940pt}{0.118pt}}
\multiput(1150.00,580.17)(41.973,10.000){2}{\rule{0.970pt}{0.400pt}}
\put(309,147){\usebox{\plotpoint}}
\multiput(309.00,147.58)(1.009,0.499){137}{\rule{0.906pt}{0.120pt}}
\multiput(309.00,146.17)(139.120,70.000){2}{\rule{0.453pt}{0.400pt}}
\multiput(450.00,217.58)(0.657,0.499){163}{\rule{0.625pt}{0.120pt}}
\multiput(450.00,216.17)(107.702,83.000){2}{\rule{0.313pt}{0.400pt}}
\multiput(559.00,300.58)(0.677,0.498){65}{\rule{0.641pt}{0.120pt}}
\multiput(559.00,299.17)(44.669,34.000){2}{\rule{0.321pt}{0.400pt}}
\multiput(605.00,334.58)(0.894,0.499){155}{\rule{0.814pt}{0.120pt}}
\multiput(605.00,333.17)(139.311,79.000){2}{\rule{0.407pt}{0.400pt}}
\multiput(746.00,413.58)(1.116,0.498){95}{\rule{0.990pt}{0.120pt}}
\multiput(746.00,412.17)(106.946,49.000){2}{\rule{0.495pt}{0.400pt}}
\multiput(855.00,462.58)(1.918,0.492){21}{\rule{1.600pt}{0.119pt}}
\multiput(855.00,461.17)(41.679,12.000){2}{\rule{0.800pt}{0.400pt}}
\multiput(900.00,474.58)(1.178,0.499){117}{\rule{1.040pt}{0.120pt}}
\multiput(900.00,473.17)(138.841,60.000){2}{\rule{0.520pt}{0.400pt}}
\multiput(1041.00,534.58)(1.567,0.498){67}{\rule{1.346pt}{0.120pt}}
\multiput(1041.00,533.17)(106.207,35.000){2}{\rule{0.673pt}{0.400pt}}
\multiput(1150.00,569.58)(1.961,0.492){21}{\rule{1.633pt}{0.119pt}}
\multiput(1150.00,568.17)(42.610,12.000){2}{\rule{0.817pt}{0.400pt}}
\put(309,616){\usebox{\plotpoint}}
\multiput(309.00,616.58)(6.628,0.492){19}{\rule{5.227pt}{0.118pt}}
\multiput(309.00,615.17)(130.151,11.000){2}{\rule{2.614pt}{0.400pt}}
\multiput(450.00,627.59)(7.163,0.488){13}{\rule{5.550pt}{0.117pt}}
\multiput(450.00,626.17)(97.481,8.000){2}{\rule{2.775pt}{0.400pt}}
\multiput(559.00,635.59)(4.107,0.482){9}{\rule{3.167pt}{0.116pt}}
\multiput(559.00,634.17)(39.427,6.000){2}{\rule{1.583pt}{0.400pt}}
\multiput(605.00,641.58)(2.741,0.497){49}{\rule{2.269pt}{0.120pt}}
\multiput(605.00,640.17)(136.290,26.000){2}{\rule{1.135pt}{0.400pt}}
\multiput(746.00,667.58)(4.303,0.493){23}{\rule{3.454pt}{0.119pt}}
\multiput(746.00,666.17)(101.831,13.000){2}{\rule{1.727pt}{0.400pt}}
\multiput(900.00,678.92)(2.973,-0.496){45}{\rule{2.450pt}{0.120pt}}
\multiput(900.00,679.17)(135.915,-24.000){2}{\rule{1.225pt}{0.400pt}}
\multiput(1041.00,656.58)(2.117,0.497){49}{\rule{1.777pt}{0.120pt}}
\multiput(1041.00,655.17)(105.312,26.000){2}{\rule{0.888pt}{0.400pt}}
\multiput(1150.00,680.92)(0.795,-0.497){55}{\rule{0.734pt}{0.120pt}}
\multiput(1150.00,681.17)(44.476,-29.000){2}{\rule{0.367pt}{0.400pt}}
\put(855.0,680.0){\rule[-0.200pt]{10.840pt}{0.400pt}}
\end{picture}
\end{center}
\caption{Errors in $S$-wave phase shifts (in radians) computed with 
the effective theory corrected through order~$a^2$ and~$a^4$. Values are for
$a\!=\!1$ and are plotted versus energy. Results are also shown
for the theory with no contact terms.}
\label{phase-fig}
\end{figure}

The phase shifts tell a similar story; see Figure~\ref{phase-fig}. At low
energies, the errors decrease steadily as the correction terms are added,
order-by-order in $a^2$. The slope of the error curve changes as
each new correction is added, getting steeper by one power of
$E\!\propto\!p^2$ each time the order of the error is increased by~$a^2$, as
expected. Similar behavior is apparent in the previous figure, which shows the
errors in the binding energies.

Both of these figures show that the correction terms have little effect at
energies~$E\!>\!1$. At these energies, the particle's
wavelength is sufficiently short that the particle can probe the detailed
structure of
$V_s(\rv)$. Effective theories are useless in this limit. To go much
beyond~$E\!=\!1$ in this example, one must somehow uncover the true
short-distance structure of the theory.  

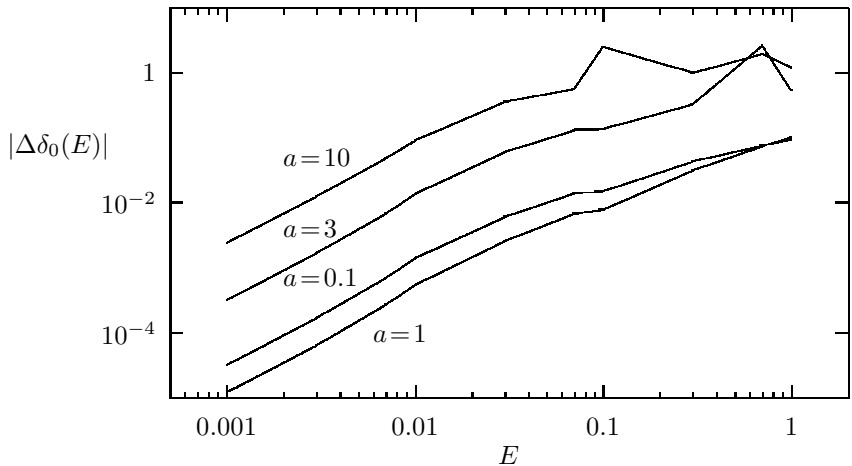
\begin{figure}
\begin{center}
% GNUPLOT: LaTeX picture
\setlength{\unitlength}{0.240900pt}
\ifx\plotpoint\undefined\newsavebox{\plotpoint}\fi
\sbox{\plotpoint}{\rule[-0.200pt]{0.400pt}{0.400pt}}%
\begin{picture}(1349,749)(0,0)
\font\gnuplot=cmr10 at 10pt
\gnuplot
\sbox{\plotpoint}{\rule[-0.200pt]{0.400pt}{0.400pt}}%
\put(220.0,215.0){\rule[-0.200pt]{4.818pt}{0.400pt}}
\put(198,215){\makebox(0,0)[r]{$10^{-4}$}}
\put(1265.0,215.0){\rule[-0.200pt]{4.818pt}{0.400pt}}
\put(220.0,420.0){\rule[-0.200pt]{4.818pt}{0.400pt}}
\put(198,420){\makebox(0,0)[r]{$10^{-2}$}}
\put(1265.0,420.0){\rule[-0.200pt]{4.818pt}{0.400pt}}
\put(220.0,624.0){\rule[-0.200pt]{4.818pt}{0.400pt}}
\put(198,624){\makebox(0,0)[r]{$1$}}
\put(1265.0,624.0){\rule[-0.200pt]{4.818pt}{0.400pt}}
\put(220.0,113.0){\rule[-0.200pt]{0.400pt}{2.409pt}}
\put(220.0,716.0){\rule[-0.200pt]{0.400pt}{2.409pt}}
\put(243.0,113.0){\rule[-0.200pt]{0.400pt}{2.409pt}}
\put(243.0,716.0){\rule[-0.200pt]{0.400pt}{2.409pt}}
\put(263.0,113.0){\rule[-0.200pt]{0.400pt}{2.409pt}}
\put(263.0,716.0){\rule[-0.200pt]{0.400pt}{2.409pt}}
\put(280.0,113.0){\rule[-0.200pt]{0.400pt}{2.409pt}}
\put(280.0,716.0){\rule[-0.200pt]{0.400pt}{2.409pt}}
\put(295.0,113.0){\rule[-0.200pt]{0.400pt}{2.409pt}}
\put(295.0,716.0){\rule[-0.200pt]{0.400pt}{2.409pt}}
\put(309.0,113.0){\rule[-0.200pt]{0.400pt}{4.818pt}}
\put(309,68){\makebox(0,0){0.001}}
\put(309.0,706.0){\rule[-0.200pt]{0.400pt}{4.818pt}}
\put(398.0,113.0){\rule[-0.200pt]{0.400pt}{2.409pt}}
\put(398.0,716.0){\rule[-0.200pt]{0.400pt}{2.409pt}}
\put(450.0,113.0){\rule[-0.200pt]{0.400pt}{2.409pt}}
\put(450.0,716.0){\rule[-0.200pt]{0.400pt}{2.409pt}}
\put(487.0,113.0){\rule[-0.200pt]{0.400pt}{2.409pt}}
\put(487.0,716.0){\rule[-0.200pt]{0.400pt}{2.409pt}}
\put(516.0,113.0){\rule[-0.200pt]{0.400pt}{2.409pt}}
\put(516.0,716.0){\rule[-0.200pt]{0.400pt}{2.409pt}}
\put(539.0,113.0){\rule[-0.200pt]{0.400pt}{2.409pt}}
\put(539.0,716.0){\rule[-0.200pt]{0.400pt}{2.409pt}}
\put(559.0,113.0){\rule[-0.200pt]{0.400pt}{2.409pt}}
\put(559.0,716.0){\rule[-0.200pt]{0.400pt}{2.409pt}}
\put(576.0,113.0){\rule[-0.200pt]{0.400pt}{2.409pt}}
\put(576.0,716.0){\rule[-0.200pt]{0.400pt}{2.409pt}}
\put(591.0,113.0){\rule[-0.200pt]{0.400pt}{2.409pt}}
\put(591.0,716.0){\rule[-0.200pt]{0.400pt}{2.409pt}}
\put(605.0,113.0){\rule[-0.200pt]{0.400pt}{4.818pt}}
\put(605,68){\makebox(0,0){0.01}}
\put(605.0,706.0){\rule[-0.200pt]{0.400pt}{4.818pt}}
\put(694.0,113.0){\rule[-0.200pt]{0.400pt}{2.409pt}}
\put(694.0,716.0){\rule[-0.200pt]{0.400pt}{2.409pt}}
\put(746.0,113.0){\rule[-0.200pt]{0.400pt}{2.409pt}}
\put(746.0,716.0){\rule[-0.200pt]{0.400pt}{2.409pt}}
\put(783.0,113.0){\rule[-0.200pt]{0.400pt}{2.409pt}}
\put(783.0,716.0){\rule[-0.200pt]{0.400pt}{2.409pt}}
\put(811.0,113.0){\rule[-0.200pt]{0.400pt}{2.409pt}}
\put(811.0,716.0){\rule[-0.200pt]{0.400pt}{2.409pt}}
\put(835.0,113.0){\rule[-0.200pt]{0.400pt}{2.409pt}}
\put(835.0,716.0){\rule[-0.200pt]{0.400pt}{2.409pt}}
\put(855.0,113.0){\rule[-0.200pt]{0.400pt}{2.409pt}}
\put(855.0,716.0){\rule[-0.200pt]{0.400pt}{2.409pt}}
\put(872.0,113.0){\rule[-0.200pt]{0.400pt}{2.409pt}}
\put(872.0,716.0){\rule[-0.200pt]{0.400pt}{2.409pt}}
\put(887.0,113.0){\rule[-0.200pt]{0.400pt}{2.409pt}}
\put(887.0,716.0){\rule[-0.200pt]{0.400pt}{2.409pt}}
\put(900.0,113.0){\rule[-0.200pt]{0.400pt}{4.818pt}}
\put(900,68){\makebox(0,0){0.1}}
\put(900.0,706.0){\rule[-0.200pt]{0.400pt}{4.818pt}}
\put(989.0,113.0){\rule[-0.200pt]{0.400pt}{2.409pt}}
\put(989.0,716.0){\rule[-0.200pt]{0.400pt}{2.409pt}}
\put(1041.0,113.0){\rule[-0.200pt]{0.400pt}{2.409pt}}
\put(1041.0,716.0){\rule[-0.200pt]{0.400pt}{2.409pt}}
\put(1078.0,113.0){\rule[-0.200pt]{0.400pt}{2.409pt}}
\put(1078.0,716.0){\rule[-0.200pt]{0.400pt}{2.409pt}}
\put(1107.0,113.0){\rule[-0.200pt]{0.400pt}{2.409pt}}
\put(1107.0,716.0){\rule[-0.200pt]{0.400pt}{2.409pt}}
\put(1130.0,113.0){\rule[-0.200pt]{0.400pt}{2.409pt}}
\put(1130.0,716.0){\rule[-0.200pt]{0.400pt}{2.409pt}}
\put(1150.0,113.0){\rule[-0.200pt]{0.400pt}{2.409pt}}
\put(1150.0,716.0){\rule[-0.200pt]{0.400pt}{2.409pt}}
\put(1167.0,113.0){\rule[-0.200pt]{0.400pt}{2.409pt}}
\put(1167.0,716.0){\rule[-0.200pt]{0.400pt}{2.409pt}}
\put(1182.0,113.0){\rule[-0.200pt]{0.400pt}{2.409pt}}
\put(1182.0,716.0){\rule[-0.200pt]{0.400pt}{2.409pt}}
\put(1196.0,113.0){\rule[-0.200pt]{0.400pt}{4.818pt}}
\put(1196,68){\makebox(0,0){1}}
\put(1196.0,706.0){\rule[-0.200pt]{0.400pt}{4.818pt}}
\put(1285.0,113.0){\rule[-0.200pt]{0.400pt}{2.409pt}}
\put(1285.0,716.0){\rule[-0.200pt]{0.400pt}{2.409pt}}
\put(220.0,113.0){\rule[-0.200pt]{256.558pt}{0.400pt}}
\put(1285.0,113.0){\rule[-0.200pt]{0.400pt}{147.672pt}}
\put(220.0,726.0){\rule[-0.200pt]{256.558pt}{0.400pt}}
\put(45,509){\makebox(0,0){$|\Delta \delta_0(E)|$}}
\put(752,23){\makebox(0,0){$E$}}
\put(398,491){\makebox(0,0)[l]{$a\!=\!10$}}
\put(398,379){\makebox(0,0)[l]{$a\!=\!3$}}
\put(398,302){\makebox(0,0)[l]{$a\!=\!0.1$}}
\put(539,215){\makebox(0,0)[l]{$a\!=\!1$}}
\put(220.0,113.0){\rule[-0.200pt]{0.400pt}{147.672pt}}
\put(309,165){\usebox{\plotpoint}}
\multiput(309.00,165.58)(0.954,0.499){145}{\rule{0.862pt}{0.120pt}}
\multiput(309.00,164.17)(139.211,74.000){2}{\rule{0.431pt}{0.400pt}}
\multiput(450.00,239.58)(0.866,0.499){123}{\rule{0.792pt}{0.120pt}}
\multiput(450.00,238.17)(107.356,63.000){2}{\rule{0.396pt}{0.400pt}}
\multiput(559.00,302.58)(0.743,0.497){59}{\rule{0.694pt}{0.120pt}}
\multiput(559.00,301.17)(44.561,31.000){2}{\rule{0.347pt}{0.400pt}}
\multiput(605.00,333.58)(1.087,0.499){127}{\rule{0.968pt}{0.120pt}}
\multiput(605.00,332.17)(138.992,65.000){2}{\rule{0.484pt}{0.400pt}}
\multiput(746.00,398.58)(1.481,0.498){71}{\rule{1.278pt}{0.120pt}}
\multiput(746.00,397.17)(106.347,37.000){2}{\rule{0.639pt}{0.400pt}}
\multiput(855.00,435.61)(9.839,0.447){3}{\rule{6.100pt}{0.108pt}}
\multiput(855.00,434.17)(32.339,3.000){2}{\rule{3.050pt}{0.400pt}}
\multiput(900.00,438.58)(1.507,0.498){91}{\rule{1.300pt}{0.120pt}}
\multiput(900.00,437.17)(138.302,47.000){2}{\rule{0.650pt}{0.400pt}}
\multiput(1041.00,485.58)(2.203,0.497){47}{\rule{1.844pt}{0.120pt}}
\multiput(1041.00,484.17)(105.173,25.000){2}{\rule{0.922pt}{0.400pt}}
\multiput(1150.00,510.58)(2.372,0.491){17}{\rule{1.940pt}{0.118pt}}
\multiput(1150.00,509.17)(41.973,10.000){2}{\rule{0.970pt}{0.400pt}}
\put(309,122){\usebox{\plotpoint}}
\multiput(309.00,122.58)(0.954,0.499){145}{\rule{0.862pt}{0.120pt}}
\multiput(309.00,121.17)(139.211,74.000){2}{\rule{0.431pt}{0.400pt}}
\multiput(450.00,196.58)(0.853,0.499){125}{\rule{0.781pt}{0.120pt}}
\multiput(450.00,195.17)(107.378,64.000){2}{\rule{0.391pt}{0.400pt}}
\multiput(559.00,260.58)(0.743,0.497){59}{\rule{0.694pt}{0.120pt}}
\multiput(559.00,259.17)(44.561,31.000){2}{\rule{0.347pt}{0.400pt}}
\multiput(605.00,291.58)(1.024,0.499){135}{\rule{0.917pt}{0.120pt}}
\multiput(605.00,290.17)(139.096,69.000){2}{\rule{0.459pt}{0.400pt}}
\multiput(746.00,360.58)(1.273,0.498){83}{\rule{1.114pt}{0.120pt}}
\multiput(746.00,359.17)(106.688,43.000){2}{\rule{0.557pt}{0.400pt}}
\multiput(855.00,403.59)(4.016,0.482){9}{\rule{3.100pt}{0.116pt}}
\multiput(855.00,402.17)(38.566,6.000){2}{\rule{1.550pt}{0.400pt}}
\multiput(900.00,409.58)(1.140,0.499){121}{\rule{1.010pt}{0.120pt}}
\multiput(900.00,408.17)(138.904,62.000){2}{\rule{0.505pt}{0.400pt}}
\multiput(1041.00,471.58)(1.442,0.498){73}{\rule{1.247pt}{0.120pt}}
\multiput(1041.00,470.17)(106.411,38.000){2}{\rule{0.624pt}{0.400pt}}
\multiput(1150.00,509.58)(1.672,0.494){25}{\rule{1.414pt}{0.119pt}}
\multiput(1150.00,508.17)(43.065,14.000){2}{\rule{0.707pt}{0.400pt}}
\put(309,267){\usebox{\plotpoint}}
\multiput(309.00,267.58)(0.954,0.499){145}{\rule{0.862pt}{0.120pt}}
\multiput(309.00,266.17)(139.211,74.000){2}{\rule{0.431pt}{0.400pt}}
\multiput(450.00,341.58)(0.881,0.499){121}{\rule{0.803pt}{0.120pt}}
\multiput(450.00,340.17)(107.333,62.000){2}{\rule{0.402pt}{0.400pt}}
\multiput(559.00,403.58)(0.743,0.497){59}{\rule{0.694pt}{0.120pt}}
\multiput(559.00,402.17)(44.561,31.000){2}{\rule{0.347pt}{0.400pt}}
\multiput(605.00,434.58)(1.071,0.499){129}{\rule{0.955pt}{0.120pt}}
\multiput(605.00,433.17)(139.019,66.000){2}{\rule{0.477pt}{0.400pt}}
\multiput(746.00,500.58)(1.613,0.498){65}{\rule{1.382pt}{0.120pt}}
\multiput(746.00,499.17)(106.131,34.000){2}{\rule{0.691pt}{0.400pt}}
\put(855,534.17){\rule{9.100pt}{0.400pt}}
\multiput(855.00,533.17)(26.112,2.000){2}{\rule{4.550pt}{0.400pt}}
\multiput(900.00,536.58)(1.819,0.498){75}{\rule{1.546pt}{0.120pt}}
\multiput(900.00,535.17)(137.791,39.000){2}{\rule{0.773pt}{0.400pt}}
\multiput(1041.00,575.58)(0.592,0.499){181}{\rule{0.574pt}{0.120pt}}
\multiput(1041.00,574.17)(107.809,92.000){2}{\rule{0.287pt}{0.400pt}}
\multiput(1150.58,664.02)(0.498,-0.773){89}{\rule{0.120pt}{0.717pt}}
\multiput(1149.17,665.51)(46.000,-69.511){2}{\rule{0.400pt}{0.359pt}}
\put(309,357){\usebox{\plotpoint}}
\multiput(309.00,357.58)(0.967,0.499){143}{\rule{0.873pt}{0.120pt}}
\multiput(309.00,356.17)(139.189,73.000){2}{\rule{0.436pt}{0.400pt}}
\multiput(450.00,430.58)(0.910,0.499){117}{\rule{0.827pt}{0.120pt}}
\multiput(450.00,429.17)(107.284,60.000){2}{\rule{0.413pt}{0.400pt}}
\multiput(559.00,490.58)(0.824,0.497){53}{\rule{0.757pt}{0.120pt}}
\multiput(559.00,489.17)(44.429,28.000){2}{\rule{0.379pt}{0.400pt}}
\multiput(605.00,518.58)(1.159,0.499){119}{\rule{1.025pt}{0.120pt}}
\multiput(605.00,517.17)(138.873,61.000){2}{\rule{0.512pt}{0.400pt}}
\multiput(746.00,579.58)(2.765,0.496){37}{\rule{2.280pt}{0.119pt}}
\multiput(746.00,578.17)(104.268,20.000){2}{\rule{1.140pt}{0.400pt}}
\multiput(855.58,599.00)(0.498,0.734){87}{\rule{0.120pt}{0.687pt}}
\multiput(854.17,599.00)(45.000,64.575){2}{\rule{0.400pt}{0.343pt}}
\multiput(900.00,663.92)(1.729,-0.498){79}{\rule{1.476pt}{0.120pt}}
\multiput(900.00,664.17)(137.937,-41.000){2}{\rule{0.738pt}{0.400pt}}
\multiput(1041.00,624.58)(1.831,0.497){57}{\rule{1.553pt}{0.120pt}}
\multiput(1041.00,623.17)(105.776,30.000){2}{\rule{0.777pt}{0.400pt}}
\multiput(1150.00,652.92)(1.053,-0.496){41}{\rule{0.936pt}{0.120pt}}
\multiput(1150.00,653.17)(44.057,-22.000){2}{\rule{0.468pt}{0.400pt}}
\end{picture}
\end{center}
\caption{The dependence on cutoff of the$S$-wave phase shift errors  plotted
against energy is shown. Errors decrease as~$a$ is reduced until
$a\!\approx\!1$, beyond which point there is no improvement. The $a^2$~theory
was used here ($d_1\!=\!0$).}
\label{a_dep-fig}
\end{figure}

I redid the analysis for a variety of different~$a$'s to explore
cutoff dependence. Figure~\ref{a_dep-fig} shows how the error in
the phase shift depends on~$a$. As $a$~is reduced from~$10\!\to\!3\!\to\!1$,
the errors decrease at all but the highest energies. For this range, $a$ is
larger than the range~$r_s$ of the true short-distance interaction, and
therefore the errors are proportional to a power of $qa$, which
decreases with decreasing~$a$. The errors
stop decreasing for $a\!<\!1$, becoming almost insensitive to $a$. From
our discussion in the previous section, this suggests that $r_s\!\approx\!1$.
The coupling constants also vary with
$a$ in the manner expected if $r_s\!=\!1$. For example, with
$d_1\!=\!0$, coupling constant $c/4\pi$ is roughly $-3$ and only
weakly dependent on~$a$ when $a\!\ge\!1$. It grows rapidly as $a$ is
reduced below one; for example,  $c\!=\!-74.0$ when  $a\!=\!0.01$.

I also looked for multiple solutions for the coupling constants.\cite{savage}
Setting $d_1\!=\!0$ and
$a\!=\!1$, for example, I found that the phase shift at
$E\!=\!10^{-10}$ is correct when
$c/4\pi$ equals~$-12.4$ or~$1.6$, as well as when it equals~$-3.18$, the
value used above. These extra solutions
give reasonably good
results for the phase shifts and binding energies, but not as good as our
original value. The reason is that with these new $c$'s the theory's
short-distance structure is significantly modified. In particular, with
$c\!=\!24.7$, there is an extra bound state with a very high binding energy
($\approx\!6$). 
We do not expect our effective theory to work well at high energies, so
there is nothing wrong with having the extra state; the usual $1S$,
$2S$\,\ldots\,states are all still present, 
each with an extra node in its wavefunction
at short distances. When
$c/4\pi= 1.6$, the $1S$ state disappears, while the other states are missing a
node at short distances. 

I find it remarkable that a simple two-parameter effective theory, tuned
using only two very low energy phase shifts, can reproduce the
full range of our data so successfully. The binding
energy of the 2$S$~state, for example, is accurately predicted by the
effective theory to better than 1\%, despite the fact that the cutoff scale
$a\!=\!1$ is almost a quarter the radius of the atom. Our most infrared data
is reproduced to better than six digits.

\subsection{Misconceptions}
It is worth pausing for a moment to address common misconceptions
concerning the procedure outlined here. One common notion is that our
effective potential will eventually become identical to the true potential,
as more and more higher-order corrections are added. This is false. There
are infinitely many theories that share the low-energy behavior of our
true theory; the low-energy data that we have, no matter how precise, does
not contain enough information to specify completely the high-energy
structure of the underlying theory. Our procedure allows us to construct one
of the infinitely many theories; it is
extremely unlikely that we would happen onto the true theory.

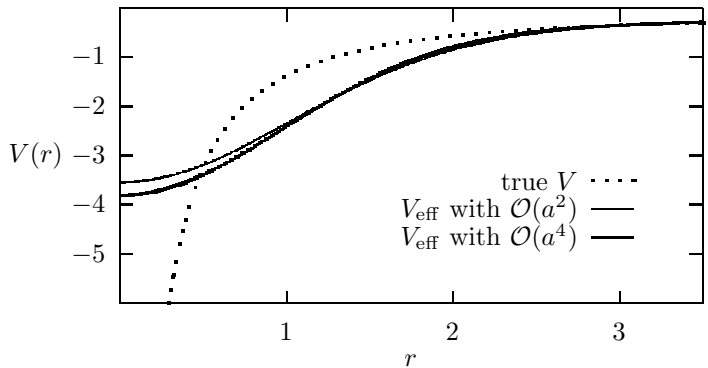
\begin{figure}
\begin{center}
% GNUPLOT: LaTeX picture
\setlength{\unitlength}{0.240900pt}
\ifx\plotpoint\undefined\newsavebox{\plotpoint}\fi
\sbox{\plotpoint}{\rule[-0.200pt]{0.400pt}{0.400pt}}%
\begin{picture}(1200,600)(0,0)
\font\gnuplot=cmr10 at 10pt
\gnuplot
\sbox{\plotpoint}{\rule[-0.200pt]{0.400pt}{0.400pt}}%
\put(220.0,190.0){\rule[-0.200pt]{4.818pt}{0.400pt}}
\put(198,190){\makebox(0,0)[r]{$-5$}}
\put(1116.0,190.0){\rule[-0.200pt]{4.818pt}{0.400pt}}
\put(220.0,268.0){\rule[-0.200pt]{4.818pt}{0.400pt}}
\put(198,268){\makebox(0,0)[r]{$-4$}}
\put(1116.0,268.0){\rule[-0.200pt]{4.818pt}{0.400pt}}
\put(220.0,345.0){\rule[-0.200pt]{4.818pt}{0.400pt}}
\put(198,345){\makebox(0,0)[r]{$-3$}}
\put(1116.0,345.0){\rule[-0.200pt]{4.818pt}{0.400pt}}
\put(220.0,422.0){\rule[-0.200pt]{4.818pt}{0.400pt}}
\put(198,422){\makebox(0,0)[r]{$-2$}}
\put(1116.0,422.0){\rule[-0.200pt]{4.818pt}{0.400pt}}
\put(220.0,500.0){\rule[-0.200pt]{4.818pt}{0.400pt}}
\put(198,500){\makebox(0,0)[r]{$-1$}}
\put(1116.0,500.0){\rule[-0.200pt]{4.818pt}{0.400pt}}
\put(482.0,113.0){\rule[-0.200pt]{0.400pt}{4.818pt}}
\put(482,68){\makebox(0,0){$1$}}
\put(482.0,557.0){\rule[-0.200pt]{0.400pt}{4.818pt}}
\put(743.0,113.0){\rule[-0.200pt]{0.400pt}{4.818pt}}
\put(743,68){\makebox(0,0){$2$}}
\put(743.0,557.0){\rule[-0.200pt]{0.400pt}{4.818pt}}
\put(1005.0,113.0){\rule[-0.200pt]{0.400pt}{4.818pt}}
\put(1005,68){\makebox(0,0){$3$}}
\put(1005.0,557.0){\rule[-0.200pt]{0.400pt}{4.818pt}}
\put(220.0,113.0){\rule[-0.200pt]{220.664pt}{0.400pt}}
\put(1136.0,113.0){\rule[-0.200pt]{0.400pt}{111.778pt}}
\put(220.0,577.0){\rule[-0.200pt]{220.664pt}{0.400pt}}
\put(89,345){\makebox(0,0){$V(r)$}}
\put(678,23){\makebox(0,0){$r$}}
\put(220.0,113.0){\rule[-0.200pt]{0.400pt}{111.778pt}}
\sbox{\plotpoint}{\rule[-0.500pt]{1.000pt}{1.000pt}}%
\put(940,306){\makebox(0,0)[r]{true $V$}}
\multiput(962,306)(20.756,0.000){4}{\usebox{\plotpoint}}
\put(1028,306){\usebox{\plotpoint}}
\multiput(296,113)(3.190,20.509){3}{\usebox{\plotpoint}}
\multiput(303,158)(4.233,20.319){2}{\usebox{\plotpoint}}
\multiput(313,206)(4.783,20.197){2}{\usebox{\plotpoint}}
\multiput(322,244)(5.619,19.980){2}{\usebox{\plotpoint}}
\put(337.30,294.91){\usebox{\plotpoint}}
\put(344.88,314.22){\usebox{\plotpoint}}
\put(353.24,333.21){\usebox{\plotpoint}}
\put(362.15,351.95){\usebox{\plotpoint}}
\put(372.24,370.07){\usebox{\plotpoint}}
\put(384.02,387.12){\usebox{\plotpoint}}
\multiput(387,391)(12.453,16.604){0}{\usebox{\plotpoint}}
\put(396.55,403.67){\usebox{\plotpoint}}
\put(410.24,419.24){\usebox{\plotpoint}}
\multiput(414,423)(15.427,13.885){0}{\usebox{\plotpoint}}
\put(425.48,433.32){\usebox{\plotpoint}}
\put(441.44,446.57){\usebox{\plotpoint}}
\multiput(442,447)(17.270,11.513){0}{\usebox{\plotpoint}}
\put(458.92,457.75){\usebox{\plotpoint}}
\multiput(461,459)(17.270,11.513){0}{\usebox{\plotpoint}}
\put(476.57,468.65){\usebox{\plotpoint}}
\multiput(479,470)(18.144,10.080){0}{\usebox{\plotpoint}}
\put(495.13,477.85){\usebox{\plotpoint}}
\multiput(498,479)(18.967,8.430){0}{\usebox{\plotpoint}}
\put(514.14,486.17){\usebox{\plotpoint}}
\multiput(516,487)(19.690,6.563){0}{\usebox{\plotpoint}}
\put(533.84,492.65){\usebox{\plotpoint}}
\multiput(535,493)(19.690,6.563){0}{\usebox{\plotpoint}}
\multiput(544,496)(19.690,6.563){0}{\usebox{\plotpoint}}
\put(553.54,499.18){\usebox{\plotpoint}}
\multiput(562,502)(20.352,4.070){0}{\usebox{\plotpoint}}
\put(573.56,504.52){\usebox{\plotpoint}}
\multiput(581,507)(20.261,4.503){0}{\usebox{\plotpoint}}
\put(593.60,509.80){\usebox{\plotpoint}}
\multiput(599,511)(20.352,4.070){0}{\usebox{\plotpoint}}
\put(613.91,514.09){\usebox{\plotpoint}}
\multiput(618,515)(20.261,4.503){0}{\usebox{\plotpoint}}
\put(634.17,518.59){\usebox{\plotpoint}}
\multiput(636,519)(20.652,2.065){0}{\usebox{\plotpoint}}
\put(654.62,521.92){\usebox{\plotpoint}}
\multiput(655,522)(20.629,2.292){0}{\usebox{\plotpoint}}
\multiput(664,523)(20.629,2.292){0}{\usebox{\plotpoint}}
\put(675.21,524.44){\usebox{\plotpoint}}
\multiput(683,526)(20.629,2.292){0}{\usebox{\plotpoint}}
\put(695.74,527.42){\usebox{\plotpoint}}
\multiput(701,528)(20.629,2.292){0}{\usebox{\plotpoint}}
\put(716.28,530.26){\usebox{\plotpoint}}
\multiput(720,531)(20.629,2.292){0}{\usebox{\plotpoint}}
\put(736.86,532.87){\usebox{\plotpoint}}
\multiput(738,533)(20.629,2.292){0}{\usebox{\plotpoint}}
\multiput(747,534)(20.756,0.000){0}{\usebox{\plotpoint}}
\put(757.55,534.06){\usebox{\plotpoint}}
\multiput(766,535)(20.629,2.292){0}{\usebox{\plotpoint}}
\put(778.18,536.35){\usebox{\plotpoint}}
\multiput(784,537)(20.652,2.065){0}{\usebox{\plotpoint}}
\put(798.82,538.54){\usebox{\plotpoint}}
\multiput(803,539)(20.756,0.000){0}{\usebox{\plotpoint}}
\put(819.50,539.83){\usebox{\plotpoint}}
\multiput(821,540)(20.652,2.065){0}{\usebox{\plotpoint}}
\multiput(831,541)(20.756,0.000){0}{\usebox{\plotpoint}}
\put(840.20,541.02){\usebox{\plotpoint}}
\multiput(849,542)(20.629,2.292){0}{\usebox{\plotpoint}}
\put(860.84,543.00){\usebox{\plotpoint}}
\multiput(868,543)(20.629,2.292){0}{\usebox{\plotpoint}}
\put(881.54,544.00){\usebox{\plotpoint}}
\multiput(886,544)(20.629,2.292){0}{\usebox{\plotpoint}}
\put(902.24,545.00){\usebox{\plotpoint}}
\multiput(905,545)(20.629,2.292){0}{\usebox{\plotpoint}}
\put(922.94,546.00){\usebox{\plotpoint}}
\multiput(923,546)(20.629,2.292){0}{\usebox{\plotpoint}}
\multiput(932,547)(20.756,0.000){0}{\usebox{\plotpoint}}
\put(943.63,547.18){\usebox{\plotpoint}}
\multiput(951,548)(20.756,0.000){0}{\usebox{\plotpoint}}
\put(964.34,548.00){\usebox{\plotpoint}}
\multiput(969,548)(20.652,2.065){0}{\usebox{\plotpoint}}
\put(985.05,549.00){\usebox{\plotpoint}}
\multiput(988,549)(20.629,2.292){0}{\usebox{\plotpoint}}
\put(1005.75,550.00){\usebox{\plotpoint}}
\multiput(1006,550)(20.756,0.000){0}{\usebox{\plotpoint}}
\multiput(1016,550)(20.629,2.292){0}{\usebox{\plotpoint}}
\put(1026.45,551.00){\usebox{\plotpoint}}
\multiput(1034,551)(20.756,0.000){0}{\usebox{\plotpoint}}
\put(1047.18,551.42){\usebox{\plotpoint}}
\multiput(1053,552)(20.756,0.000){0}{\usebox{\plotpoint}}
\put(1067.91,552.00){\usebox{\plotpoint}}
\multiput(1071,552)(20.629,2.292){0}{\usebox{\plotpoint}}
\put(1088.61,553.00){\usebox{\plotpoint}}
\multiput(1090,553)(20.756,0.000){0}{\usebox{\plotpoint}}
\multiput(1099,553)(20.756,0.000){0}{\usebox{\plotpoint}}
\put(1109.36,553.15){\usebox{\plotpoint}}
\multiput(1117,554)(20.756,0.000){0}{\usebox{\plotpoint}}
\put(1130.07,554.00){\usebox{\plotpoint}}
\put(1136,554){\usebox{\plotpoint}}
\sbox{\plotpoint}{\rule[-0.200pt]{0.400pt}{0.400pt}}%
\put(940,261){\makebox(0,0)[r]{$V_\eff$ with $\order(a^2)$}}
\put(962.0,261.0){\rule[-0.200pt]{15.899pt}{0.400pt}}
\put(220,302){\usebox{\plotpoint}}
\put(220,301.67){\rule{2.168pt}{0.400pt}}
\multiput(220.00,301.17)(4.500,1.000){2}{\rule{1.084pt}{0.400pt}}
\put(239,302.67){\rule{2.168pt}{0.400pt}}
\multiput(239.00,302.17)(4.500,1.000){2}{\rule{1.084pt}{0.400pt}}
\put(248,303.67){\rule{2.168pt}{0.400pt}}
\multiput(248.00,303.17)(4.500,1.000){2}{\rule{1.084pt}{0.400pt}}
\put(257,304.67){\rule{2.168pt}{0.400pt}}
\multiput(257.00,304.17)(4.500,1.000){2}{\rule{1.084pt}{0.400pt}}
\put(266,306.17){\rule{2.100pt}{0.400pt}}
\multiput(266.00,305.17)(5.641,2.000){2}{\rule{1.050pt}{0.400pt}}
\put(276,308.17){\rule{1.900pt}{0.400pt}}
\multiput(276.00,307.17)(5.056,2.000){2}{\rule{0.950pt}{0.400pt}}
\put(285,310.17){\rule{1.900pt}{0.400pt}}
\multiput(285.00,309.17)(5.056,2.000){2}{\rule{0.950pt}{0.400pt}}
\put(294,312.17){\rule{1.900pt}{0.400pt}}
\multiput(294.00,311.17)(5.056,2.000){2}{\rule{0.950pt}{0.400pt}}
\multiput(303.00,314.61)(2.025,0.447){3}{\rule{1.433pt}{0.108pt}}
\multiput(303.00,313.17)(7.025,3.000){2}{\rule{0.717pt}{0.400pt}}
\multiput(313.00,317.61)(1.802,0.447){3}{\rule{1.300pt}{0.108pt}}
\multiput(313.00,316.17)(6.302,3.000){2}{\rule{0.650pt}{0.400pt}}
\multiput(322.00,320.61)(1.802,0.447){3}{\rule{1.300pt}{0.108pt}}
\multiput(322.00,319.17)(6.302,3.000){2}{\rule{0.650pt}{0.400pt}}
\multiput(331.00,323.61)(1.802,0.447){3}{\rule{1.300pt}{0.108pt}}
\multiput(331.00,322.17)(6.302,3.000){2}{\rule{0.650pt}{0.400pt}}
\multiput(340.00,326.60)(1.358,0.468){5}{\rule{1.100pt}{0.113pt}}
\multiput(340.00,325.17)(7.717,4.000){2}{\rule{0.550pt}{0.400pt}}
\multiput(350.00,330.61)(1.802,0.447){3}{\rule{1.300pt}{0.108pt}}
\multiput(350.00,329.17)(6.302,3.000){2}{\rule{0.650pt}{0.400pt}}
\multiput(359.00,333.60)(1.212,0.468){5}{\rule{1.000pt}{0.113pt}}
\multiput(359.00,332.17)(6.924,4.000){2}{\rule{0.500pt}{0.400pt}}
\multiput(368.00,337.60)(1.212,0.468){5}{\rule{1.000pt}{0.113pt}}
\multiput(368.00,336.17)(6.924,4.000){2}{\rule{0.500pt}{0.400pt}}
\multiput(377.00,341.59)(1.044,0.477){7}{\rule{0.900pt}{0.115pt}}
\multiput(377.00,340.17)(8.132,5.000){2}{\rule{0.450pt}{0.400pt}}
\multiput(387.00,346.60)(1.212,0.468){5}{\rule{1.000pt}{0.113pt}}
\multiput(387.00,345.17)(6.924,4.000){2}{\rule{0.500pt}{0.400pt}}
\multiput(396.00,350.60)(1.212,0.468){5}{\rule{1.000pt}{0.113pt}}
\multiput(396.00,349.17)(6.924,4.000){2}{\rule{0.500pt}{0.400pt}}
\multiput(405.00,354.59)(0.933,0.477){7}{\rule{0.820pt}{0.115pt}}
\multiput(405.00,353.17)(7.298,5.000){2}{\rule{0.410pt}{0.400pt}}
\multiput(414.00,359.59)(1.044,0.477){7}{\rule{0.900pt}{0.115pt}}
\multiput(414.00,358.17)(8.132,5.000){2}{\rule{0.450pt}{0.400pt}}
\multiput(424.00,364.59)(0.933,0.477){7}{\rule{0.820pt}{0.115pt}}
\multiput(424.00,363.17)(7.298,5.000){2}{\rule{0.410pt}{0.400pt}}
\multiput(433.00,369.59)(0.933,0.477){7}{\rule{0.820pt}{0.115pt}}
\multiput(433.00,368.17)(7.298,5.000){2}{\rule{0.410pt}{0.400pt}}
\multiput(442.00,374.59)(0.933,0.477){7}{\rule{0.820pt}{0.115pt}}
\multiput(442.00,373.17)(7.298,5.000){2}{\rule{0.410pt}{0.400pt}}
\multiput(451.00,379.59)(1.044,0.477){7}{\rule{0.900pt}{0.115pt}}
\multiput(451.00,378.17)(8.132,5.000){2}{\rule{0.450pt}{0.400pt}}
\multiput(461.00,384.59)(0.933,0.477){7}{\rule{0.820pt}{0.115pt}}
\multiput(461.00,383.17)(7.298,5.000){2}{\rule{0.410pt}{0.400pt}}
\multiput(470.00,389.59)(0.933,0.477){7}{\rule{0.820pt}{0.115pt}}
\multiput(470.00,388.17)(7.298,5.000){2}{\rule{0.410pt}{0.400pt}}
\multiput(479.00,394.59)(0.933,0.477){7}{\rule{0.820pt}{0.115pt}}
\multiput(479.00,393.17)(7.298,5.000){2}{\rule{0.410pt}{0.400pt}}
\multiput(488.00,399.59)(1.044,0.477){7}{\rule{0.900pt}{0.115pt}}
\multiput(488.00,398.17)(8.132,5.000){2}{\rule{0.450pt}{0.400pt}}
\multiput(498.00,404.59)(0.933,0.477){7}{\rule{0.820pt}{0.115pt}}
\multiput(498.00,403.17)(7.298,5.000){2}{\rule{0.410pt}{0.400pt}}
\multiput(507.00,409.59)(0.933,0.477){7}{\rule{0.820pt}{0.115pt}}
\multiput(507.00,408.17)(7.298,5.000){2}{\rule{0.410pt}{0.400pt}}
\multiput(516.00,414.59)(0.933,0.477){7}{\rule{0.820pt}{0.115pt}}
\multiput(516.00,413.17)(7.298,5.000){2}{\rule{0.410pt}{0.400pt}}
\multiput(525.00,419.59)(1.044,0.477){7}{\rule{0.900pt}{0.115pt}}
\multiput(525.00,418.17)(8.132,5.000){2}{\rule{0.450pt}{0.400pt}}
\multiput(535.00,424.59)(0.933,0.477){7}{\rule{0.820pt}{0.115pt}}
\multiput(535.00,423.17)(7.298,5.000){2}{\rule{0.410pt}{0.400pt}}
\multiput(544.00,429.59)(0.933,0.477){7}{\rule{0.820pt}{0.115pt}}
\multiput(544.00,428.17)(7.298,5.000){2}{\rule{0.410pt}{0.400pt}}
\multiput(553.00,434.59)(0.933,0.477){7}{\rule{0.820pt}{0.115pt}}
\multiput(553.00,433.17)(7.298,5.000){2}{\rule{0.410pt}{0.400pt}}
\multiput(562.00,439.59)(1.044,0.477){7}{\rule{0.900pt}{0.115pt}}
\multiput(562.00,438.17)(8.132,5.000){2}{\rule{0.450pt}{0.400pt}}
\multiput(572.00,444.60)(1.212,0.468){5}{\rule{1.000pt}{0.113pt}}
\multiput(572.00,443.17)(6.924,4.000){2}{\rule{0.500pt}{0.400pt}}
\multiput(581.00,448.59)(0.933,0.477){7}{\rule{0.820pt}{0.115pt}}
\multiput(581.00,447.17)(7.298,5.000){2}{\rule{0.410pt}{0.400pt}}
\multiput(590.00,453.60)(1.212,0.468){5}{\rule{1.000pt}{0.113pt}}
\multiput(590.00,452.17)(6.924,4.000){2}{\rule{0.500pt}{0.400pt}}
\multiput(599.00,457.60)(1.358,0.468){5}{\rule{1.100pt}{0.113pt}}
\multiput(599.00,456.17)(7.717,4.000){2}{\rule{0.550pt}{0.400pt}}
\multiput(609.00,461.59)(0.933,0.477){7}{\rule{0.820pt}{0.115pt}}
\multiput(609.00,460.17)(7.298,5.000){2}{\rule{0.410pt}{0.400pt}}
\multiput(618.00,466.60)(1.212,0.468){5}{\rule{1.000pt}{0.113pt}}
\multiput(618.00,465.17)(6.924,4.000){2}{\rule{0.500pt}{0.400pt}}
\multiput(627.00,470.60)(1.212,0.468){5}{\rule{1.000pt}{0.113pt}}
\multiput(627.00,469.17)(6.924,4.000){2}{\rule{0.500pt}{0.400pt}}
\multiput(636.00,474.60)(1.358,0.468){5}{\rule{1.100pt}{0.113pt}}
\multiput(636.00,473.17)(7.717,4.000){2}{\rule{0.550pt}{0.400pt}}
\multiput(646.00,478.61)(1.802,0.447){3}{\rule{1.300pt}{0.108pt}}
\multiput(646.00,477.17)(6.302,3.000){2}{\rule{0.650pt}{0.400pt}}
\multiput(655.00,481.60)(1.212,0.468){5}{\rule{1.000pt}{0.113pt}}
\multiput(655.00,480.17)(6.924,4.000){2}{\rule{0.500pt}{0.400pt}}
\multiput(664.00,485.60)(1.212,0.468){5}{\rule{1.000pt}{0.113pt}}
\multiput(664.00,484.17)(6.924,4.000){2}{\rule{0.500pt}{0.400pt}}
\multiput(673.00,489.61)(2.025,0.447){3}{\rule{1.433pt}{0.108pt}}
\multiput(673.00,488.17)(7.025,3.000){2}{\rule{0.717pt}{0.400pt}}
\multiput(683.00,492.61)(1.802,0.447){3}{\rule{1.300pt}{0.108pt}}
\multiput(683.00,491.17)(6.302,3.000){2}{\rule{0.650pt}{0.400pt}}
\multiput(692.00,495.61)(1.802,0.447){3}{\rule{1.300pt}{0.108pt}}
\multiput(692.00,494.17)(6.302,3.000){2}{\rule{0.650pt}{0.400pt}}
\multiput(701.00,498.60)(1.212,0.468){5}{\rule{1.000pt}{0.113pt}}
\multiput(701.00,497.17)(6.924,4.000){2}{\rule{0.500pt}{0.400pt}}
\put(710,502.17){\rule{2.100pt}{0.400pt}}
\multiput(710.00,501.17)(5.641,2.000){2}{\rule{1.050pt}{0.400pt}}
\multiput(720.00,504.61)(1.802,0.447){3}{\rule{1.300pt}{0.108pt}}
\multiput(720.00,503.17)(6.302,3.000){2}{\rule{0.650pt}{0.400pt}}
\multiput(729.00,507.61)(1.802,0.447){3}{\rule{1.300pt}{0.108pt}}
\multiput(729.00,506.17)(6.302,3.000){2}{\rule{0.650pt}{0.400pt}}
\put(738,510.17){\rule{1.900pt}{0.400pt}}
\multiput(738.00,509.17)(5.056,2.000){2}{\rule{0.950pt}{0.400pt}}
\multiput(747.00,512.61)(2.025,0.447){3}{\rule{1.433pt}{0.108pt}}
\multiput(747.00,511.17)(7.025,3.000){2}{\rule{0.717pt}{0.400pt}}
\put(757,515.17){\rule{1.900pt}{0.400pt}}
\multiput(757.00,514.17)(5.056,2.000){2}{\rule{0.950pt}{0.400pt}}
\put(766,517.17){\rule{1.900pt}{0.400pt}}
\multiput(766.00,516.17)(5.056,2.000){2}{\rule{0.950pt}{0.400pt}}
\put(775,519.17){\rule{1.900pt}{0.400pt}}
\multiput(775.00,518.17)(5.056,2.000){2}{\rule{0.950pt}{0.400pt}}
\put(784,521.17){\rule{2.100pt}{0.400pt}}
\multiput(784.00,520.17)(5.641,2.000){2}{\rule{1.050pt}{0.400pt}}
\put(794,523.17){\rule{1.900pt}{0.400pt}}
\multiput(794.00,522.17)(5.056,2.000){2}{\rule{0.950pt}{0.400pt}}
\put(803,525.17){\rule{1.900pt}{0.400pt}}
\multiput(803.00,524.17)(5.056,2.000){2}{\rule{0.950pt}{0.400pt}}
\put(812,527.17){\rule{1.900pt}{0.400pt}}
\multiput(812.00,526.17)(5.056,2.000){2}{\rule{0.950pt}{0.400pt}}
\put(821,529.17){\rule{2.100pt}{0.400pt}}
\multiput(821.00,528.17)(5.641,2.000){2}{\rule{1.050pt}{0.400pt}}
\put(831,530.67){\rule{2.168pt}{0.400pt}}
\multiput(831.00,530.17)(4.500,1.000){2}{\rule{1.084pt}{0.400pt}}
\put(840,532.17){\rule{1.900pt}{0.400pt}}
\multiput(840.00,531.17)(5.056,2.000){2}{\rule{0.950pt}{0.400pt}}
\put(849,533.67){\rule{2.168pt}{0.400pt}}
\multiput(849.00,533.17)(4.500,1.000){2}{\rule{1.084pt}{0.400pt}}
\put(858,534.67){\rule{2.409pt}{0.400pt}}
\multiput(858.00,534.17)(5.000,1.000){2}{\rule{1.204pt}{0.400pt}}
\put(868,535.67){\rule{2.168pt}{0.400pt}}
\multiput(868.00,535.17)(4.500,1.000){2}{\rule{1.084pt}{0.400pt}}
\put(877,537.17){\rule{1.900pt}{0.400pt}}
\multiput(877.00,536.17)(5.056,2.000){2}{\rule{0.950pt}{0.400pt}}
\put(886,538.67){\rule{2.168pt}{0.400pt}}
\multiput(886.00,538.17)(4.500,1.000){2}{\rule{1.084pt}{0.400pt}}
\put(895,539.67){\rule{2.409pt}{0.400pt}}
\multiput(895.00,539.17)(5.000,1.000){2}{\rule{1.204pt}{0.400pt}}
\put(905,540.67){\rule{2.168pt}{0.400pt}}
\multiput(905.00,540.17)(4.500,1.000){2}{\rule{1.084pt}{0.400pt}}
\put(914,541.67){\rule{2.168pt}{0.400pt}}
\multiput(914.00,541.17)(4.500,1.000){2}{\rule{1.084pt}{0.400pt}}
\put(923,542.67){\rule{2.168pt}{0.400pt}}
\multiput(923.00,542.17)(4.500,1.000){2}{\rule{1.084pt}{0.400pt}}
\put(229.0,303.0){\rule[-0.200pt]{2.409pt}{0.400pt}}
\put(942,543.67){\rule{2.168pt}{0.400pt}}
\multiput(942.00,543.17)(4.500,1.000){2}{\rule{1.084pt}{0.400pt}}
\put(951,544.67){\rule{2.168pt}{0.400pt}}
\multiput(951.00,544.17)(4.500,1.000){2}{\rule{1.084pt}{0.400pt}}
\put(960,545.67){\rule{2.168pt}{0.400pt}}
\multiput(960.00,545.17)(4.500,1.000){2}{\rule{1.084pt}{0.400pt}}
\put(932.0,544.0){\rule[-0.200pt]{2.409pt}{0.400pt}}
\put(979,546.67){\rule{2.168pt}{0.400pt}}
\multiput(979.00,546.17)(4.500,1.000){2}{\rule{1.084pt}{0.400pt}}
\put(969.0,547.0){\rule[-0.200pt]{2.409pt}{0.400pt}}
\put(997,547.67){\rule{2.168pt}{0.400pt}}
\multiput(997.00,547.17)(4.500,1.000){2}{\rule{1.084pt}{0.400pt}}
\put(1006,548.67){\rule{2.409pt}{0.400pt}}
\multiput(1006.00,548.17)(5.000,1.000){2}{\rule{1.204pt}{0.400pt}}
\put(988.0,548.0){\rule[-0.200pt]{2.168pt}{0.400pt}}
\put(1025,549.67){\rule{2.168pt}{0.400pt}}
\multiput(1025.00,549.17)(4.500,1.000){2}{\rule{1.084pt}{0.400pt}}
\put(1016.0,550.0){\rule[-0.200pt]{2.168pt}{0.400pt}}
\put(1053,550.67){\rule{2.168pt}{0.400pt}}
\multiput(1053.00,550.17)(4.500,1.000){2}{\rule{1.084pt}{0.400pt}}
\put(1034.0,551.0){\rule[-0.200pt]{4.577pt}{0.400pt}}
\put(1071,551.67){\rule{2.168pt}{0.400pt}}
\multiput(1071.00,551.17)(4.500,1.000){2}{\rule{1.084pt}{0.400pt}}
\put(1062.0,552.0){\rule[-0.200pt]{2.168pt}{0.400pt}}
\put(1099,552.67){\rule{2.168pt}{0.400pt}}
\multiput(1099.00,552.17)(4.500,1.000){2}{\rule{1.084pt}{0.400pt}}
\put(1080.0,553.0){\rule[-0.200pt]{4.577pt}{0.400pt}}
\put(1108.0,554.0){\rule[-0.200pt]{6.745pt}{0.400pt}}
\sbox{\plotpoint}{\rule[-0.400pt]{0.800pt}{0.800pt}}%
\put(940,216){\makebox(0,0)[r]{$V_\eff$ with $\order(a^4)$}}
\put(962.0,216.0){\rule[-0.400pt]{15.899pt}{0.800pt}}
\put(220,282){\usebox{\plotpoint}}
\put(229,280.84){\rule{2.409pt}{0.800pt}}
\multiput(229.00,280.34)(5.000,1.000){2}{\rule{1.204pt}{0.800pt}}
\put(239,281.84){\rule{2.168pt}{0.800pt}}
\multiput(239.00,281.34)(4.500,1.000){2}{\rule{1.084pt}{0.800pt}}
\put(248,282.84){\rule{2.168pt}{0.800pt}}
\multiput(248.00,282.34)(4.500,1.000){2}{\rule{1.084pt}{0.800pt}}
\put(257,284.34){\rule{2.168pt}{0.800pt}}
\multiput(257.00,283.34)(4.500,2.000){2}{\rule{1.084pt}{0.800pt}}
\put(266,285.84){\rule{2.409pt}{0.800pt}}
\multiput(266.00,285.34)(5.000,1.000){2}{\rule{1.204pt}{0.800pt}}
\put(276,287.84){\rule{2.168pt}{0.800pt}}
\multiput(276.00,286.34)(4.500,3.000){2}{\rule{1.084pt}{0.800pt}}
\put(285,290.34){\rule{2.168pt}{0.800pt}}
\multiput(285.00,289.34)(4.500,2.000){2}{\rule{1.084pt}{0.800pt}}
\put(294,292.84){\rule{2.168pt}{0.800pt}}
\multiput(294.00,291.34)(4.500,3.000){2}{\rule{1.084pt}{0.800pt}}
\put(303,295.84){\rule{2.409pt}{0.800pt}}
\multiput(303.00,294.34)(5.000,3.000){2}{\rule{1.204pt}{0.800pt}}
\put(313,298.84){\rule{2.168pt}{0.800pt}}
\multiput(313.00,297.34)(4.500,3.000){2}{\rule{1.084pt}{0.800pt}}
\put(322,302.34){\rule{2.000pt}{0.800pt}}
\multiput(322.00,300.34)(4.849,4.000){2}{\rule{1.000pt}{0.800pt}}
\put(331,306.34){\rule{2.000pt}{0.800pt}}
\multiput(331.00,304.34)(4.849,4.000){2}{\rule{1.000pt}{0.800pt}}
\put(340,310.34){\rule{2.200pt}{0.800pt}}
\multiput(340.00,308.34)(5.434,4.000){2}{\rule{1.100pt}{0.800pt}}
\multiput(350.00,315.38)(1.096,0.560){3}{\rule{1.640pt}{0.135pt}}
\multiput(350.00,312.34)(5.596,5.000){2}{\rule{0.820pt}{0.800pt}}
\put(359,319.34){\rule{2.000pt}{0.800pt}}
\multiput(359.00,317.34)(4.849,4.000){2}{\rule{1.000pt}{0.800pt}}
\multiput(368.00,324.38)(1.096,0.560){3}{\rule{1.640pt}{0.135pt}}
\multiput(368.00,321.34)(5.596,5.000){2}{\rule{0.820pt}{0.800pt}}
\multiput(377.00,329.38)(1.264,0.560){3}{\rule{1.800pt}{0.135pt}}
\multiput(377.00,326.34)(6.264,5.000){2}{\rule{0.900pt}{0.800pt}}
\multiput(387.00,334.38)(1.096,0.560){3}{\rule{1.640pt}{0.135pt}}
\multiput(387.00,331.34)(5.596,5.000){2}{\rule{0.820pt}{0.800pt}}
\multiput(396.00,339.38)(1.096,0.560){3}{\rule{1.640pt}{0.135pt}}
\multiput(396.00,336.34)(5.596,5.000){2}{\rule{0.820pt}{0.800pt}}
\multiput(405.00,344.39)(0.797,0.536){5}{\rule{1.400pt}{0.129pt}}
\multiput(405.00,341.34)(6.094,6.000){2}{\rule{0.700pt}{0.800pt}}
\multiput(414.00,350.38)(1.264,0.560){3}{\rule{1.800pt}{0.135pt}}
\multiput(414.00,347.34)(6.264,5.000){2}{\rule{0.900pt}{0.800pt}}
\multiput(424.00,355.39)(0.797,0.536){5}{\rule{1.400pt}{0.129pt}}
\multiput(424.00,352.34)(6.094,6.000){2}{\rule{0.700pt}{0.800pt}}
\multiput(433.00,361.39)(0.797,0.536){5}{\rule{1.400pt}{0.129pt}}
\multiput(433.00,358.34)(6.094,6.000){2}{\rule{0.700pt}{0.800pt}}
\multiput(442.00,367.38)(1.096,0.560){3}{\rule{1.640pt}{0.135pt}}
\multiput(442.00,364.34)(5.596,5.000){2}{\rule{0.820pt}{0.800pt}}
\multiput(451.00,372.39)(0.909,0.536){5}{\rule{1.533pt}{0.129pt}}
\multiput(451.00,369.34)(6.817,6.000){2}{\rule{0.767pt}{0.800pt}}
\multiput(461.00,378.39)(0.797,0.536){5}{\rule{1.400pt}{0.129pt}}
\multiput(461.00,375.34)(6.094,6.000){2}{\rule{0.700pt}{0.800pt}}
\multiput(470.00,384.39)(0.797,0.536){5}{\rule{1.400pt}{0.129pt}}
\multiput(470.00,381.34)(6.094,6.000){2}{\rule{0.700pt}{0.800pt}}
\multiput(479.00,390.39)(0.797,0.536){5}{\rule{1.400pt}{0.129pt}}
\multiput(479.00,387.34)(6.094,6.000){2}{\rule{0.700pt}{0.800pt}}
\multiput(488.00,396.38)(1.264,0.560){3}{\rule{1.800pt}{0.135pt}}
\multiput(488.00,393.34)(6.264,5.000){2}{\rule{0.900pt}{0.800pt}}
\multiput(498.00,401.39)(0.797,0.536){5}{\rule{1.400pt}{0.129pt}}
\multiput(498.00,398.34)(6.094,6.000){2}{\rule{0.700pt}{0.800pt}}
\multiput(507.00,407.39)(0.797,0.536){5}{\rule{1.400pt}{0.129pt}}
\multiput(507.00,404.34)(6.094,6.000){2}{\rule{0.700pt}{0.800pt}}
\multiput(516.00,413.38)(1.096,0.560){3}{\rule{1.640pt}{0.135pt}}
\multiput(516.00,410.34)(5.596,5.000){2}{\rule{0.820pt}{0.800pt}}
\multiput(525.00,418.39)(0.909,0.536){5}{\rule{1.533pt}{0.129pt}}
\multiput(525.00,415.34)(6.817,6.000){2}{\rule{0.767pt}{0.800pt}}
\multiput(535.00,424.38)(1.096,0.560){3}{\rule{1.640pt}{0.135pt}}
\multiput(535.00,421.34)(5.596,5.000){2}{\rule{0.820pt}{0.800pt}}
\multiput(544.00,429.39)(0.797,0.536){5}{\rule{1.400pt}{0.129pt}}
\multiput(544.00,426.34)(6.094,6.000){2}{\rule{0.700pt}{0.800pt}}
\multiput(553.00,435.38)(1.096,0.560){3}{\rule{1.640pt}{0.135pt}}
\multiput(553.00,432.34)(5.596,5.000){2}{\rule{0.820pt}{0.800pt}}
\multiput(562.00,440.38)(1.264,0.560){3}{\rule{1.800pt}{0.135pt}}
\multiput(562.00,437.34)(6.264,5.000){2}{\rule{0.900pt}{0.800pt}}
\multiput(572.00,445.38)(1.096,0.560){3}{\rule{1.640pt}{0.135pt}}
\multiput(572.00,442.34)(5.596,5.000){2}{\rule{0.820pt}{0.800pt}}
\multiput(581.00,450.38)(1.096,0.560){3}{\rule{1.640pt}{0.135pt}}
\multiput(581.00,447.34)(5.596,5.000){2}{\rule{0.820pt}{0.800pt}}
\multiput(590.00,455.38)(1.096,0.560){3}{\rule{1.640pt}{0.135pt}}
\multiput(590.00,452.34)(5.596,5.000){2}{\rule{0.820pt}{0.800pt}}
\multiput(599.00,460.38)(1.264,0.560){3}{\rule{1.800pt}{0.135pt}}
\multiput(599.00,457.34)(6.264,5.000){2}{\rule{0.900pt}{0.800pt}}
\put(609,464.34){\rule{2.000pt}{0.800pt}}
\multiput(609.00,462.34)(4.849,4.000){2}{\rule{1.000pt}{0.800pt}}
\multiput(618.00,469.38)(1.096,0.560){3}{\rule{1.640pt}{0.135pt}}
\multiput(618.00,466.34)(5.596,5.000){2}{\rule{0.820pt}{0.800pt}}
\put(627,473.34){\rule{2.000pt}{0.800pt}}
\multiput(627.00,471.34)(4.849,4.000){2}{\rule{1.000pt}{0.800pt}}
\put(636,477.34){\rule{2.200pt}{0.800pt}}
\multiput(636.00,475.34)(5.434,4.000){2}{\rule{1.100pt}{0.800pt}}
\put(646,481.34){\rule{2.000pt}{0.800pt}}
\multiput(646.00,479.34)(4.849,4.000){2}{\rule{1.000pt}{0.800pt}}
\put(655,485.34){\rule{2.000pt}{0.800pt}}
\multiput(655.00,483.34)(4.849,4.000){2}{\rule{1.000pt}{0.800pt}}
\put(664,488.84){\rule{2.168pt}{0.800pt}}
\multiput(664.00,487.34)(4.500,3.000){2}{\rule{1.084pt}{0.800pt}}
\put(673,492.34){\rule{2.200pt}{0.800pt}}
\multiput(673.00,490.34)(5.434,4.000){2}{\rule{1.100pt}{0.800pt}}
\put(683,495.84){\rule{2.168pt}{0.800pt}}
\multiput(683.00,494.34)(4.500,3.000){2}{\rule{1.084pt}{0.800pt}}
\put(692,498.84){\rule{2.168pt}{0.800pt}}
\multiput(692.00,497.34)(4.500,3.000){2}{\rule{1.084pt}{0.800pt}}
\put(701,501.84){\rule{2.168pt}{0.800pt}}
\multiput(701.00,500.34)(4.500,3.000){2}{\rule{1.084pt}{0.800pt}}
\put(710,504.84){\rule{2.409pt}{0.800pt}}
\multiput(710.00,503.34)(5.000,3.000){2}{\rule{1.204pt}{0.800pt}}
\put(720,507.84){\rule{2.168pt}{0.800pt}}
\multiput(720.00,506.34)(4.500,3.000){2}{\rule{1.084pt}{0.800pt}}
\put(729,510.84){\rule{2.168pt}{0.800pt}}
\multiput(729.00,509.34)(4.500,3.000){2}{\rule{1.084pt}{0.800pt}}
\put(738,513.34){\rule{2.168pt}{0.800pt}}
\multiput(738.00,512.34)(4.500,2.000){2}{\rule{1.084pt}{0.800pt}}
\put(747,515.84){\rule{2.409pt}{0.800pt}}
\multiput(747.00,514.34)(5.000,3.000){2}{\rule{1.204pt}{0.800pt}}
\put(757,518.34){\rule{2.168pt}{0.800pt}}
\multiput(757.00,517.34)(4.500,2.000){2}{\rule{1.084pt}{0.800pt}}
\put(766,520.34){\rule{2.168pt}{0.800pt}}
\multiput(766.00,519.34)(4.500,2.000){2}{\rule{1.084pt}{0.800pt}}
\put(775,522.34){\rule{2.168pt}{0.800pt}}
\multiput(775.00,521.34)(4.500,2.000){2}{\rule{1.084pt}{0.800pt}}
\put(784,524.34){\rule{2.409pt}{0.800pt}}
\multiput(784.00,523.34)(5.000,2.000){2}{\rule{1.204pt}{0.800pt}}
\put(794,526.34){\rule{2.168pt}{0.800pt}}
\multiput(794.00,525.34)(4.500,2.000){2}{\rule{1.084pt}{0.800pt}}
\put(803,527.84){\rule{2.168pt}{0.800pt}}
\multiput(803.00,527.34)(4.500,1.000){2}{\rule{1.084pt}{0.800pt}}
\put(812,529.34){\rule{2.168pt}{0.800pt}}
\multiput(812.00,528.34)(4.500,2.000){2}{\rule{1.084pt}{0.800pt}}
\put(821,531.34){\rule{2.409pt}{0.800pt}}
\multiput(821.00,530.34)(5.000,2.000){2}{\rule{1.204pt}{0.800pt}}
\put(831,532.84){\rule{2.168pt}{0.800pt}}
\multiput(831.00,532.34)(4.500,1.000){2}{\rule{1.084pt}{0.800pt}}
\put(840,533.84){\rule{2.168pt}{0.800pt}}
\multiput(840.00,533.34)(4.500,1.000){2}{\rule{1.084pt}{0.800pt}}
\put(849,535.34){\rule{2.168pt}{0.800pt}}
\multiput(849.00,534.34)(4.500,2.000){2}{\rule{1.084pt}{0.800pt}}
\put(858,536.84){\rule{2.409pt}{0.800pt}}
\multiput(858.00,536.34)(5.000,1.000){2}{\rule{1.204pt}{0.800pt}}
\put(868,537.84){\rule{2.168pt}{0.800pt}}
\multiput(868.00,537.34)(4.500,1.000){2}{\rule{1.084pt}{0.800pt}}
\put(877,538.84){\rule{2.168pt}{0.800pt}}
\multiput(877.00,538.34)(4.500,1.000){2}{\rule{1.084pt}{0.800pt}}
\put(886,539.84){\rule{2.168pt}{0.800pt}}
\multiput(886.00,539.34)(4.500,1.000){2}{\rule{1.084pt}{0.800pt}}
\put(895,540.84){\rule{2.409pt}{0.800pt}}
\multiput(895.00,540.34)(5.000,1.000){2}{\rule{1.204pt}{0.800pt}}
\put(905,541.84){\rule{2.168pt}{0.800pt}}
\multiput(905.00,541.34)(4.500,1.000){2}{\rule{1.084pt}{0.800pt}}
\put(914,542.84){\rule{2.168pt}{0.800pt}}
\multiput(914.00,542.34)(4.500,1.000){2}{\rule{1.084pt}{0.800pt}}
\put(220.0,282.0){\rule[-0.400pt]{2.168pt}{0.800pt}}
\put(932,543.84){\rule{2.409pt}{0.800pt}}
\multiput(932.00,543.34)(5.000,1.000){2}{\rule{1.204pt}{0.800pt}}
\put(942,544.84){\rule{2.168pt}{0.800pt}}
\multiput(942.00,544.34)(4.500,1.000){2}{\rule{1.084pt}{0.800pt}}
\put(923.0,545.0){\rule[-0.400pt]{2.168pt}{0.800pt}}
\put(960,545.84){\rule{2.168pt}{0.800pt}}
\multiput(960.00,545.34)(4.500,1.000){2}{\rule{1.084pt}{0.800pt}}
\put(951.0,547.0){\rule[-0.400pt]{2.168pt}{0.800pt}}
\put(979,546.84){\rule{2.168pt}{0.800pt}}
\multiput(979.00,546.34)(4.500,1.000){2}{\rule{1.084pt}{0.800pt}}
\put(988,547.84){\rule{2.168pt}{0.800pt}}
\multiput(988.00,547.34)(4.500,1.000){2}{\rule{1.084pt}{0.800pt}}
\put(969.0,548.0){\rule[-0.400pt]{2.409pt}{0.800pt}}
\put(1016,548.84){\rule{2.168pt}{0.800pt}}
\multiput(1016.00,548.34)(4.500,1.000){2}{\rule{1.084pt}{0.800pt}}
\put(997.0,550.0){\rule[-0.400pt]{4.577pt}{0.800pt}}
\put(1034,549.84){\rule{2.168pt}{0.800pt}}
\multiput(1034.00,549.34)(4.500,1.000){2}{\rule{1.084pt}{0.800pt}}
\put(1025.0,551.0){\rule[-0.400pt]{2.168pt}{0.800pt}}
\put(1062,550.84){\rule{2.168pt}{0.800pt}}
\multiput(1062.00,550.34)(4.500,1.000){2}{\rule{1.084pt}{0.800pt}}
\put(1043.0,552.0){\rule[-0.400pt]{4.577pt}{0.800pt}}
\put(1090,551.84){\rule{2.168pt}{0.800pt}}
\multiput(1090.00,551.34)(4.500,1.000){2}{\rule{1.084pt}{0.800pt}}
\put(1071.0,553.0){\rule[-0.400pt]{4.577pt}{0.800pt}}
\put(1127,552.84){\rule{2.168pt}{0.800pt}}
\multiput(1127.00,552.34)(4.500,1.000){2}{\rule{1.084pt}{0.800pt}}
\put(1099.0,554.0){\rule[-0.400pt]{6.745pt}{0.800pt}}
\end{picture}
\end{center}
\caption{The true potential is plotted versus~$r$ together with the $a^2$ and
$a^4$ effective potentials ($a\!=\!1$).}
\label{V_vs_Veff-fig}
\end{figure}

In Figure~\ref{V_vs_Veff-fig} I compare the true potential, $V(\rv)$, 
with the $V_\eff$ having
only the $a^2$ correction, and with that having both $a^2$ and $a^4$
corrections. The two $V_\eff$'s look almost identical, and show no sign of
converging to $V$. Nevertheless the $a^4$~theory gives much
more accurate results than the $a^2$~theory.

Another misconception, suggested by the discussion leading
to~\eq{Vs-expansion}, is that the Fourier transform of $V_\eff$ converges to
that of $V$ for momenta $q\!\ll\!1/a$. This is not the case
because, as we discussed, the effective potential also corrects for
high-momentum intermediate states omitted by the cutoff, and so is related in
a nonlinear way to the true potential.

Another observation that is sometimes made is that our procedure is mere
curve fitting\,---\,if you add more parameters, you get better answers.
Indeed it is curve fitting, but highly optimized curve fitting and
systematic: errors are removed order-by-order in $a$. At each order,
renormalization theory tells us how many parameters are needed, and what
form the corrections should have. That is, it tells us 
which parameters are worth
tuning. So while one could treat our $a^2$ theory as a two-parameter theory,
varying $c$ and $a$ to get the best fit, renormalization theory says 
that a much better two-parameter theory is obtained by
freezing~$a$ at an appropriate value, and then 
tuning $c$ and $d_1$ in
our $a^4$ theory. And it tells us roughly how much better as well.

\subsection{Operators and the Operator Product Expansion}
So far we have used our effective theory to compute binding energies and
phase shifts. We now examine quantities that depend in detail on
the wavefunctions. Consider, for example, the matrix element
$\langle n | \pfour |n\rangle$, which might be important if we wished to
include relativistic corrections in our potential model. In
Table~\ref{p4-tab} I list values of this matrix element for several
$S$-states both for the true theory, and for our corrected theory (with
$a\!=\!1$). The values disagree by more than a factor of two, even for very
low-energy states,  despite the fact that the two theories agree
on the corresponding binding energies to several digits.

\begin{table}
\caption{Expectation values of $\pfour$ for $S$-states in the
true theory, and in the effective theory ($a\!=\!1$). These are compared 
with matrix elements of the renormalized operator in the effective
theory that corresponds to
$\pv^4$ in the true theory.}
\begin{center}
\begin{tabular}{|rlll|}
\hline 
level & $\langle \pfour \rangle$ & $\langle \pfour \rangle_\eff$ & 
$\langle Z\pfour+\gamma\delta^3_a/a+\cdots\rangle_\eff$\tstrut \\ \hline
$1S$ & 69         & 5.9   &  28\tstrut    \\
$2S$ & 5.50       & 1.7   &  5.34    \\
$3S$ & 1.309      & 0.4   &  1.306    \\
$4S$ & 0.5070     & 0.17  &  0.5068    \\
$5S$ & 0.24740    & 0.08  &  0.24738    \\
$6S$ & 0.138784   & 0.05  &  0.138780    \\ 
% $10S$& 0.0282154  & 0.010 &    tuned          \\
% $15S$& 0.0479190  & 0.017 &   tuned          \\
% $20S$& 0.0194328  & 0.007 &    tuned         \\
\hline
\end{tabular}
\end{center}
\label{p4-tab}
\end{table}

The problem is that the operator in the effective theory 
that corresponds to
$\pfour$ in the true theory is not~$\pfour$. As is true of the
hamiltonian, there are local corrections to $\pfour$ in the effective
theory. Thus, for any $S$~state, we expect
\be 
\langle \pfour \rangle_\true = Z\,\langle \pfour \rangle_\eff
+ \frac{\gamma}{a}\,\langle \delta^3_a(\rv)\rangle_\eff
+ \eta\,a\,\langle \nabla^2\delta^3_a(\rv) \rangle_\eff
+ \order(a^3)
\label{p4-eq}
\ee
where $Z$, $\gamma$ and $\eta$ are dimensionless constants that are independent
of the state. I tuned these constants so that
\eq{p4-eq} gave correct values for the $10S$, $15S$ and $20S$ bound states,
and found 
$Z\!=\!1$, $\gamma\!=\!-96.2$ and $\eta\!=\!-140.6$ when $a\!=\!1$. 
This formula gives excellent results, as
shown in Table~\ref{p4-tab}.
Just as for the binding energies, the relative error decreases steadily with
decreasing energy.

This example illustrates the subtlety of the equivalence between the effective
theory and the true theory. Although the infrared spectra and phase shifts
are quite similar, the short-distance structure of corresponding wavefunctions
in the two theories may be completely different. This means that matrix
elements of operators sensitive to short distances, like~$\pfour$, are not
equal, even for very low-energy states. Renormalization
theory, however,
tells us how to define correction terms for the operators in the effective
theory that correct for the differences in short-distance behavior between
the two theories. Just as in the hamiltonian, the corrections are local
operators with state-independent coupling constants, and, again, only a
finite number of terms is needed to any given order in~$a$.

\begin{table}
\caption{Wavefunctions evaluated at $r\!=\!0$ for $S$-states in the
true theory, and in the effective theory ($a\!=\!1$). These are compared 
with
the renormalized expression in the effective theory that corresponds to the
wavefunction at the origin in the true theory.}
\begin{center}
\begin{tabular}{|rlll|}
\hline 
level & $\psi(0)$ & $\psi_\eff(0)$ &
$\overline{\gamma}\int\psi_\eff\delta^3_a+\cdots$ \tstrut \\ \hline
$1S$ & 1.50 & 0.53  & $-3.4$  \tstrut \\
$2S$ & 0.383  & 0.19  & 0.369  \\
$3S$ & 0.1837  & 0.09  & 0.1830  \\
$4S$ & 0.11353  & 0.06  & 0.11344  \\
$5S$ & 0.079005  & 0.04  & 0.078986  \\
$6S$ & 0.059031  & 0.03  & 0.059025  \\
% $10S$ & 0.0264958 & 0.0132285 & 0.0264958 \\
% $15S$ & 0.0344604 & 0.0172034 & 0.0344604 \\
%$20S$ & 0.0219212 & 0.0109432 & 0.0219212 \\
\hline
\end{tabular}
\end{center}
\label{psi0-tab}
\end{table}

Another example of the same idea is provided by the wavefunction at the origin,
which is often needed in phenomenological work. In 
Table~\ref{psi0-tab} I list
values of the wave function at the origin for several $S$-wave bound states.
Again the effective theory gives results that are completely different from the
true theory. Renormalization theory, however, implies that
\bearray
\psi_\true(r\!=\!0) &=& \overline{\gamma} \,\int\!{d^3r}\,
\psi_\eff\,\delta_a^3(\rv) \nl
&+&\overline{\eta}\,a^2\,\int\!{d^3r}\,\psi_\eff\,\nabla^2\delta_a^3(\rv)
\nl
&+& \order(a^4)
\eearray
where $\overline{\gamma}$ and $\overline{\eta}$ are
dimensionless, state-independent constants. For $a\!=\!1$, I found
$\overline{\gamma}\!=\!-28.53$ and
$\overline{\eta}\!=\!-3.615$ by tuning this formula to give correct results for
the $20S$ and $15S$ bound states. With these values, the results from
the effective theory for $\psi_\true(0)$ are given in the last column of
Table~\ref{psi0-tab}. Again the agreement is excellent for all but the
most ultraviolet of the states, and it gets better as the states are more
infrared\,---\,just as expected.

The last formula is a special case of one valid for
any $r\!\le\!a$:
\bearray
\psi_\true(r) &=& \overline{\gamma}(r) \,\int\!{d^3r}\,
\psi_\eff\,\delta_a^3(\rv) \nl
&+&\overline{\eta}(r)\,a^2\,\int\!{
d^3r}\,\psi_\eff\,\nabla^2\delta_a^3(\rv)
\nl
&+& \order(a^4), \label{psi(r)}
\eearray
where coefficient functions $\overline{\gamma}(r)$ and $\overline{\eta}(r)$
are state-independent. This formula is striking. It says that given only
the two coefficient functions,  we can use the effective theory to compute
the small~$r$ 
behavior of {\em any\/} low-energy $S$-state, bound or otherwise.
All the information needed concerning the short-distance behavior of the true
theory is summarized in the coefficient functions; everything we need to know
about the long-distance structure of the state is given by two numbers, the
values of the two integrals. This separation into short-distance and
long-distance factors is often referred to as ``factorization.''

Factorization is most useful when either the
short-distance or long-distance physics is particularly simple. For example,
\eq{psi(r)} is valid for the quark-antiquark wavefunction describing
an $\psi/J$~meson. The long-distance structure of the wavefunction cannot be
easily computed using QCD, but the short-distance coefficient functions are
readily calculated using perturbation theory, since QCD becomes perturbative
at short distances. Thus we know the small~$r$ behavior of the wavefunction
up to two parameters\,---\,the values of the integrals over the
unknown wavefunction. This is enough information to allow us to compute the
momentum-dependence of form factors involving the~$\Upsilon$ at high
momentum.\cite{lepage-brodsky}

\eq{psi(r)} may be modified to include a
term~$\overline{Z}\,\psi_\eff(r)$ on the right-hand side. Then the
relation works for large $r$'s as well, since
$\psi_\eff(r)\!\propto\!\psi(r)$ for $r\!\gg\!a$.

Formula~(\ref{psi(r)}) is an example of the ``operator product expansion''
or OPE, which is an important tool in modern field theory.\footnote{The
operators forming the ``product'' here are the field
operators associated with the two bound particles. In quantum field theory, the
wavefunction is a matrix element of the product of these operators.}
It is reminiscent of a Taylor expansion: the
small~$x$ behavior of an analytic function~$f(x)$ may be specified by  a set
of known functions, $1,x,x^2\ldots$, and a set of constants,
$f(0),f^\prime(0)\ldots$. But our expansion is not a Taylor expansion.
The wavefunctions in our true theory are not
analytic at $r\!=\!0$: $\nabla^2\psi$ is infinite there because, as it
happens, $V(0)\!=\!-\infty$. Indeed the
wavefunction diverges at the origin if, for example, the potential
includes an attractive $1/r^2$ potential.
\eq{psi(r)}~still works in such cases;
the divergent behavior is captured by the coefficient functions.
In this particular case one can show that the leading coefficient
function contain a powerlaw divergence of the form $(a/r)^\xi$. Parameter
$\xi$ is referred to as an ``anomalous dimension.''

\subsection{Perturbative Derivation}\label{pert-deriv}

Effective theories are useful in two common situations. In one, the
short-distance behavior of the true theory is unknown, as above, but
low-energy data is available from which we can systematically construct the
low-energy effective theory. In the other, the short-distance
behavior is completely understood, but a low-energy approximation is sought
in order to simplify calculations of low-energy behavior. This latter
situation is encountered in applications of both~QCD and~QED.
In this section we illustrate the second situation:  We 
construct an effective theory to model the low-energy
behavior of a theory whose short-distance, high-momentum behavior is
perturbative. 
We can calculate the coupling constants in the effective theory using
perturbation theory because the short-distance dynamics is
perturbative. Our simple analysis illustrates how and why effective
theories work. 

To keep things as simple as possible, we  take the Schr\"odinger-Coulomb
problem as our underlying theory,
\be
H = \frac{\pv^2}{2m} - \frac{\alpha}{r}
\ee
with $m\!=\!100$ and $\alpha\!=\!.01$. We model the low-energy behavior of
this theory using our standard effective theory:
\be
H_\eff = \frac{\pv^2}{2m} - \frac{\alpha}{r}\,\erf(r/\sqrt{2}\,a) 
- 2\pi\alpha\,c\,a^2\,\delta^3_a(\rv).
\ee
The challenge is to determine the value of coupling constant~$c$ using
perturbation theory.

The technique by which we determine~$c$ is called ``perturbative
matching,'' and is standard in QED  and QCD applications. We compute the
scattering amplitude~$T(\pv,\qv)$, 
where $\pv$ is the initial momentum and $\qv$
the momentum transfer, using perturbation theory in the underlying theory. We
then compute the same scattering amplitude using perturbation theory in the
effective theory, to obtain an amplitude~$T_\eff(\pv,\qv,c,a)$ 
that also depends
on the unknown coupling constant~$c$ and the cutoff~$a$. We determine~$c$ by
requiring
\be
T(\pv,\qv) = T_\eff(\pv,\qv,c,a)\times\left(1+\order((p\,a)^4)\right).
\ee
Given~$a$, this equation can be solved for~$c$ in the limit $p\,a\!\to\!0$,
order-by-order in~$\alpha$. 

Note that perturbation theory doesn't
converge for either of these amplitudes in the limit $p\!\to\!0$, because of
the (infrared) Coulomb threshold singularity.\footnote{Perturbation theory in
a potential~$V(r)$ generally fails 
near threshold if either potential $V$ or potential $-V$
has bound states.} Coupling constant~$c$,
however, has a convergent perturbation series, since it is sensitive only to
the short-distance behavior in the theory. Thus, for our
purposes, we can compute the scattering amplitudes using ordinary
perturbation theory, ignoring the infrared problems of the expansion. When
we solve for~$c$, the infrared parts of $T$ and~$T_\eff$ cancel
completely, leaving behind a convergent, infrared-free perturbation
expansion.  This works because, although
the infrared parts of the perturbative
expansions for $T$ or~$T_\eff$ may be completely incorrect, the ultraviolet
parts are
fine, and that is all we need to compute~$c$. This point is essential
in dealing with QCD where, because of confinement, perturbation theory is
always inapplicable to low-energy on-shell quark scattering\,---\,except in
calculations of the sort we are doing here.

In lowest-order perturbation theory (the Born approximation), the two
amplitudes are
\be
T^{(1)}(\qv) = -\frac{4\pi\alpha}{q^2}
\ee
and
\bearray
T^{(1)}_\eff(\qv) &=&
-\frac{4\pi\alpha}{q^2}\,\e^{-q^2a^2/2}\,\left(1+c\,q^2a^2/2\right)
\nl
&=&-\frac{4\pi\alpha}{q^2}\,\left(1 + (c-1)\,q^2a^2/2 + \order(q^4a^4)\right).
\label{T1eff}
\eearray
Thus the two amplitudes match to the requisite order if $c\!=\!1$. The
lowest-order analysis is complete. We can verify our result using the same
computer code used in the previous sections, with suitably modified
potentials. I calculated the error in the $1S$~binding  energy when
$a\!=\!.01$ using the effective theory with different values of~$c$. The
result,
\be
\mbox{error in $1S$ energy} = \cases{.04\% & \mbox{for $c\!=\!0$}\cr
           .0004\% & \mbox{for $c\!=\!1$}\cr},
\ee
shows substantial improvement when the lowest-order contact term is
included.

In second-order perturbation theory, the scattering amplitude in the full
theory is
\be
T^{(2)} =
\int\frac{d^3k}{(2\pi)^3}\,T^{(1)}(\kv-\qv)\,\frac{2m}{p^2-(\kv+\pv)^2+{\rm
i}\epsilon}\,T^{(1)}(\kv),
\ee
while $T_\eff^{(2)}$ is given by the same expression but with
$T^{(1)}\!\to\!T^{(1)}_\eff$. We must compare these two amplitudes. When 
loop momentum~$k$ is small compared with~$1/a$, the integrands of the two
amplitudes become almost identical since, by construction,
\be
T^{(1)}(\kv) = T^{(1)}_\eff(\kv)\left(1+(k^4a^4)\right).
\ee
The two integrands, however, differ significantly 
when~$k$ is of order~$1/a$ or larger. 
Thus we can neglect~$p$ and~$q$ relative to~$k$ when
computing the difference between $T^{(2)}_\eff$ and
$T^{(2)}$, thereby
significantly simplifying the calculation:
\bearray
T^{(2)}_\eff - T^{(2)} &\approx& \int \frac{d^3k}{(2\pi)^3}\,\left[
\left(\frac{-4\pi\alpha}{k^2}\,\frac{-2m}{k^2}\,\frac{-4\pi\alpha}{k^2}
\right)\,\e^{-k^2a^2}(1+k^2a^2/2)^2 \right. \nl
&&\left.\rule{5em}{0ex}-\left(\frac{-4\pi\alpha}{k^2}\,\frac{-2m}{k^2}\,
\frac{-4\pi\alpha}{k^2}
\right)\right]\nl
&=& -2\pi\alpha^2\,a\,m\,0.940
\label{T2-diff}
\eearray
The second-order amplitudes do not agree, and therefore we need an extra
correction in the effective theory. Since the difference between the
amplitudes is momentum independent, we can correct for it by modifying the
value of coupling constant~$c$ in the effective theory:
\be
c = 1 + 0.94\,\alpha\,a\,m.
\ee
Then there is an $\order(\alpha^2)$~contribution from
the tree-level amplitude, \eq{T1eff}, that cancels the error
in~$T^{(2)}_\eff$.  
The coupling constant is said to have been ``renormalized'' by one-loop
quantum fluctuations. Again we can verify our one-loop correction numerically;
 the result,
\be
\mbox{error in $1S$ energy} = 
\cases{4\times10^{-2}\,\% & \mbox{for $c\!=\!0$}\cr
       4\times10^{-4}\,\% & \mbox{for $c\!=\!1$}\cr
       8\times10^{-7}\,\% & \mbox{for $c\!=\!1.0094$}\cr},
\ee
is greatly improved.

Our analysis illustrates why the renormalization of $c$ depends on high-energy
details of the theory: in \eq{T2-diff}, the entire contribution
is from loop momenta of order $1/a$ or larger. This analysis also shows why
the form of the correction terms is universal. The difference $T^{(2)}_\eff -
T^{(2)}$ comes from loop momenta $k\approx 1/a \gg p,q$. Thus in general 
we can Taylor expand the integrand in powers of the external momenta
to obtain ultimately
\be
T^{(2)}_\eff - T^{(2)} = c_0 + c_2\,q^2a^2 +\cdots.
\ee
The $c_0$ term is corrected by a $\delta(\rv)$ potential, the $c_2$~term by
a $\nabla^2\delta(\rv)$ potential, and so on. The form of these potential is
dictated by Taylor's theorem, and is independent of the actual
short-distance physics.

Note that the integral in \eq{T2-diff} becomes infrared divergent if
one expands the integrand beyond leading order in~$p/k$ or~$q/k$.
This is
because we have only matched the $T^{(1)}$'s through order $q^2a^2$,
while the next-to-leading error in $T^{(2)}_\eff$ contributes in
order~$q^4a^4$. This is inconsistent, and consistency is essential.
In general
the expansion of a one-loop correction works if carried out to the
same  order in~$a$ as the tree-level analysis, but not beyond that order. 
The appearance of an infrared
divergence in the calculation of a coupling constant is a sure sign that
something has been omitted from the effective theory at lower order.

Our earlier analyses of operators like~$\pfour$, and of the wavefunction at
small~$r$ can be repeated for the perturbative theory, using perturbation
theory to calculate the coupling constants and coefficient functions. I invite
you to try it.

\section{Rigorous Potential Models}

Potential models are widely used as approximations to more complex quantum
field theories. For example, the QCD dynamics of a heavy-quark meson like
the $\psi/J$ or the $\Upsilon$ is described by a phenomenological quark
potential model. Similarly the dynamics of the electron and positron in
positronium are well approximated by a simple Schr\"odinger equation.
Another example, which we discuss in a later section, is the nucleon-nucleon
force, which is often described in terms of a potential that somehow must
come from QCD. In most such applications the potential model is regarded as
a useful approximation, but not one that might serve as the basis for a
rigorous, systematic analysis. In fact, potential models can often be given a
rigorous formulation as effective theories.

\subsection{Positronium}

The defining characteristic of a potential model is that it treats a system
as though it has a fixed number of constituents. Taking positronium as our
example, the potential model describes the atom as a bound state of an
electron and a positron. In reality, however, positronium is far more
complex. In addition to the $e\ebar$~component, a positronium state has many
other components, each with its own wavefunction:
\be
|{\rm Ps}\rangle = \sum_{e\ebar} |e\ebar\rangle\,\psi_{e\ebar} +
\sum_{e\ebar\gamma} |e\ebar\gamma\rangle\,\psi_{e\ebar\gamma} +
\sum_{e\ebar\gamma\gamma}
|e\ebar\gamma\gamma\rangle\,\psi_{e\ebar\gamma\gamma} +\cdots.
\ee 
The probability for finding the atom in any one of these states can be
computed from the corresponding wavefunction in the usual way, with the
constraint that the sum of all probabilities is one:
\be \label{psi-norm}
 1 = \sum_{e\ebar} |\psi_{e\ebar}|^2 +
\sum_{e\ebar\gamma} |\psi_{e\ebar\gamma}|^2 +
\sum_{e\ebar\gamma\gamma} |\psi_{e\ebar\gamma\gamma}|^2 +\cdots.
\ee 
One wonders why, and to what extent, we are allowed to ignore all but the
first term in this ``Fock state'' expansion of the positronium state.

The bulk of the energy in a positronium atom is in the rest masses of the
electron and positron. The kinetic energies and three momenta of these
particles are much smaller\,---\,typically 
\be
K_e \approx \alpha^2 m \quad\mbox{and}\quad p_e \approx \alpha m
\approx 1/(\mbox{atom size}),
\ee
respectively. The electron and positron constantly exchange photons, which
provide the binding. The exchanged photons have momenta~$q_\gamma$ that are
typically the same size as the electron's three momentum:
\be
q_\gamma \approx p_e  .
\ee
The photon's energy, however, is the same size as its three momentum, since
it is massless, and therefore its energy is much larger than the electron's
kinetic energy ($1/\alpha$ times larger):
\be
E_\gamma = q_\gamma \gg K_e.
\ee
This means that the exchanged photon is highly virtual, and therefore the
exchange is effectively instantaneous on the
time scale important to the electron and positron.
Consequently photon exchange can be
modelled with an instantaneous potential: the Coulomb potential to leading
order in $v_e/c$, and the Breit interaction in the next order.

A potential model exists for positronium because of the disparity between
the electron-positron kinetic energy and the photon's energy. We can
formally derive the potential model as an effective theory by introducing
an ultraviolet cutoff that restricts only particle energies. If we choose
$\Lambda_E$ between $K_e$ and $E_\gamma$, then the exchanged photons are
excluded from the effective theory. Their effect is reintroduced through
correction terms, but, since the cutoff is on energy and not three
momentum, the correction terms are local in~$t$ but
nonlocal in~$r$. That is, the correction terms are instantaneous
potentials~$V(\pv,\rv)$. We have derived the potential model 
using renormalization theory.

The analysis of positronium is not complete because it is possible to have
very low momentum photons whose energy is below the cutoff:
\be
q < \Lambda_E \ll p_e .
\ee
These photons, however, have wavelengths that are much larger than the size
of the atom, and so their interactions can be expanded, in the effective
theory, using a multipole expansion.
This means that they couple only
weakly to the atom; the probability of finding a (soft) photon with the
$e\ebar$ in a positronium atom is
\be
P(e\ebar\gamma) \approx \alpha\,v_e^2 \approx \alpha^3.
\ee
Positronium can be described accurately
by a potential model only insofar as contributions from such states are
negligible. The $e\ebar\gamma$ states in the effective theory have energy of
order $\alpha^2 m$, and so the shift positronium binding energies by
$P(e\ebar\gamma)\,\alpha^2m\approx \alpha^5 m$, which is relatively very
small. The Lamb shift is the most famous example of such a contribution.

There are two ways of systematically incorporating soft photons. 
The standard approach is to use the cutoff effective theory to compute
electron-positron potentials that include the contributions 
due to soft photons.
Since the dynamical time scale for these photons is the same as for
the~$e\ebar$, the potentials are energy dependent and have divergent expansions
in perturbation theory. Techniques exist for dealing with the divergent
expansions for Coulombic systems like positronium. The extent of the energy
dependence in the potential is directly related to the probability carried by
Fock states with soft photons. This is obvious from the standard normalization
condition for the wavefunction when the potential is energy dependent:
\be
1 = \int d^3r\,\psi^\dagger(\rv)\left(1 - \partial V/\partial E\right)
\psi(\rv).
\ee
The second term on the right-hand side gives the probability carried by all
Fock states other than the $e\ebar$~state; that is, it equals the sum of
all terms on the right-hand side of \eq{psi-norm} after the first.

A second approach worth investigating is to treat the
effective theory as a coupled channel problem, where now the wave function
has two or more components,
\be
\Psi(\rv) = \left( \begin{array}{c} \psi_{e\ebar}(\rv) \\
\psi_{e\ebar\gamma}(\rv,q) \\ \vdots \\ \end{array} \right),
\ee
one for each Fock state that is important to the analysis. For most
high-precision QED work only the $e\ebar$ and 
$e\ebar\gamma$ states are relevant;
additional soft photons are too highly suppressed to contribute. The
components of this wavefunction satisfy coupled Schr\"odinger equations
whose  potentials are instantaneous (energy independent) and calculable from
{\em convergent\/} perturbation series in QED. The formalism is
dramatically simplified by the multipole expansion, which is valid because
of the cutoff. This expansion means that the $e\ebar\gamma$ wavefunction,
for example, is a function of the $e\ebar$~separation, and of the
magnitude~$q$ of the photon momentum; the wavefunction is independent of the
direction of~$\qv$. This last fact makes a numerical treatment feasible.

Effective field theories already play a big role in high-precision
studies of positronium and other simple QED
atoms.\cite{labelle} 
NRQED, for example, is an effective
field theory in which the electron and positron dynamics is
nonrelativistic. In the most common approach, the boundstate equation in
NRQED is a Schr\"odinger equation with energy-dependent potentials
describing soft photon contributions, as above. This equation is usually
solved perturbatively. A different approach would be to compute
effective potentials for the Schr\"odinger equation using QED or NRQED
perturbation theory and the renormalization techniques outlined in
these lectures. Then the boundstate equations could be solved 
numerically, as described in
the earlier parts of these lectures. Once the potentials are determined, the
properties of any state can be extracted from a simple numerical
calculation. It is perhaps remarkable that there are effective
potentials which include all the effects of full QED. If these
potentials were computed, they could be used for multielectron atoms
or in solidstate analyses\,---\,a relatively painless way of
introducing QED effects.

\subsection{Heavy Quark Mesons}

Much of the discussion above for positronium can also be applied to mesons,
like the $\Upsilon$, composed of heavy quarks. The $b$~quark and antiquark
in an $\Upsilon$ are very heavy, and therefore they are moderately
nonrelativistic: $v_b^2\approx 0.1$. Thus the dominant gluons binding the
quarks have energies that are much larger than the quarks' kinetic energy,
and so can be rigorously modelled using an instantaneous potential. This is
the basis for the phenomenological quark model, which has been very
successful in describing the~$\Upsilon$ and its excitations. But, as in
the case of positronium, the quark model is not the whole story. States
with soft gluons in addition to the $b\bbar$ also contribute. Again the
gluon coupling to the quarks can be expanded in a multipole expansion, and
therefore the soft gluons are suppressed: the probability of the $b\bbar g$
state is $\alpha_s v^2\!\approx\!0.1$. 
Nevertheless there are important phenomena in
which these additional Fock states play a leading role.\cite{bbl} 
It would be
worthwhile to explore a coupled-channel approach to formulating the dynamics
of this Fock state, similar to that outlined above for positronium.

\subsection{Potential Models in General}

Potential models arise as effective theories, low-energy approximations to
more fundamental quantum field theories, when the important energies in a
system are much smaller than the three~momenta. Then, when
modelling that system, it is natural to impose a stronger cutoff on energies
than on three~momenta in the effective theory. This results in
effective interactions that are potentials\,---\,that is, they are
local in time but not
in space. A system of interacting nonrelativistic particles always has
smaller kinetic energies than three~momenta, smaller by~$v/c$, and so is a
prime candidate for a potential model. 

The utility of a  potential model
depends upon the importance of retardation corrections. Consider, for
example, the case where the interaction between the nonrelativistic
particles is mediated by a massless particle, like the photon in
positronium. Here exchanges in which the massless particle is very soft are
allowed by the energy cutoff, and lead to retardation effects that can
greatly complicate the potential.  Nevertheless, as in positronium, these
interactions can often be simplified using multipole expansions, and are
often strongly suppressed.  

Retardation effects are much smaller if the
exchanged particle is massive. Indeed, there are no
retardation effects at all if the mass of the exchanged particle is larger
than the energy cutoff in the effective theory; the exchanged particle is
always highly virtual, no matter what its three~momentum, and so can
always be described by an instantaneous potential. In such cases
simple potential
models provide a framework for calculations of arbitrary precision. We now
turn to such a system.
\section{Nucleon-Nucleon Interactions}

Another example of an effective theory  is the pion-nucleon theory of
low-energy nuclear interactions. In this section I discuss the design of an
effective theory that describes low-energy nuclear interactions. I tune the
couplings to reproduce phase shifts for $np$~scattering. I also review
a problem encountered in a recent analysis of this system that uses
dimensional regularization. Finally I discuss briefly the general
implications of
effective theory for low-energy nuclear physics.

The analysis I present here is pedagogical. 
Others have completed analyses that are more thorough,
although slightly
different in detail from what follows here.\cite{ordonez96}

\subsection{What is a Hadron?}

The true theory of the nuclear force is QCD: nucleons are made of quarks and
they interact by exchanging quarks and gluons. Nevertheless the
electric form factor of the proton, for example, is given roughly by
\be
F_p(Q^2) \approx \left(\frac{1}{1+(Q/800\,\mbox{MeV})^2}\right)^2
\ee
which goes to $1$, the value for a pointlike particle, for momentum transfers
$Q\!\ll\!800$\,MeV. This and other experimental results suggest that nucleons
appear approximately point-like to probes with momenta less than
500--1000\,MeV. Therefore an effective theory that describes nucleons and
other hadrons in this low-momentum region should treat hadrons as elementary
particles. Quark substructure cannot be resolved at such
momenta. Since nucleon momenta are of order 300\,MeV or less in nuclear
matter,  an effective theory of this sort might provide an excellent
framework for the systematic study of traditional low-energy nuclear physics.

Our effective theory, therefore, should have a cutoff 
of order the color resolution scale,
$\Lambda_c\approx500$--1000\,MeV. 
(I switch here to designating the
cutoff in momentum space, $\Lambda$, rather than in coordinate space,
$a\!=\!1/\Lambda$,
because the former is more conventional.) Heavy
hadrons, like the proton, neutron, $\Delta$, $\rho$, and so on,  have masses of
order the cutoff or larger and so cannot be relativistic. These are most
efficiently treated as pointlike nonrelativistic particles, with kinetic
energies given by
\be
K = \frac{\pv^2}{2m} - \frac{\pfour}{8m^3} +\cdots.
\ee
Because of the cutoff, the effective theory contains no pair creation for
these hadrons; particle and antiparticle decouple. 

The light hadrons\,---\,that is, the pion and possibly also the
kaon\,---\,can be relativistic and so must be included as pointlike 
particles described by a full
relativistic quantum field theory. This field theory inherits a chiral
symmetry from QCD that implies that all interactions  are suppressed by
powers of $p/\Lambda_c$ or $m_\pi/\Lambda_c$. 
Consequently, even though pions
are strongly interacting particles, their interactions at low momenta are
weak, and can be analyzed using perturbation theory. The Feynman rules for
this theory are given in many standard texts.\cite{chi-pth}

The usual chiral theory is a quantum field theory. 
Potential models, however, provide a natural framework for
describing nonrelativistic systems, as we discussed in the previous
section. This is particularly true for the
nucleon-nucleon problem, since the pion and kaon masses are larger than the
typical nucleon energy in nuclear matter. Thus retardation due to soft pion
and kaon exchange plays only a small role in our effective nucleon-nucleon
theory, and can be hidden inside energy-independent potentials.
Our effective theory for low-energy nucleon-nucleon physics, therefore,
is specified by a hamiltonian:
\be \label{Hnn}
H = \sum_{\rm nucleons} \left(\frac{\pv^2}{2m}-\frac{\pfour}{8m^3}\right)
+V(\rv,\pv,\sigmav_i,\tauv_i,\Lv)
\ee
where the potential is a function of the positions, momenta, spins, isospins
and angular momentum of the two nucleons.

\subsection{The Nucleon-Nucleon Potential}

The potential in our nucleon-nucleon hamiltonian can be determined by
``matching'' scattering amplitudes computed in the potential model with the
same amplitudes computed using the chiral field theory, as in
Section~\ref{pert-deriv}. There are both long-range parts, involving the
exchange of one or more pions, and short-range or ``contact'' potentials like
the delta function potentials in Section~\ref{ren-schrod}.

The leading-order long-range potential is due to the 
exchange of a single pion; it gives
\be
V_{1\pi} = \alpha_\pi\,\tauv_1\!\cdot\!\tauv_2\,
\frac{\sigmav_1\!\cdot\!\nablav\,\sigmav_2\!\cdot\!\nablav}{m_\pi^2}\,
{v_\Lambda(r)}
\ee
where $v_\Lambda(r)$ is a Yukawa potential at large~$r$,
\be
v_\Lambda(r)\,\stackrel{{r\,\rm large}}{\longrightarrow}\, \frac{\e^{-m_\pi
r}}{r},
\ee
and the $\sigmav$'s and $\tauv$'s are Pauli matrix operators for the
nucleon spin and isospin respectively.  Chiral symmetry implies that the
dimensionless coupling
$\alpha_\pi$ is related to the pion decay constant,
$f_\pi\!\approx\!93$\,MeV, the nucleon's axial vector coupling,
$g_A\,\approx\,1.25$, and the pion mass~$m_\pi$:
\be
\alpha_\pi \approx \frac{g_A^2\,m_\pi^2}{16\pi\,f_\pi^2} \approx
\mbox{0.06--0.08}.
\ee
The coupling is small because it is proportional to
$m_\pi^2/\Lambda^2_c$, $\Lambda_c$ being of order
$4\sqrt{\pi}f_\pi$. Projecting the one-pion potential onto different
spin-orbit eigenstates, we obtain
\be
V_{1\pi} \to -\alpha_\pi \left[\,b \,v_\Lambda(r) + b_T\,v_T(r)\right]
\ee
where $b$ and $b_T$ are constants that depend upon the channel, and
\bearray
v_T(r) &\equiv& \frac{1}{m_\pi^2}\left(v^{\prime\prime}_\Lambda -
\frac{v^\prime_\Lambda}{r}\right) \\
&\stackrel{{r\,\rm large}}{\longrightarrow} & \frac{\e^{-m_\pi r}}{r}
\,\left[1+\frac{3}{m_\pi r}+\frac{3}{m_\pi^2r^2}\right].
\eearray
Here I have dropped a part of the potential that is indistinguishable from
the contact terms we add below. We introduce an ultraviolet cutoff in the
Fourier transform, as in Section~\ref{eff-theory-sec}, to obtain
\be
v_\Lambda(r) \equiv \frac{1}{2r}\left[\e^{-m_\pi r}\erfc\!\left(
-(r\Lambda - m_\pi/\Lambda)/\sqrt{2}\right) - (r\!\to\!-r)\right],
\ee
where $\erfc(x)=1-\erf(x)$.

Higher-order (in $p/\Lambda_c$) contributions to the long-range potential
come from two-pion exchange. I will not include these here;
they are described in the literature.\cite{ordonez96}

The short-distance terms in the potential are smeared delta functions,
with $a\!=\!1/\Lambda$,
multiplied by powers of the momentum operator $\pv\!=\!\nablav/{\rm
i}$, and the spin
and isopin operators. It is convenient for our analysis to classify these by
the spin-orbit channels that they affect. 

The ${}^1S_0$ and ${}^3S_1$ channels have  $\Lambda^{-2}$ terms with
no~$\nablav$'s; there is one for each channel, each with its own coupling
constant. 
In order~$\Lambda^{-4}$ there are two contact potentials, one for each
channel, proportional to $\pv^2\delta_{1/\Lambda}^3(\rv) +
\delta_{1/\Lambda}^3(\rv)\pv^2$. The 
form of this potential can be simplified by adding
$-2\,\pv\cdot\delta_{1/\Lambda}^3(\rv)\pv$, which doesn't couple to
$S$-states in this 
order; the resulting potential is then proportional to
$(\nabla^2\delta^3_{1/\Lambda})$, just as in Section~\ref{eff-theory-sec}. 

The $P$~states do not couple to contact terms  in~$\order(\Lambda^{-2})$.
Thus our effective theory predicts that $P$~states are more accurately
described by just the long-range potential than $S$-states. The leading
$P$-wave contact terms are $\order(\Lambda^{-4})$, and, after
projecting out the 
orbital and spin angular momenta, are all proportional to
$\pv\cdot\delta_{1/\Lambda}^3(\rv)\pv$. This operator can be
simplified by adding 
$-1/2(\pv^2\delta^3_{1/\Lambda}(\rv) +
\delta^3_{1/\Lambda}(\rv)\pv^2)$, which doesn't couple to 
$P$~states in this order, to obtain again a potential proportional to
$(\nabla^2\delta^3_{1/\Lambda})$. There is one such potential for
each of the four 
spin-orbit $P$-wave channels, and each has its own coupling constant.

There are no contact terms in~$\order(\Lambda^{-2},\Lambda^{-4})$ that 
couple just to
$D$~states; thus $D$~states should be even more accurately described by 
the long-range potential than $P$~states. The only contact term in
order~$\Lambda^{-4}$ that affects a
$D$~states is one that couples the ${}^3D_1$ and ${}^3S_1$ states.

There are nine contact terms in all through order~$\Lambda^{-4}$: two
$\Lambda^{-2}$~terms and two $\Lambda^{-4}$~terms for the $S$-states, four
$\Lambda^{-4}$~terms for the $P$~states, and one
$\Lambda^{-4}$ term coupling $S$~and $D$~states. Each of these has its
own coupling constant. This seems like a lot of coupling constants, but it
is only one or two per channel.

Note that our effective theory does not include contributions from the
exchange of heavy mesons like the~$\rho$ or~$\omega$. This is neither an
oversight nor an error. It makes no sense, physically or
mathematically, to include such exchanges in our effective theory. For
example, a~$\rho$ exchanged by nonrelativistic nucleons is off energy-shell
by an amount of order its mass, and therefore, by the uncertainty principle,
can exist only for a short time\,---\,far too short a time for its
constituent quark and antiquark to realize that they are in a~$\rho$. Even
if it made physical sense to think of $\rho$~exchange here,
its contribution would be indistinguishable from a nucleon-nucleon contact
term; nonrelativistic nucleons cannot tell the difference.
We must include the contact terms in any case; thus 
it is most efficient to use
them to account for any heavy-meson exchange as well.

Our effective potential also is instantaneous; it has no energy-dependent
terms. We design our potentials by matching on-shell scattering amplitudes
from the potential theory with those from the chiral field theory. Since
the amplitudes are on-shell, all energies can be expressed in terms of
three~momenta and so the potentials are naturally and rigorously
energy-independent.\footnote{Thus, for example, the energy-dependent term in
Eq.~(17) of reference~\cite{ordonez96} vanishes on-shell and so would not be
included in our tree-level potential. The effects of pion retardation are
small, and in our formalism would be automatically included with the
one-loop contributions to the potential.} I find that this simplifies the
analysis and design of effective theories.

\subsection{Relativistic Kinetic Energies}

Our discussion has focused on corrections to the potential in effective
theories. In general there are corrections to the kinetic energy as well.
Thus, in our nucleon-nucleon theory, there is a $-\pfour/4m^3$ correction
due to relativity. This term is too singular to include in ordinary
Schr\"odinger equations, but can be handled in our effective theories
because of the cutoff. It is important, however, that the cutoff be low
enough to prevent the $\pfour$~term, which is negative, from overwhelming
the $\pv^2$~term and destabilizing the effective theory. 

While it is quite
possible to include the $\pfour$~correction directly, I have found it
numerically simpler to transform it into a potential, albeit an
energy-dependent one. Starting with the Schr\"odinger equation, which has
the form
\be
\left[E - V + \frac{\nabla^2}{2m_r} + \frac{\nabla^4}{32m_r^3}\right]\psi 
= 0
\ee
where $m_r$ is the reduced mass, we introduce a transformed wave
function~$\phi$ where $\psi\!\equiv\!\Omega \phi$ and
\be
\Omega = 1 + \frac{2m_r(E-V)-\nabla^2}{32m_r^2}.
\ee
Multiplying the Schr\"odinger equation on the left by $\Omega$, we obtain
\bearray
0 &=& \Omega\left[E - V + \frac{\nabla^2}{2m_r} +
                  \frac{\nabla^4}{32m_r^3}\right]\Omega\phi\nl
  &=& \left[E - V + \frac{(E-V)^2}{8m_r} + \frac{\nabla^2}{2m_r}
            + \order\left(\frac{p^6}{m_r^5}\right)\right]\phi.
\eearray
Solving the last equation, without the $p^6$~terms, is equivalent to solving
the original Schr\"odinger equation up to corrections of order $p^6$, which
are higher-order and therefore negligible here. We can recover the original
wavefunction~$\psi$ by undoing the transformation:
\bearray
\psi &=& \Omega \phi \nl
&=& \left[1 + \frac{2m_r(E-V)-\nabla^2}{32m_r^2}\right]\phi \\
&=& \left[1+\frac{E-V}{8m_r}+\order\left(\frac{p^4}{m_r^4}\right)\right]\phi.
\eearray
I used the Schr\"odinger equation to obtain the last equation; the
$p^4$~term is higher-order and can be neglected here. Note that the correct
normalization for  $\phi$, given that the $\psi$'s are normalized to one, 
is
\be
\int d^3r\, \phi^\dagger(\rv) \left[1 + \frac{E-V}{4m_r}\right] \phi(\rv) = 1.
\ee

The procedure used in the previous paragraph to transform~$\pfour$
into a potential is often described as ``using the equations of motion.'' 
In effect we replaced $\pfour$ by $4m_r^2(V-E)^2$. We could have guessed
the result by noting that the Schr\"odinger equation, the ``equation of
motion'' here, implies that $\pv^2\approx 2m_r(V-E)$ when acting on the
wavefunction of an energy eigenstate. Thus it is plausible that $\pfour$ is
the square of the same operator. Transformations of the wavefunction are a
powerful tool for simplifying effective field theories. They also
underscore the fact that there are infinitely many theories that have
equivalent low-energy behavior. Useful transformations~$\Omega$ typically
are close to unity at low energies, so that complicated cross terms
are high-order and can be neglected, as above.

I have included this discussion to show how one deals with
relativistic corrections to the kinetic energy. However, I have
checked and found that
$\pv^4$~corrections have little effect on the quantities I discuss
below. Consequently I chose to omit them from that analysis.

\subsection{Comparison with Data}

Having outlined the design of a rigorous potential model for nonrelativistic 
nucleon-nucleon interactions, we now compare this model with experimental
data. For simplicity, I restricted my study to $np$~scattering, thereby
avoiding the problem of  Coulomb distortion in
$pp$~scattering. 
I chose to fit phase shifts for different channels,
rather than the full scattering amplitude, because this greatly simplifies the
analysis: there are only one or two couplings that need be tuned for each
channel, and tuning  can be done one channel at a time. The disadvantage is
that the phase shifts are not really data; they must be extracted from the
data using some sort of theoretically motivated procedure. This makes it
difficult to assign experimental errors to the phase shifts. My goal, however,
is more to illustrate the procedure than to obtain definitive theoretical
results. The effective theory would fit exact data just as well as it
fits the phase shift solutions I use here. I use phase-shifts from the
multi-energy fits by the Nijmegen group.\cite{nijmegen} Multi-energy
fits are preferable here since they enforce a smooth energy dependence.

We begin with the ${}^1S_0$ phase shifts. The potential, including one-pion
exchange and contact terms through order~$\Lambda^{-4}$, is
\be
V(r) = -\alpha_\pi v_\Lambda(r) + c\,\frac{\delta_{1/\Lambda}^3(\rv)}{\Lambda^{2}} -
d\,\frac{\nabla^2\delta_{1/\Lambda}^3(\rv)}{\Lambda^{4}},
\ee
where I use $\alpha_\pi\!=\!0.075$ and $m_\pi\!=\!140$\,MeV. Coupling
constants $c$ and
$d$ must be tuned to the data. I used the Nijmegen phase shifts at
$E\!=\!0.05$\,MeV and 0.1\,MeV 
to tune $c$ and~$d$ for $\Lambda$'s ranging from
50\,MeV to~1000\,MeV. 
Here and elsewhere in this section $E$ refers to the (nonrelativistic)
center-of-mass energy.

\begin{figure}
\begin{center}
% GNUPLOT: LaTeX picture
\setlength{\unitlength}{0.240900pt}
\ifx\plotpoint\undefined\newsavebox{\plotpoint}\fi
\sbox{\plotpoint}{\rule[-0.200pt]{0.400pt}{0.400pt}}%
\begin{picture}(1349,749)(0,0)
\font\gnuplot=cmr10 at 10pt
\gnuplot
\sbox{\plotpoint}{\rule[-0.200pt]{0.400pt}{0.400pt}}%
\put(220.0,113.0){\rule[-0.200pt]{4.818pt}{0.400pt}}
\put(198,113){\makebox(0,0)[r]{$10^{-6}$}}
\put(1265.0,113.0){\rule[-0.200pt]{4.818pt}{0.400pt}}
\put(220.0,317.0){\rule[-0.200pt]{4.818pt}{0.400pt}}
\put(198,317){\makebox(0,0)[r]{$10^{-4}$}}
\put(1265.0,317.0){\rule[-0.200pt]{4.818pt}{0.400pt}}
\put(220.0,522.0){\rule[-0.200pt]{4.818pt}{0.400pt}}
\put(198,522){\makebox(0,0)[r]{$10^{-2}$}}
\put(1265.0,522.0){\rule[-0.200pt]{4.818pt}{0.400pt}}
\put(220.0,726.0){\rule[-0.200pt]{4.818pt}{0.400pt}}
\put(198,726){\makebox(0,0)[r]{$1$}}
\put(1265.0,726.0){\rule[-0.200pt]{4.818pt}{0.400pt}}
\put(220.0,113.0){\rule[-0.200pt]{0.400pt}{4.818pt}}
\put(220,68){\makebox(0,0){0.1}}
\put(220.0,706.0){\rule[-0.200pt]{0.400pt}{4.818pt}}
\put(317.0,113.0){\rule[-0.200pt]{0.400pt}{2.409pt}}
\put(317.0,716.0){\rule[-0.200pt]{0.400pt}{2.409pt}}
\put(374.0,113.0){\rule[-0.200pt]{0.400pt}{2.409pt}}
\put(374.0,716.0){\rule[-0.200pt]{0.400pt}{2.409pt}}
\put(414.0,113.0){\rule[-0.200pt]{0.400pt}{2.409pt}}
\put(414.0,716.0){\rule[-0.200pt]{0.400pt}{2.409pt}}
\put(446.0,113.0){\rule[-0.200pt]{0.400pt}{2.409pt}}
\put(446.0,716.0){\rule[-0.200pt]{0.400pt}{2.409pt}}
\put(471.0,113.0){\rule[-0.200pt]{0.400pt}{2.409pt}}
\put(471.0,716.0){\rule[-0.200pt]{0.400pt}{2.409pt}}
\put(493.0,113.0){\rule[-0.200pt]{0.400pt}{2.409pt}}
\put(493.0,716.0){\rule[-0.200pt]{0.400pt}{2.409pt}}
\put(511.0,113.0){\rule[-0.200pt]{0.400pt}{2.409pt}}
\put(511.0,716.0){\rule[-0.200pt]{0.400pt}{2.409pt}}
\put(528.0,113.0){\rule[-0.200pt]{0.400pt}{2.409pt}}
\put(528.0,716.0){\rule[-0.200pt]{0.400pt}{2.409pt}}
\put(543.0,113.0){\rule[-0.200pt]{0.400pt}{4.818pt}}
\put(543,68){\makebox(0,0){1}}
\put(543.0,706.0){\rule[-0.200pt]{0.400pt}{4.818pt}}
\put(640.0,113.0){\rule[-0.200pt]{0.400pt}{2.409pt}}
\put(640.0,716.0){\rule[-0.200pt]{0.400pt}{2.409pt}}
\put(697.0,113.0){\rule[-0.200pt]{0.400pt}{2.409pt}}
\put(697.0,716.0){\rule[-0.200pt]{0.400pt}{2.409pt}}
\put(737.0,113.0){\rule[-0.200pt]{0.400pt}{2.409pt}}
\put(737.0,716.0){\rule[-0.200pt]{0.400pt}{2.409pt}}
\put(768.0,113.0){\rule[-0.200pt]{0.400pt}{2.409pt}}
\put(768.0,716.0){\rule[-0.200pt]{0.400pt}{2.409pt}}
\put(794.0,113.0){\rule[-0.200pt]{0.400pt}{2.409pt}}
\put(794.0,716.0){\rule[-0.200pt]{0.400pt}{2.409pt}}
\put(815.0,113.0){\rule[-0.200pt]{0.400pt}{2.409pt}}
\put(815.0,716.0){\rule[-0.200pt]{0.400pt}{2.409pt}}
\put(834.0,113.0){\rule[-0.200pt]{0.400pt}{2.409pt}}
\put(834.0,716.0){\rule[-0.200pt]{0.400pt}{2.409pt}}
\put(850.0,113.0){\rule[-0.200pt]{0.400pt}{2.409pt}}
\put(850.0,716.0){\rule[-0.200pt]{0.400pt}{2.409pt}}
\put(865.0,113.0){\rule[-0.200pt]{0.400pt}{4.818pt}}
\put(865,68){\makebox(0,0){10}}
\put(865.0,706.0){\rule[-0.200pt]{0.400pt}{4.818pt}}
\put(962.0,113.0){\rule[-0.200pt]{0.400pt}{2.409pt}}
\put(962.0,716.0){\rule[-0.200pt]{0.400pt}{2.409pt}}
\put(1019.0,113.0){\rule[-0.200pt]{0.400pt}{2.409pt}}
\put(1019.0,716.0){\rule[-0.200pt]{0.400pt}{2.409pt}}
\put(1059.0,113.0){\rule[-0.200pt]{0.400pt}{2.409pt}}
\put(1059.0,716.0){\rule[-0.200pt]{0.400pt}{2.409pt}}
\put(1091.0,113.0){\rule[-0.200pt]{0.400pt}{2.409pt}}
\put(1091.0,716.0){\rule[-0.200pt]{0.400pt}{2.409pt}}
\put(1116.0,113.0){\rule[-0.200pt]{0.400pt}{2.409pt}}
\put(1116.0,716.0){\rule[-0.200pt]{0.400pt}{2.409pt}}
\put(1138.0,113.0){\rule[-0.200pt]{0.400pt}{2.409pt}}
\put(1138.0,716.0){\rule[-0.200pt]{0.400pt}{2.409pt}}
\put(1157.0,113.0){\rule[-0.200pt]{0.400pt}{2.409pt}}
\put(1157.0,716.0){\rule[-0.200pt]{0.400pt}{2.409pt}}
\put(1173.0,113.0){\rule[-0.200pt]{0.400pt}{2.409pt}}
\put(1173.0,716.0){\rule[-0.200pt]{0.400pt}{2.409pt}}
\put(1188.0,113.0){\rule[-0.200pt]{0.400pt}{4.818pt}}
\put(1188,68){\makebox(0,0){100}}
\put(1188.0,706.0){\rule[-0.200pt]{0.400pt}{4.818pt}}
\put(1285.0,113.0){\rule[-0.200pt]{0.400pt}{2.409pt}}
\put(1285.0,716.0){\rule[-0.200pt]{0.400pt}{2.409pt}}
\put(220.0,113.0){\rule[-0.200pt]{256.558pt}{0.400pt}}
\put(1285.0,113.0){\rule[-0.200pt]{0.400pt}{147.672pt}}
\put(220.0,726.0){\rule[-0.200pt]{256.558pt}{0.400pt}}
\put(45,419){\makebox(0,0){$|\Delta \delta_0(E)|$}}
\put(752,23){\makebox(0,0){$E$ (MeV)}}
\put(317,624){\makebox(0,0)[l]{$\Lambda\!=\!50$\,MeV}}
\put(374,506){\makebox(0,0)[l]{$100$\,MeV}}
\put(374,335){\makebox(0,0)[l]{$330$\,MeV}}
\put(374,175){\makebox(0,0)[l]{$400$\,MeV}}
\put(220.0,113.0){\rule[-0.200pt]{0.400pt}{147.672pt}}
\put(317,483){\usebox{\plotpoint}}
\multiput(317.00,483.58)(0.717,0.499){177}{\rule{0.673pt}{0.120pt}}
\multiput(317.00,482.17)(127.602,90.000){2}{\rule{0.337pt}{0.400pt}}
\multiput(446.00,573.58)(1.035,0.498){91}{\rule{0.926pt}{0.120pt}}
\multiput(446.00,572.17)(95.079,47.000){2}{\rule{0.463pt}{0.400pt}}
\multiput(543.00,620.58)(1.394,0.498){67}{\rule{1.209pt}{0.120pt}}
\multiput(543.00,619.17)(94.492,35.000){2}{\rule{0.604pt}{0.400pt}}
\multiput(640.00,655.58)(2.227,0.497){55}{\rule{1.866pt}{0.120pt}}
\multiput(640.00,654.17)(124.128,29.000){2}{\rule{0.933pt}{0.400pt}}
\multiput(768.00,684.59)(7.358,0.485){11}{\rule{5.643pt}{0.117pt}}
\multiput(768.00,683.17)(85.288,7.000){2}{\rule{2.821pt}{0.400pt}}
\multiput(865.00,689.94)(14.080,-0.468){5}{\rule{9.800pt}{0.113pt}}
\multiput(865.00,690.17)(76.660,-4.000){2}{\rule{4.900pt}{0.400pt}}
\multiput(962.00,685.92)(2.609,-0.497){47}{\rule{2.164pt}{0.120pt}}
\multiput(962.00,686.17)(124.509,-25.000){2}{\rule{1.082pt}{0.400pt}}
\multiput(1091.58,659.09)(0.499,-0.753){191}{\rule{0.120pt}{0.702pt}}
\multiput(1090.17,660.54)(97.000,-144.543){2}{\rule{0.400pt}{0.351pt}}
\multiput(1188.58,516.00)(0.499,1.241){111}{\rule{0.120pt}{1.089pt}}
\multiput(1187.17,516.00)(57.000,138.739){2}{\rule{0.400pt}{0.545pt}}
\put(317,344){\usebox{\plotpoint}}
\multiput(317.00,344.58)(0.672,0.499){189}{\rule{0.637pt}{0.120pt}}
\multiput(317.00,343.17)(127.677,96.000){2}{\rule{0.319pt}{0.400pt}}
\multiput(446.00,440.58)(0.884,0.499){107}{\rule{0.805pt}{0.120pt}}
\multiput(446.00,439.17)(95.328,55.000){2}{\rule{0.403pt}{0.400pt}}
\multiput(543.00,495.58)(1.035,0.498){91}{\rule{0.926pt}{0.120pt}}
\multiput(543.00,494.17)(95.079,47.000){2}{\rule{0.463pt}{0.400pt}}
\multiput(640.00,542.58)(1.212,0.498){103}{\rule{1.066pt}{0.120pt}}
\multiput(640.00,541.17)(125.787,53.000){2}{\rule{0.533pt}{0.400pt}}
\multiput(768.00,595.58)(1.686,0.497){55}{\rule{1.438pt}{0.120pt}}
\multiput(768.00,594.17)(94.016,29.000){2}{\rule{0.719pt}{0.400pt}}
\multiput(865.00,624.58)(3.827,0.493){23}{\rule{3.085pt}{0.119pt}}
\multiput(865.00,623.17)(90.598,13.000){2}{\rule{1.542pt}{0.400pt}}
\multiput(962.00,635.92)(1.321,-0.498){95}{\rule{1.153pt}{0.120pt}}
\multiput(962.00,636.17)(126.607,-49.000){2}{\rule{0.577pt}{0.400pt}}
\multiput(1091.00,588.58)(0.884,0.499){107}{\rule{0.805pt}{0.120pt}}
\multiput(1091.00,587.17)(95.328,55.000){2}{\rule{0.403pt}{0.400pt}}
\multiput(1188.00,643.58)(0.841,0.498){65}{\rule{0.771pt}{0.120pt}}
\multiput(1188.00,642.17)(55.401,34.000){2}{\rule{0.385pt}{0.400pt}}
\put(317,181){\usebox{\plotpoint}}
\multiput(317.00,181.58)(0.679,0.499){187}{\rule{0.643pt}{0.120pt}}
\multiput(317.00,180.17)(127.665,95.000){2}{\rule{0.322pt}{0.400pt}}
\multiput(446.00,276.58)(1.081,0.498){87}{\rule{0.962pt}{0.120pt}}
\multiput(446.00,275.17)(95.003,45.000){2}{\rule{0.481pt}{0.400pt}}
\multiput(543.00,321.58)(1.686,0.497){55}{\rule{1.438pt}{0.120pt}}
\multiput(543.00,320.17)(94.016,29.000){2}{\rule{0.719pt}{0.400pt}}
\multiput(640.00,348.92)(1.496,-0.498){83}{\rule{1.291pt}{0.120pt}}
\multiput(640.00,349.17)(125.321,-43.000){2}{\rule{0.645pt}{0.400pt}}
\multiput(768.58,307.00)(0.499,0.748){191}{\rule{0.120pt}{0.698pt}}
\multiput(767.17,307.00)(97.000,143.551){2}{\rule{0.400pt}{0.349pt}}
\multiput(865.00,452.58)(0.759,0.499){125}{\rule{0.706pt}{0.120pt}}
\multiput(865.00,451.17)(95.534,64.000){2}{\rule{0.353pt}{0.400pt}}
\multiput(962.00,516.58)(1.245,0.498){101}{\rule{1.092pt}{0.120pt}}
\multiput(962.00,515.17)(126.733,52.000){2}{\rule{0.546pt}{0.400pt}}
\multiput(1091.00,566.93)(10.729,-0.477){7}{\rule{7.860pt}{0.115pt}}
\multiput(1091.00,567.17)(80.686,-5.000){2}{\rule{3.930pt}{0.400pt}}
\multiput(1188.00,563.59)(3.291,0.489){15}{\rule{2.633pt}{0.118pt}}
\multiput(1188.00,562.17)(51.534,9.000){2}{\rule{1.317pt}{0.400pt}}
\put(317,148){\usebox{\plotpoint}}
\multiput(317.00,148.58)(0.694,0.499){183}{\rule{0.655pt}{0.120pt}}
\multiput(317.00,147.17)(127.641,93.000){2}{\rule{0.327pt}{0.400pt}}
\multiput(446.00,241.58)(1.883,0.497){49}{\rule{1.592pt}{0.120pt}}
\multiput(446.00,240.17)(93.695,26.000){2}{\rule{0.796pt}{0.400pt}}
\multiput(543.00,267.58)(4.552,0.492){19}{\rule{3.627pt}{0.118pt}}
\multiput(543.00,266.17)(89.471,11.000){2}{\rule{1.814pt}{0.400pt}}
\multiput(640.58,278.00)(0.499,0.582){253}{\rule{0.120pt}{0.566pt}}
\multiput(639.17,278.00)(128.000,147.826){2}{\rule{0.400pt}{0.283pt}}
\multiput(768.00,427.58)(0.724,0.499){131}{\rule{0.679pt}{0.120pt}}
\multiput(768.00,426.17)(95.590,67.000){2}{\rule{0.340pt}{0.400pt}}
\multiput(865.00,494.58)(0.884,0.499){107}{\rule{0.805pt}{0.120pt}}
\multiput(865.00,493.17)(95.328,55.000){2}{\rule{0.403pt}{0.400pt}}
\multiput(962.00,549.58)(1.096,0.499){115}{\rule{0.975pt}{0.120pt}}
\multiput(962.00,548.17)(126.977,59.000){2}{\rule{0.487pt}{0.400pt}}
\multiput(1091.00,608.58)(1.576,0.497){59}{\rule{1.352pt}{0.120pt}}
\multiput(1091.00,607.17)(94.195,31.000){2}{\rule{0.676pt}{0.400pt}}
\multiput(1188.00,639.58)(2.666,0.492){19}{\rule{2.173pt}{0.118pt}}
\multiput(1188.00,638.17)(52.490,11.000){2}{\rule{1.086pt}{0.400pt}}
\end{picture}
\end{center}
\caption{Errors in the ${}^1S_0$ phase shifts (in radians) 
versus energy for the
full effective theory with different values of the cutoff~$\Lambda$.}
\label{1s0_lamdep-fig}
\end{figure}
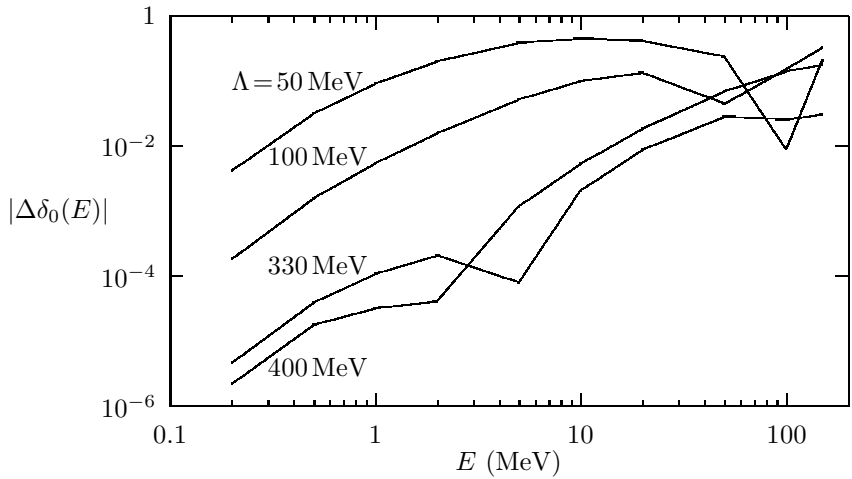

\begin{figure}
\begin{center}
% GNUPLOT: LaTeX picture
\setlength{\unitlength}{0.240900pt}
\ifx\plotpoint\undefined\newsavebox{\plotpoint}\fi
\sbox{\plotpoint}{\rule[-0.200pt]{0.400pt}{0.400pt}}%
\begin{picture}(1349,749)(0,0)
\font\gnuplot=cmr10 at 10pt
\gnuplot
\sbox{\plotpoint}{\rule[-0.200pt]{0.400pt}{0.400pt}}%
\put(220.0,171.0){\rule[-0.200pt]{4.818pt}{0.400pt}}
\put(198,171){\makebox(0,0)[r]{$10^{-6}$}}
\put(1265.0,171.0){\rule[-0.200pt]{4.818pt}{0.400pt}}
\put(220.0,337.0){\rule[-0.200pt]{4.818pt}{0.400pt}}
\put(198,337){\makebox(0,0)[r]{$10^{-4}$}}
\put(1265.0,337.0){\rule[-0.200pt]{4.818pt}{0.400pt}}
\put(220.0,502.0){\rule[-0.200pt]{4.818pt}{0.400pt}}
\put(198,502){\makebox(0,0)[r]{$10^{-2}$}}
\put(1265.0,502.0){\rule[-0.200pt]{4.818pt}{0.400pt}}
\put(220.0,668.0){\rule[-0.200pt]{4.818pt}{0.400pt}}
\put(198,668){\makebox(0,0)[r]{$1$}}
\put(1265.0,668.0){\rule[-0.200pt]{4.818pt}{0.400pt}}
\put(220.0,113.0){\rule[-0.200pt]{0.400pt}{4.818pt}}
\put(220,68){\makebox(0,0){0.1}}
\put(220.0,706.0){\rule[-0.200pt]{0.400pt}{4.818pt}}
\put(317.0,113.0){\rule[-0.200pt]{0.400pt}{2.409pt}}
\put(317.0,716.0){\rule[-0.200pt]{0.400pt}{2.409pt}}
\put(374.0,113.0){\rule[-0.200pt]{0.400pt}{2.409pt}}
\put(374.0,716.0){\rule[-0.200pt]{0.400pt}{2.409pt}}
\put(414.0,113.0){\rule[-0.200pt]{0.400pt}{2.409pt}}
\put(414.0,716.0){\rule[-0.200pt]{0.400pt}{2.409pt}}
\put(446.0,113.0){\rule[-0.200pt]{0.400pt}{2.409pt}}
\put(446.0,716.0){\rule[-0.200pt]{0.400pt}{2.409pt}}
\put(471.0,113.0){\rule[-0.200pt]{0.400pt}{2.409pt}}
\put(471.0,716.0){\rule[-0.200pt]{0.400pt}{2.409pt}}
\put(493.0,113.0){\rule[-0.200pt]{0.400pt}{2.409pt}}
\put(493.0,716.0){\rule[-0.200pt]{0.400pt}{2.409pt}}
\put(511.0,113.0){\rule[-0.200pt]{0.400pt}{2.409pt}}
\put(511.0,716.0){\rule[-0.200pt]{0.400pt}{2.409pt}}
\put(528.0,113.0){\rule[-0.200pt]{0.400pt}{2.409pt}}
\put(528.0,716.0){\rule[-0.200pt]{0.400pt}{2.409pt}}
\put(543.0,113.0){\rule[-0.200pt]{0.400pt}{4.818pt}}
\put(543,68){\makebox(0,0){1}}
\put(543.0,706.0){\rule[-0.200pt]{0.400pt}{4.818pt}}
\put(640.0,113.0){\rule[-0.200pt]{0.400pt}{2.409pt}}
\put(640.0,716.0){\rule[-0.200pt]{0.400pt}{2.409pt}}
\put(697.0,113.0){\rule[-0.200pt]{0.400pt}{2.409pt}}
\put(697.0,716.0){\rule[-0.200pt]{0.400pt}{2.409pt}}
\put(737.0,113.0){\rule[-0.200pt]{0.400pt}{2.409pt}}
\put(737.0,716.0){\rule[-0.200pt]{0.400pt}{2.409pt}}
\put(768.0,113.0){\rule[-0.200pt]{0.400pt}{2.409pt}}
\put(768.0,716.0){\rule[-0.200pt]{0.400pt}{2.409pt}}
\put(794.0,113.0){\rule[-0.200pt]{0.400pt}{2.409pt}}
\put(794.0,716.0){\rule[-0.200pt]{0.400pt}{2.409pt}}
\put(815.0,113.0){\rule[-0.200pt]{0.400pt}{2.409pt}}
\put(815.0,716.0){\rule[-0.200pt]{0.400pt}{2.409pt}}
\put(834.0,113.0){\rule[-0.200pt]{0.400pt}{2.409pt}}
\put(834.0,716.0){\rule[-0.200pt]{0.400pt}{2.409pt}}
\put(850.0,113.0){\rule[-0.200pt]{0.400pt}{2.409pt}}
\put(850.0,716.0){\rule[-0.200pt]{0.400pt}{2.409pt}}
\put(865.0,113.0){\rule[-0.200pt]{0.400pt}{4.818pt}}
\put(865,68){\makebox(0,0){10}}
\put(865.0,706.0){\rule[-0.200pt]{0.400pt}{4.818pt}}
\put(962.0,113.0){\rule[-0.200pt]{0.400pt}{2.409pt}}
\put(962.0,716.0){\rule[-0.200pt]{0.400pt}{2.409pt}}
\put(1019.0,113.0){\rule[-0.200pt]{0.400pt}{2.409pt}}
\put(1019.0,716.0){\rule[-0.200pt]{0.400pt}{2.409pt}}
\put(1059.0,113.0){\rule[-0.200pt]{0.400pt}{2.409pt}}
\put(1059.0,716.0){\rule[-0.200pt]{0.400pt}{2.409pt}}
\put(1091.0,113.0){\rule[-0.200pt]{0.400pt}{2.409pt}}
\put(1091.0,716.0){\rule[-0.200pt]{0.400pt}{2.409pt}}
\put(1116.0,113.0){\rule[-0.200pt]{0.400pt}{2.409pt}}
\put(1116.0,716.0){\rule[-0.200pt]{0.400pt}{2.409pt}}
\put(1138.0,113.0){\rule[-0.200pt]{0.400pt}{2.409pt}}
\put(1138.0,716.0){\rule[-0.200pt]{0.400pt}{2.409pt}}
\put(1157.0,113.0){\rule[-0.200pt]{0.400pt}{2.409pt}}
\put(1157.0,716.0){\rule[-0.200pt]{0.400pt}{2.409pt}}
\put(1173.0,113.0){\rule[-0.200pt]{0.400pt}{2.409pt}}
\put(1173.0,716.0){\rule[-0.200pt]{0.400pt}{2.409pt}}
\put(1188.0,113.0){\rule[-0.200pt]{0.400pt}{4.818pt}}
\put(1188,68){\makebox(0,0){100}}
\put(1188.0,706.0){\rule[-0.200pt]{0.400pt}{4.818pt}}
\put(1285.0,113.0){\rule[-0.200pt]{0.400pt}{2.409pt}}
\put(1285.0,716.0){\rule[-0.200pt]{0.400pt}{2.409pt}}
\put(220.0,113.0){\rule[-0.200pt]{256.558pt}{0.400pt}}
\put(1285.0,113.0){\rule[-0.200pt]{0.400pt}{147.672pt}}
\put(220.0,726.0){\rule[-0.200pt]{256.558pt}{0.400pt}}
\put(45,419){\makebox(0,0){$|\Delta \delta_0(E)|$}}
\put(752,23){\makebox(0,0){$E$ (MeV)}}
\put(374,630){\makebox(0,0)[l]{$V_{1\pi}$}}
\put(374,484){\makebox(0,0)[l]{$\Lambda^{-2}$}}
\put(374,337){\makebox(0,0)[l]{$\Lambda^{-4}$}}
\put(220.0,113.0){\rule[-0.200pt]{0.400pt}{147.672pt}}
\put(317,664){\usebox{\plotpoint}}
\multiput(317.00,664.60)(18.759,0.468){5}{\rule{13.000pt}{0.113pt}}
\multiput(317.00,663.17)(102.018,4.000){2}{\rule{6.500pt}{0.400pt}}
\put(446,667.67){\rule{23.367pt}{0.400pt}}
\multiput(446.00,667.17)(48.500,1.000){2}{\rule{11.684pt}{0.400pt}}
\put(543,667.17){\rule{19.500pt}{0.400pt}}
\multiput(543.00,668.17)(56.527,-2.000){2}{\rule{9.750pt}{0.400pt}}
\multiput(640.00,665.94)(18.613,-0.468){5}{\rule{12.900pt}{0.113pt}}
\multiput(640.00,666.17)(101.225,-4.000){2}{\rule{6.450pt}{0.400pt}}
\multiput(768.00,661.93)(10.729,-0.477){7}{\rule{7.860pt}{0.115pt}}
\multiput(768.00,662.17)(80.686,-5.000){2}{\rule{3.930pt}{0.400pt}}
\multiput(865.00,656.93)(6.371,-0.488){13}{\rule{4.950pt}{0.117pt}}
\multiput(865.00,657.17)(86.726,-8.000){2}{\rule{2.475pt}{0.400pt}}
\multiput(962.00,648.92)(2.971,-0.496){41}{\rule{2.445pt}{0.120pt}}
\multiput(962.00,649.17)(123.924,-22.000){2}{\rule{1.223pt}{0.400pt}}
\multiput(1091.00,626.92)(0.714,-0.499){133}{\rule{0.671pt}{0.120pt}}
\multiput(1091.00,627.17)(95.608,-68.000){2}{\rule{0.335pt}{0.400pt}}
\multiput(1188.00,560.58)(0.663,0.498){83}{\rule{0.630pt}{0.120pt}}
\multiput(1188.00,559.17)(55.692,43.000){2}{\rule{0.315pt}{0.400pt}}
\put(317,508){\usebox{\plotpoint}}
\multiput(317.00,508.58)(2.244,0.497){55}{\rule{1.879pt}{0.120pt}}
\multiput(317.00,507.17)(125.099,29.000){2}{\rule{0.940pt}{0.400pt}}
\multiput(446.00,537.58)(2.739,0.495){33}{\rule{2.256pt}{0.119pt}}
\multiput(446.00,536.17)(92.318,18.000){2}{\rule{1.128pt}{0.400pt}}
\multiput(543.00,555.58)(3.302,0.494){27}{\rule{2.687pt}{0.119pt}}
\multiput(543.00,554.17)(91.424,15.000){2}{\rule{1.343pt}{0.400pt}}
\multiput(640.00,570.58)(3.423,0.495){35}{\rule{2.795pt}{0.119pt}}
\multiput(640.00,569.17)(122.199,19.000){2}{\rule{1.397pt}{0.400pt}}
\multiput(768.00,589.58)(3.545,0.494){25}{\rule{2.871pt}{0.119pt}}
\multiput(768.00,588.17)(91.040,14.000){2}{\rule{1.436pt}{0.400pt}}
\multiput(865.00,603.58)(3.545,0.494){25}{\rule{2.871pt}{0.119pt}}
\multiput(865.00,602.17)(91.040,14.000){2}{\rule{1.436pt}{0.400pt}}
\multiput(962.00,617.58)(4.114,0.494){29}{\rule{3.325pt}{0.119pt}}
\multiput(962.00,616.17)(122.099,16.000){2}{\rule{1.663pt}{0.400pt}}
\multiput(1091.00,633.58)(4.158,0.492){21}{\rule{3.333pt}{0.119pt}}
\multiput(1091.00,632.17)(90.081,12.000){2}{\rule{1.667pt}{0.400pt}}
\multiput(1188.00,645.59)(3.730,0.488){13}{\rule{2.950pt}{0.117pt}}
\multiput(1188.00,644.17)(50.877,8.000){2}{\rule{1.475pt}{0.400pt}}
\put(317,226){\usebox{\plotpoint}}
\multiput(317.00,226.58)(0.839,0.499){151}{\rule{0.770pt}{0.120pt}}
\multiput(317.00,225.17)(127.402,77.000){2}{\rule{0.385pt}{0.400pt}}
\multiput(446.00,303.58)(1.318,0.498){71}{\rule{1.149pt}{0.120pt}}
\multiput(446.00,302.17)(94.616,37.000){2}{\rule{0.574pt}{0.400pt}}
\multiput(543.00,340.58)(2.133,0.496){43}{\rule{1.787pt}{0.120pt}}
\multiput(543.00,339.17)(93.291,23.000){2}{\rule{0.893pt}{0.400pt}}
\multiput(640.00,361.92)(1.841,-0.498){67}{\rule{1.563pt}{0.120pt}}
\multiput(640.00,362.17)(124.756,-35.000){2}{\rule{0.781pt}{0.400pt}}
\multiput(768.58,328.00)(0.499,0.608){191}{\rule{0.120pt}{0.587pt}}
\multiput(767.17,328.00)(97.000,116.782){2}{\rule{0.400pt}{0.293pt}}
\multiput(865.00,446.58)(0.935,0.498){101}{\rule{0.846pt}{0.120pt}}
\multiput(865.00,445.17)(95.244,52.000){2}{\rule{0.423pt}{0.400pt}}
\multiput(962.00,498.58)(1.544,0.498){81}{\rule{1.329pt}{0.120pt}}
\multiput(962.00,497.17)(126.242,42.000){2}{\rule{0.664pt}{0.400pt}}
\multiput(1091.00,538.94)(14.080,-0.468){5}{\rule{9.800pt}{0.113pt}}
\multiput(1091.00,539.17)(76.660,-4.000){2}{\rule{4.900pt}{0.400pt}}
\multiput(1188.00,536.59)(4.306,0.485){11}{\rule{3.357pt}{0.117pt}}
\multiput(1188.00,535.17)(50.032,7.000){2}{\rule{1.679pt}{0.400pt}}
\end{picture}
\end{center}
\caption{Errors in the ${}^1S_0$ phase shifts (in radians) 
 versus energy for the
effective theory through orders $\Lambda^{-2}$
and~$\Lambda^{-4}$. Results from the theory with just pion exchange
($V_{1\pi}$) are also shown. The cutoff was $\Lambda\!=\!330$\,MeV in each
case.}
\label{1s0_lam2lam4-fig}
\end{figure}
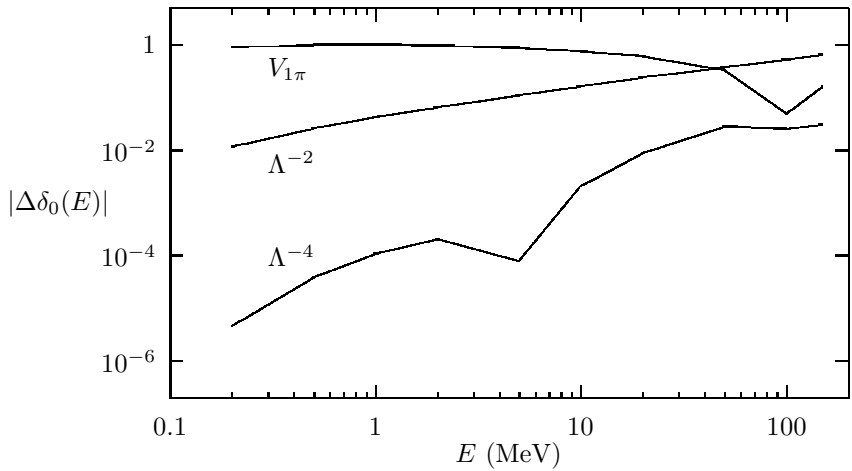

In Figure~\ref{1s0_lamdep-fig} I show the errors obtained in the phase
shift as a function of center-of-mass energy $E$ for a variety of
different values of the cutoff~$\Lambda$. As in our synthetic example, the
errors start out large 
when $\Lambda$ is small, and decrease steadily until
$\Lambda\!\approx\!300$\,MeV. No further improvement  is seen for
larger~$\Lambda$'s. 

In Figure~\ref{1s0_lam2lam4-fig} I compare the
errors obtained from the
$\Lambda^{-2}$~theory ($d\!=\!0$) with those from the
$\Lambda^{-4}$~theory. At low energies the errors in the
$\Lambda^{-2}$~theory grow approximately like~$E^{0.5}$ with energy. The
errors in the $\Lambda^{-4}$~theory are much smaller, but grow
like~$E^{1.5}$\,---\,that is faster by two powers of~$p$, exactly as
expected. 
So here, as in our synthetic data, we
see clear evidence that the theory is improved by adding higher-order
corrections, and that it is improved in just the manner
predicted. This convinces me that the effective theory is working.

The curves in the logarithmic plots of $\Delta\delta_0(E)$
versus~$E$ dip at various different points. These dips result when the
error changes sign. They should be ignored.

I also show errors in the last figure for the effective theory with
one-pion exchange but no contact terms ($V_{1\pi}$). Evidently
one-pion exchange contributes little to the phase shift at most
energies. This is surprising since the leading contact term and the
one-pion potential are both~$\order(1/\Lambda^2)$. The difference is
that one-pion exchange is order~$m_\pi^2/\Lambda^2$, while the
contact term is order~$q^2/\Lambda^2$. Thus the data suggest that 
momenta $q\!\gg\!m_\pi$ play an
important role during the scattering process, even for low incident
energies. This is possible if the interaction is fairly strong;
the phase shifts below 10\,MeV depend quite nonlinearly on the coupling
constants which indicates that this is case.

\begin{figure}
\begin{center}
% GNUPLOT: LaTeX picture
\setlength{\unitlength}{0.240900pt}
\ifx\plotpoint\undefined\newsavebox{\plotpoint}\fi
\sbox{\plotpoint}{\rule[-0.200pt]{0.400pt}{0.400pt}}%
\begin{picture}(1349,900)(0,0)
\font\gnuplot=cmr10 at 10pt
\gnuplot
\sbox{\plotpoint}{\rule[-0.200pt]{0.400pt}{0.400pt}}%
\put(220.0,222.0){\rule[-0.200pt]{4.818pt}{0.400pt}}
\put(198,222){\makebox(0,0)[r]{$0$}}
\put(1265.0,222.0){\rule[-0.200pt]{4.818pt}{0.400pt}}
\put(220.0,331.0){\rule[-0.200pt]{4.818pt}{0.400pt}}
\put(198,331){\makebox(0,0)[r]{$0.2$}}
\put(1265.0,331.0){\rule[-0.200pt]{4.818pt}{0.400pt}}
\put(220.0,440.0){\rule[-0.200pt]{4.818pt}{0.400pt}}
\put(198,440){\makebox(0,0)[r]{$0.4$}}
\put(1265.0,440.0){\rule[-0.200pt]{4.818pt}{0.400pt}}
\put(220.0,550.0){\rule[-0.200pt]{4.818pt}{0.400pt}}
\put(198,550){\makebox(0,0)[r]{$0.6$}}
\put(1265.0,550.0){\rule[-0.200pt]{4.818pt}{0.400pt}}
\put(220.0,659.0){\rule[-0.200pt]{4.818pt}{0.400pt}}
\put(198,659){\makebox(0,0)[r]{$0.8$}}
\put(1265.0,659.0){\rule[-0.200pt]{4.818pt}{0.400pt}}
\put(220.0,768.0){\rule[-0.200pt]{4.818pt}{0.400pt}}
\put(198,768){\makebox(0,0)[r]{$1$}}
\put(1265.0,768.0){\rule[-0.200pt]{4.818pt}{0.400pt}}
\put(220.0,113.0){\rule[-0.200pt]{0.400pt}{4.818pt}}
\put(220,68){\makebox(0,0){$0.1$}}
\put(220.0,857.0){\rule[-0.200pt]{0.400pt}{4.818pt}}
\put(317.0,113.0){\rule[-0.200pt]{0.400pt}{2.409pt}}
\put(317.0,867.0){\rule[-0.200pt]{0.400pt}{2.409pt}}
\put(374.0,113.0){\rule[-0.200pt]{0.400pt}{2.409pt}}
\put(374.0,867.0){\rule[-0.200pt]{0.400pt}{2.409pt}}
\put(414.0,113.0){\rule[-0.200pt]{0.400pt}{2.409pt}}
\put(414.0,867.0){\rule[-0.200pt]{0.400pt}{2.409pt}}
\put(446.0,113.0){\rule[-0.200pt]{0.400pt}{2.409pt}}
\put(446.0,867.0){\rule[-0.200pt]{0.400pt}{2.409pt}}
\put(471.0,113.0){\rule[-0.200pt]{0.400pt}{2.409pt}}
\put(471.0,867.0){\rule[-0.200pt]{0.400pt}{2.409pt}}
\put(493.0,113.0){\rule[-0.200pt]{0.400pt}{2.409pt}}
\put(493.0,867.0){\rule[-0.200pt]{0.400pt}{2.409pt}}
\put(511.0,113.0){\rule[-0.200pt]{0.400pt}{2.409pt}}
\put(511.0,867.0){\rule[-0.200pt]{0.400pt}{2.409pt}}
\put(528.0,113.0){\rule[-0.200pt]{0.400pt}{2.409pt}}
\put(528.0,867.0){\rule[-0.200pt]{0.400pt}{2.409pt}}
\put(543.0,113.0){\rule[-0.200pt]{0.400pt}{4.818pt}}
\put(543,68){\makebox(0,0){$1$}}
\put(543.0,857.0){\rule[-0.200pt]{0.400pt}{4.818pt}}
\put(640.0,113.0){\rule[-0.200pt]{0.400pt}{2.409pt}}
\put(640.0,867.0){\rule[-0.200pt]{0.400pt}{2.409pt}}
\put(697.0,113.0){\rule[-0.200pt]{0.400pt}{2.409pt}}
\put(697.0,867.0){\rule[-0.200pt]{0.400pt}{2.409pt}}
\put(737.0,113.0){\rule[-0.200pt]{0.400pt}{2.409pt}}
\put(737.0,867.0){\rule[-0.200pt]{0.400pt}{2.409pt}}
\put(768.0,113.0){\rule[-0.200pt]{0.400pt}{2.409pt}}
\put(768.0,867.0){\rule[-0.200pt]{0.400pt}{2.409pt}}
\put(794.0,113.0){\rule[-0.200pt]{0.400pt}{2.409pt}}
\put(794.0,867.0){\rule[-0.200pt]{0.400pt}{2.409pt}}
\put(815.0,113.0){\rule[-0.200pt]{0.400pt}{2.409pt}}
\put(815.0,867.0){\rule[-0.200pt]{0.400pt}{2.409pt}}
\put(834.0,113.0){\rule[-0.200pt]{0.400pt}{2.409pt}}
\put(834.0,867.0){\rule[-0.200pt]{0.400pt}{2.409pt}}
\put(850.0,113.0){\rule[-0.200pt]{0.400pt}{2.409pt}}
\put(850.0,867.0){\rule[-0.200pt]{0.400pt}{2.409pt}}
\put(865.0,113.0){\rule[-0.200pt]{0.400pt}{4.818pt}}
\put(865,68){\makebox(0,0){$10$}}
\put(865.0,857.0){\rule[-0.200pt]{0.400pt}{4.818pt}}
\put(962.0,113.0){\rule[-0.200pt]{0.400pt}{2.409pt}}
\put(962.0,867.0){\rule[-0.200pt]{0.400pt}{2.409pt}}
\put(1019.0,113.0){\rule[-0.200pt]{0.400pt}{2.409pt}}
\put(1019.0,867.0){\rule[-0.200pt]{0.400pt}{2.409pt}}
\put(1059.0,113.0){\rule[-0.200pt]{0.400pt}{2.409pt}}
\put(1059.0,867.0){\rule[-0.200pt]{0.400pt}{2.409pt}}
\put(1091.0,113.0){\rule[-0.200pt]{0.400pt}{2.409pt}}
\put(1091.0,867.0){\rule[-0.200pt]{0.400pt}{2.409pt}}
\put(1116.0,113.0){\rule[-0.200pt]{0.400pt}{2.409pt}}
\put(1116.0,867.0){\rule[-0.200pt]{0.400pt}{2.409pt}}
\put(1138.0,113.0){\rule[-0.200pt]{0.400pt}{2.409pt}}
\put(1138.0,867.0){\rule[-0.200pt]{0.400pt}{2.409pt}}
\put(1157.0,113.0){\rule[-0.200pt]{0.400pt}{2.409pt}}
\put(1157.0,867.0){\rule[-0.200pt]{0.400pt}{2.409pt}}
\put(1173.0,113.0){\rule[-0.200pt]{0.400pt}{2.409pt}}
\put(1173.0,867.0){\rule[-0.200pt]{0.400pt}{2.409pt}}
\put(1188.0,113.0){\rule[-0.200pt]{0.400pt}{4.818pt}}
\put(1188,68){\makebox(0,0){$100$}}
\put(1188.0,857.0){\rule[-0.200pt]{0.400pt}{4.818pt}}
\put(1285.0,113.0){\rule[-0.200pt]{0.400pt}{2.409pt}}
\put(1285.0,867.0){\rule[-0.200pt]{0.400pt}{2.409pt}}
\put(220.0,113.0){\rule[-0.200pt]{256.558pt}{0.400pt}}
\put(1285.0,113.0){\rule[-0.200pt]{0.400pt}{184.048pt}}
\put(220.0,877.0){\rule[-0.200pt]{256.558pt}{0.400pt}}
\put(45,495){\makebox(0,0){$\delta_0(E)$}}
\put(752,23){\makebox(0,0){$E$ (MeV)}}
\put(220.0,113.0){\rule[-0.200pt]{0.400pt}{184.048pt}}
\sbox{\plotpoint}{\rule[-0.500pt]{1.000pt}{1.000pt}}%
\put(511,440){\makebox(0,0)[r]{data}}
\put(555,440){\circle{24}}
\put(317,746){\circle{24}}
\put(446,813){\circle{24}}
\put(543,836){\circle{24}}
\put(640,834){\circle{24}}
\put(768,793){\circle{24}}
\put(865,732){\circle{24}}
\put(962,643){\circle{24}}
\put(1091,477){\circle{24}}
\put(1188,307){\circle{24}}
\put(1245,180){\circle{24}}
\put(511,395){\makebox(0,0)[r]{$\Lambda^{-2}$ theory}}
\multiput(533,395)(20.756,0.000){4}{\usebox{\plotpoint}}
\put(599,395){\usebox{\plotpoint}}
\put(317,747){\usebox{\plotpoint}}
\multiput(317,747)(18.361,9.679){8}{\usebox{\plotpoint}}
\multiput(446,815)(20.099,5.180){4}{\usebox{\plotpoint}}
\multiput(543,840)(20.756,0.000){5}{\usebox{\plotpoint}}
\multiput(640,840)(19.980,-5.619){6}{\usebox{\plotpoint}}
\multiput(768,804)(18.135,-10.096){6}{\usebox{\plotpoint}}
\multiput(865,750)(16.420,-12.696){6}{\usebox{\plotpoint}}
\multiput(962,675)(15.138,-14.199){8}{\usebox{\plotpoint}}
\multiput(1091,554)(15.531,-13.769){6}{\usebox{\plotpoint}}
\multiput(1188,468)(16.849,-12.120){4}{\usebox{\plotpoint}}
\put(1245,427){\usebox{\plotpoint}}
\sbox{\plotpoint}{\rule[-0.200pt]{0.400pt}{0.400pt}}%
\put(511,350){\makebox(0,0)[r]{$\Lambda\!=\!275$\,MeV}}
\put(533.0,350.0){\rule[-0.200pt]{15.899pt}{0.400pt}}
\put(317,746){\usebox{\plotpoint}}
\multiput(317.00,746.58)(0.965,0.499){131}{\rule{0.870pt}{0.120pt}}
\multiput(317.00,745.17)(127.194,67.000){2}{\rule{0.435pt}{0.400pt}}
\multiput(446.00,813.58)(2.133,0.496){43}{\rule{1.787pt}{0.120pt}}
\multiput(446.00,812.17)(93.291,23.000){2}{\rule{0.893pt}{0.400pt}}
\put(543,834.17){\rule{19.500pt}{0.400pt}}
\multiput(543.00,835.17)(56.527,-2.000){2}{\rule{9.750pt}{0.400pt}}
\multiput(640.00,832.92)(1.609,-0.498){77}{\rule{1.380pt}{0.120pt}}
\multiput(640.00,833.17)(125.136,-40.000){2}{\rule{0.690pt}{0.400pt}}
\multiput(768.00,792.92)(0.809,-0.499){117}{\rule{0.747pt}{0.120pt}}
\multiput(768.00,793.17)(95.450,-60.000){2}{\rule{0.373pt}{0.400pt}}
\multiput(865.00,732.92)(0.557,-0.499){171}{\rule{0.546pt}{0.120pt}}
\multiput(865.00,733.17)(95.867,-87.000){2}{\rule{0.273pt}{0.400pt}}
\multiput(962.58,644.62)(0.499,-0.593){255}{\rule{0.120pt}{0.574pt}}
\multiput(961.17,645.81)(129.000,-151.808){2}{\rule{0.400pt}{0.287pt}}
\multiput(1091.58,491.58)(0.499,-0.603){191}{\rule{0.120pt}{0.582pt}}
\multiput(1090.17,492.79)(97.000,-115.791){2}{\rule{0.400pt}{0.291pt}}
\multiput(1188.00,375.92)(0.570,-0.498){97}{\rule{0.556pt}{0.120pt}}
\multiput(1188.00,376.17)(55.846,-50.000){2}{\rule{0.278pt}{0.400pt}}
\sbox{\plotpoint}{\rule[-0.400pt]{0.800pt}{0.800pt}}%
\put(511,305){\makebox(0,0)[r]{$\Lambda\!=\!330$\,MeV}}
\put(533.0,305.0){\rule[-0.400pt]{15.899pt}{0.800pt}}
\put(317,746){\usebox{\plotpoint}}
\multiput(317.00,747.41)(0.966,0.501){127}{\rule{1.740pt}{0.121pt}}
\multiput(317.00,744.34)(125.388,67.000){2}{\rule{0.870pt}{0.800pt}}
\multiput(446.00,814.41)(2.162,0.505){39}{\rule{3.574pt}{0.122pt}}
\multiput(446.00,811.34)(89.582,23.000){2}{\rule{1.787pt}{0.800pt}}
\put(543,833.34){\rule{23.367pt}{0.800pt}}
\multiput(543.00,834.34)(48.500,-2.000){2}{\rule{11.684pt}{0.800pt}}
\multiput(640.00,832.09)(1.578,-0.502){75}{\rule{2.698pt}{0.121pt}}
\multiput(640.00,832.34)(122.401,-41.000){2}{\rule{1.349pt}{0.800pt}}
\multiput(768.00,791.09)(0.784,-0.502){117}{\rule{1.452pt}{0.121pt}}
\multiput(768.00,791.34)(93.987,-62.000){2}{\rule{0.726pt}{0.800pt}}
\multiput(865.00,729.09)(0.527,-0.501){177}{\rule{1.043pt}{0.121pt}}
\multiput(865.00,729.34)(94.834,-92.000){2}{\rule{0.522pt}{0.800pt}}
\multiput(963.41,633.61)(0.501,-0.686){251}{\rule{0.121pt}{1.298pt}}
\multiput(960.34,636.31)(129.000,-174.307){2}{\rule{0.800pt}{0.649pt}}
\multiput(1092.41,455.42)(0.501,-0.868){187}{\rule{0.121pt}{1.586pt}}
\multiput(1089.34,458.71)(97.000,-164.709){2}{\rule{0.800pt}{0.793pt}}
\multiput(1189.41,287.52)(0.502,-0.854){107}{\rule{0.121pt}{1.561pt}}
\multiput(1186.34,290.76)(57.000,-93.759){2}{\rule{0.800pt}{0.781pt}}
\sbox{\plotpoint}{\rule[-0.600pt]{1.200pt}{1.200pt}}%
\put(511,260){\makebox(0,0)[r]{$\Lambda\!=\!400$\,MeV}}
\put(533.0,260.0){\rule[-0.600pt]{15.899pt}{1.200pt}}
\put(317,746){\usebox{\plotpoint}}
\multiput(317.00,748.24)(0.961,0.500){124}{\rule{2.610pt}{0.120pt}}
\multiput(317.00,743.51)(123.582,67.000){2}{\rule{1.305pt}{1.200pt}}
\multiput(446.00,815.24)(2.136,0.501){36}{\rule{5.361pt}{0.121pt}}
\multiput(446.00,810.51)(85.873,23.000){2}{\rule{2.680pt}{1.200pt}}
\put(543,832.51){\rule{23.367pt}{1.200pt}}
\multiput(543.00,833.51)(48.500,-2.000){2}{\rule{11.684pt}{1.200pt}}
\multiput(640.00,831.26)(1.566,-0.500){72}{\rule{4.046pt}{0.121pt}}
\multiput(640.00,831.51)(119.602,-41.000){2}{\rule{2.023pt}{1.200pt}}
\multiput(768.00,790.26)(0.754,-0.500){118}{\rule{2.119pt}{0.120pt}}
\multiput(768.00,790.51)(92.602,-64.000){2}{\rule{1.059pt}{1.200pt}}
\multiput(865.00,726.26)(0.501,-0.500){182}{\rule{1.513pt}{0.120pt}}
\multiput(865.00,726.51)(93.861,-96.000){2}{\rule{0.756pt}{1.200pt}}
\multiput(964.24,624.26)(0.500,-0.750){248}{\rule{0.120pt}{2.105pt}}
\multiput(959.51,628.63)(129.000,-189.632){2}{\rule{1.200pt}{1.052pt}}
\multiput(1093.24,427.02)(0.500,-1.077){184}{\rule{0.120pt}{2.886pt}}
\multiput(1088.51,433.01)(97.000,-203.011){2}{\rule{1.200pt}{1.443pt}}
\multiput(1190.24,215.80)(0.500,-1.301){80}{\rule{0.121pt}{3.420pt}}
\multiput(1185.51,222.90)(45.000,-109.902){2}{\rule{1.200pt}{1.710pt}}
\end{picture}
\end{center}
\caption{The ${}^1S_0$ phase shifts (in radians) versus energy for the full
effective theory with various values of the cutoff~$\Lambda$. Results
for the $\Lambda^{-2}$ version of the theory  ($d\!=\!0$) 
are also shown ($\Lambda\!=\!200$\,MeV).}
\label{1s0_phase-fig}
\end{figure}
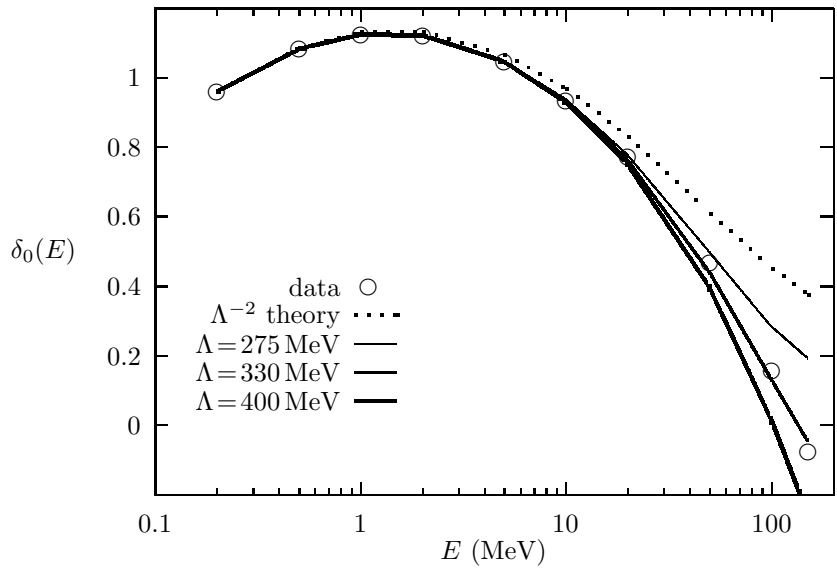

In Figure~\ref{1s0_phase-fig} I
show the actual phase shifts, together with the data. The results from
the full theory are almost independent of the cutoff. The variation at
high energies gives an indication of the size of
the $\Lambda^{-6}$~corrections yet to be included.

It is slightly mysterious that the full effective theory stops improving
when $\Lambda\!\approx\!300$--400\,MeV. We had
expected  hadron structure and other QCD effects to appear
at~500--1000\,MeV, not~300\,MeV. One possible explanation is that we have
omitted two-pion exchange from our potential. Two-pion contributions are
nominally the same order as the $\Lambda^{-4}$~contact term;
to be consistent, we should include them in our
analysis. We don't expect a particularly 
large contribution from two-pion exchange since
one-pion exchange is so small. Nevertheless the errors are small
at $\Lambda\!=\!300$\,MeV, and so we do not need much from two pions.

\begin{figure}
\begin{center}
% GNUPLOT: LaTeX picture
\setlength{\unitlength}{0.240900pt}
\ifx\plotpoint\undefined\newsavebox{\plotpoint}\fi
\sbox{\plotpoint}{\rule[-0.200pt]{0.400pt}{0.400pt}}%
\begin{picture}(1349,749)(0,0)
\font\gnuplot=cmr10 at 10pt
\gnuplot
\sbox{\plotpoint}{\rule[-0.200pt]{0.400pt}{0.400pt}}%
\put(220.0,171.0){\rule[-0.200pt]{4.818pt}{0.400pt}}
\put(198,171){\makebox(0,0)[r]{$10^{-6}$}}
\put(1265.0,171.0){\rule[-0.200pt]{4.818pt}{0.400pt}}
\put(220.0,337.0){\rule[-0.200pt]{4.818pt}{0.400pt}}
\put(198,337){\makebox(0,0)[r]{$10^{-4}$}}
\put(1265.0,337.0){\rule[-0.200pt]{4.818pt}{0.400pt}}
\put(220.0,502.0){\rule[-0.200pt]{4.818pt}{0.400pt}}
\put(198,502){\makebox(0,0)[r]{$10^{-2}$}}
\put(1265.0,502.0){\rule[-0.200pt]{4.818pt}{0.400pt}}
\put(220.0,668.0){\rule[-0.200pt]{4.818pt}{0.400pt}}
\put(198,668){\makebox(0,0)[r]{$1$}}
\put(1265.0,668.0){\rule[-0.200pt]{4.818pt}{0.400pt}}
\put(220.0,113.0){\rule[-0.200pt]{0.400pt}{4.818pt}}
\put(220,68){\makebox(0,0){0.1}}
\put(220.0,706.0){\rule[-0.200pt]{0.400pt}{4.818pt}}
\put(317.0,113.0){\rule[-0.200pt]{0.400pt}{2.409pt}}
\put(317.0,716.0){\rule[-0.200pt]{0.400pt}{2.409pt}}
\put(374.0,113.0){\rule[-0.200pt]{0.400pt}{2.409pt}}
\put(374.0,716.0){\rule[-0.200pt]{0.400pt}{2.409pt}}
\put(414.0,113.0){\rule[-0.200pt]{0.400pt}{2.409pt}}
\put(414.0,716.0){\rule[-0.200pt]{0.400pt}{2.409pt}}
\put(446.0,113.0){\rule[-0.200pt]{0.400pt}{2.409pt}}
\put(446.0,716.0){\rule[-0.200pt]{0.400pt}{2.409pt}}
\put(471.0,113.0){\rule[-0.200pt]{0.400pt}{2.409pt}}
\put(471.0,716.0){\rule[-0.200pt]{0.400pt}{2.409pt}}
\put(493.0,113.0){\rule[-0.200pt]{0.400pt}{2.409pt}}
\put(493.0,716.0){\rule[-0.200pt]{0.400pt}{2.409pt}}
\put(511.0,113.0){\rule[-0.200pt]{0.400pt}{2.409pt}}
\put(511.0,716.0){\rule[-0.200pt]{0.400pt}{2.409pt}}
\put(528.0,113.0){\rule[-0.200pt]{0.400pt}{2.409pt}}
\put(528.0,716.0){\rule[-0.200pt]{0.400pt}{2.409pt}}
\put(543.0,113.0){\rule[-0.200pt]{0.400pt}{4.818pt}}
\put(543,68){\makebox(0,0){1}}
\put(543.0,706.0){\rule[-0.200pt]{0.400pt}{4.818pt}}
\put(640.0,113.0){\rule[-0.200pt]{0.400pt}{2.409pt}}
\put(640.0,716.0){\rule[-0.200pt]{0.400pt}{2.409pt}}
\put(697.0,113.0){\rule[-0.200pt]{0.400pt}{2.409pt}}
\put(697.0,716.0){\rule[-0.200pt]{0.400pt}{2.409pt}}
\put(737.0,113.0){\rule[-0.200pt]{0.400pt}{2.409pt}}
\put(737.0,716.0){\rule[-0.200pt]{0.400pt}{2.409pt}}
\put(768.0,113.0){\rule[-0.200pt]{0.400pt}{2.409pt}}
\put(768.0,716.0){\rule[-0.200pt]{0.400pt}{2.409pt}}
\put(794.0,113.0){\rule[-0.200pt]{0.400pt}{2.409pt}}
\put(794.0,716.0){\rule[-0.200pt]{0.400pt}{2.409pt}}
\put(815.0,113.0){\rule[-0.200pt]{0.400pt}{2.409pt}}
\put(815.0,716.0){\rule[-0.200pt]{0.400pt}{2.409pt}}
\put(834.0,113.0){\rule[-0.200pt]{0.400pt}{2.409pt}}
\put(834.0,716.0){\rule[-0.200pt]{0.400pt}{2.409pt}}
\put(850.0,113.0){\rule[-0.200pt]{0.400pt}{2.409pt}}
\put(850.0,716.0){\rule[-0.200pt]{0.400pt}{2.409pt}}
\put(865.0,113.0){\rule[-0.200pt]{0.400pt}{4.818pt}}
\put(865,68){\makebox(0,0){10}}
\put(865.0,706.0){\rule[-0.200pt]{0.400pt}{4.818pt}}
\put(962.0,113.0){\rule[-0.200pt]{0.400pt}{2.409pt}}
\put(962.0,716.0){\rule[-0.200pt]{0.400pt}{2.409pt}}
\put(1019.0,113.0){\rule[-0.200pt]{0.400pt}{2.409pt}}
\put(1019.0,716.0){\rule[-0.200pt]{0.400pt}{2.409pt}}
\put(1059.0,113.0){\rule[-0.200pt]{0.400pt}{2.409pt}}
\put(1059.0,716.0){\rule[-0.200pt]{0.400pt}{2.409pt}}
\put(1091.0,113.0){\rule[-0.200pt]{0.400pt}{2.409pt}}
\put(1091.0,716.0){\rule[-0.200pt]{0.400pt}{2.409pt}}
\put(1116.0,113.0){\rule[-0.200pt]{0.400pt}{2.409pt}}
\put(1116.0,716.0){\rule[-0.200pt]{0.400pt}{2.409pt}}
\put(1138.0,113.0){\rule[-0.200pt]{0.400pt}{2.409pt}}
\put(1138.0,716.0){\rule[-0.200pt]{0.400pt}{2.409pt}}
\put(1157.0,113.0){\rule[-0.200pt]{0.400pt}{2.409pt}}
\put(1157.0,716.0){\rule[-0.200pt]{0.400pt}{2.409pt}}
\put(1173.0,113.0){\rule[-0.200pt]{0.400pt}{2.409pt}}
\put(1173.0,716.0){\rule[-0.200pt]{0.400pt}{2.409pt}}
\put(1188.0,113.0){\rule[-0.200pt]{0.400pt}{4.818pt}}
\put(1188,68){\makebox(0,0){100}}
\put(1188.0,706.0){\rule[-0.200pt]{0.400pt}{4.818pt}}
\put(1285.0,113.0){\rule[-0.200pt]{0.400pt}{2.409pt}}
\put(1285.0,716.0){\rule[-0.200pt]{0.400pt}{2.409pt}}
\put(220.0,113.0){\rule[-0.200pt]{256.558pt}{0.400pt}}
\put(1285.0,113.0){\rule[-0.200pt]{0.400pt}{147.672pt}}
\put(220.0,726.0){\rule[-0.200pt]{256.558pt}{0.400pt}}
\put(45,419){\makebox(0,0){$|\Delta \delta_0(E)|$}}
\put(752,23){\makebox(0,0){$E$ (MeV)}}
\put(220.0,113.0){\rule[-0.200pt]{0.400pt}{147.672pt}}
\sbox{\plotpoint}{\rule[-0.500pt]{1.000pt}{1.000pt}}%
\put(1138,293){\makebox(0,0)[r]{no pion, $\Lambda\!=\!100$\,MeV}}
\multiput(1160,293)(41.511,0.000){2}{\usebox{\plotpoint}}
\put(1226,293){\usebox{\plotpoint}}
\put(317,365){\usebox{\plotpoint}}
\multiput(317,365)(35.644,21.276){4}{\usebox{\plotpoint}}
\multiput(446,442)(37.804,17.148){3}{\usebox{\plotpoint}}
\multiput(543,486)(38.651,15.142){2}{\usebox{\plotpoint}}
\multiput(640,524)(39.532,12.663){3}{\usebox{\plotpoint}}
\multiput(768,565)(40.483,9.182){3}{\usebox{\plotpoint}}
\multiput(865,587)(41.292,4.257){2}{\usebox{\plotpoint}}
\multiput(962,597)(39.819,-11.730){3}{\usebox{\plotpoint}}
\multiput(1091,559)(38.236,16.161){3}{\usebox{\plotpoint}}
\put(1208.33,609.99){\usebox{\plotpoint}}
\put(1245,628){\usebox{\plotpoint}}
\sbox{\plotpoint}{\rule[-0.200pt]{0.400pt}{0.400pt}}%
\put(1138,248){\makebox(0,0)[r]{no pion, $\Lambda\!=\!150$\,MeV}}
\put(1160.0,248.0){\rule[-0.200pt]{15.899pt}{0.400pt}}
\put(317,322){\usebox{\plotpoint}}
\multiput(317.00,322.58)(0.839,0.499){151}{\rule{0.770pt}{0.120pt}}
\multiput(317.00,321.17)(127.402,77.000){2}{\rule{0.385pt}{0.400pt}}
\multiput(446.00,399.58)(1.132,0.498){83}{\rule{1.002pt}{0.120pt}}
\multiput(446.00,398.17)(94.920,43.000){2}{\rule{0.501pt}{0.400pt}}
\multiput(543.00,442.58)(1.318,0.498){71}{\rule{1.149pt}{0.120pt}}
\multiput(543.00,441.17)(94.616,37.000){2}{\rule{0.574pt}{0.400pt}}
\multiput(640.00,479.58)(1.650,0.498){75}{\rule{1.413pt}{0.120pt}}
\multiput(640.00,478.17)(125.068,39.000){2}{\rule{0.706pt}{0.400pt}}
\multiput(768.00,518.58)(2.591,0.495){35}{\rule{2.142pt}{0.119pt}}
\multiput(768.00,517.17)(92.554,19.000){2}{\rule{1.071pt}{0.400pt}}
\put(865,536.67){\rule{23.367pt}{0.400pt}}
\multiput(865.00,536.17)(48.500,1.000){2}{\rule{11.684pt}{0.400pt}}
\multiput(962.00,538.58)(3.115,0.496){39}{\rule{2.557pt}{0.119pt}}
\multiput(962.00,537.17)(123.693,21.000){2}{\rule{1.279pt}{0.400pt}}
\multiput(1091.00,559.58)(0.884,0.499){107}{\rule{0.805pt}{0.120pt}}
\multiput(1091.00,558.17)(95.328,55.000){2}{\rule{0.403pt}{0.400pt}}
\multiput(1188.00,614.58)(1.440,0.496){37}{\rule{1.240pt}{0.119pt}}
\multiput(1188.00,613.17)(54.426,20.000){2}{\rule{0.620pt}{0.400pt}}
\sbox{\plotpoint}{\rule[-0.400pt]{0.800pt}{0.800pt}}%
\put(1138,203){\makebox(0,0)[r]{no pion, $\Lambda\!=\!200$\,MeV}}
\put(1160.0,203.0){\rule[-0.400pt]{15.899pt}{0.800pt}}
\put(317,327){\usebox{\plotpoint}}
\multiput(317.00,328.41)(0.840,0.501){147}{\rule{1.540pt}{0.121pt}}
\multiput(317.00,325.34)(125.803,77.000){2}{\rule{0.770pt}{0.800pt}}
\multiput(446.00,405.41)(1.137,0.502){79}{\rule{2.005pt}{0.121pt}}
\multiput(446.00,402.34)(92.839,43.000){2}{\rule{1.002pt}{0.800pt}}
\multiput(543.00,448.41)(1.325,0.503){67}{\rule{2.297pt}{0.121pt}}
\multiput(543.00,445.34)(92.232,37.000){2}{\rule{1.149pt}{0.800pt}}
\multiput(640.00,485.41)(1.578,0.502){75}{\rule{2.698pt}{0.121pt}}
\multiput(640.00,482.34)(122.401,41.000){2}{\rule{1.349pt}{0.800pt}}
\multiput(768.00,526.41)(2.162,0.505){39}{\rule{3.574pt}{0.122pt}}
\multiput(768.00,523.34)(89.582,23.000){2}{\rule{1.787pt}{0.800pt}}
\multiput(865.00,549.41)(4.333,0.511){17}{\rule{6.667pt}{0.123pt}}
\multiput(865.00,546.34)(83.163,12.000){2}{\rule{3.333pt}{0.800pt}}
\multiput(962.00,558.09)(0.829,-0.501){149}{\rule{1.523pt}{0.121pt}}
\multiput(962.00,558.34)(125.839,-78.000){2}{\rule{0.762pt}{0.800pt}}
\multiput(1092.41,482.00)(0.501,0.619){187}{\rule{0.121pt}{1.190pt}}
\multiput(1089.34,482.00)(97.000,117.531){2}{\rule{0.800pt}{0.595pt}}
\multiput(1188.00,603.41)(1.069,0.504){47}{\rule{1.889pt}{0.121pt}}
\multiput(1188.00,600.34)(53.080,27.000){2}{\rule{0.944pt}{0.800pt}}
\sbox{\plotpoint}{\rule[-0.500pt]{1.000pt}{1.000pt}}%
\put(1138,158){\makebox(0,0)[r]{pion, $\Lambda\!=\!330$\,MeV}}
\multiput(1160,158)(20.756,0.000){4}{\usebox{\plotpoint}}
\put(1226,158){\usebox{\plotpoint}}
\put(317,226){\usebox{\plotpoint}}
\multiput(317,226)(17.822,10.638){8}{\usebox{\plotpoint}}
\multiput(446,303)(19.393,7.397){5}{\usebox{\plotpoint}}
\multiput(543,340)(20.196,4.789){5}{\usebox{\plotpoint}}
\multiput(640,363)(20.021,-5.474){6}{\usebox{\plotpoint}}
\multiput(768,328)(13.180,16.034){7}{\usebox{\plotpoint}}
\multiput(865,446)(18.293,9.806){6}{\usebox{\plotpoint}}
\multiput(962,498)(19.736,6.426){6}{\usebox{\plotpoint}}
\multiput(1091,540)(20.738,-0.855){5}{\usebox{\plotpoint}}
\multiput(1188,536)(20.601,2.530){3}{\usebox{\plotpoint}}
\put(1245,543){\usebox{\plotpoint}}
\end{picture}
\end{center}
\caption{Errors in ${}^1S_0$ phase shifts (in radians) versus energy 
for the effective theory with no pion exchange, with various
values of the cutoff~$\Lambda$. Results for the full theory, including
pion exchange, are also shown.}
\label{1s0_nopion-fig}
\end{figure}
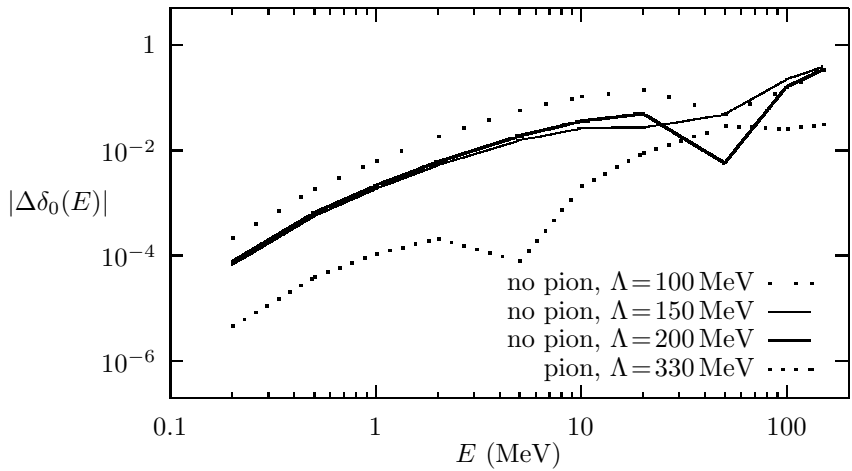

One way to test whether our analysis should be sensitive to a small two-pion
contribution is see what happens if we drop one-pion exchange.
I retuned the effective theory to the data, keeping only
the contact terms.  The results are summarized in
Figure~\ref{1s0_nopion-fig}. I compare the phase shift 
errors from the theory without pions, for different $\Lambda$'s,
to the best fit that includes pion exchange.
The theory without pions
stops improving at $\Lambda\!\approx\!150$\,MeV, rather than at 300\,MeV,
and so is substantially less accurate at {\em all\/} energies
below~100\,MeV. This indicates that there is new physics in the data,
physics that cannot be modelled by the contact potentials, at a momentum
scale of order~150\,MeV. The new physics, of course, is one-pion exchange.
The contact potentials are unable to model one-pion exchange accurately
once momenta of order $m_\pi$ are admitted into the effective theory,
despite the fact that the one-pion contribution is relatively quite
small. We can move the cutoff to 300--400\,MeV only by including one-pion
exchange in the potential; perhaps we can move the cutoff further by
including two-pion exchange.

There are other possible explanations for the  300\,MeV scale in our
analysis. For example, the inverse radius of a proton is about
240\,MeV. If quarks can be resolved at this scale, then it is likely
that we will be unable to raise the cutoff without switching to QCD.
The radius of the proton is not necessarily the same as the color
resolution scale since some of the radius could be due to a pion
cloud. Other issues that I haven't
examined with sufficient thoroughness include: tuning $\alpha_\pi$; the
differences in mass and $\alpha_\pi$ between charged and neutral
pions; contributions from~$\Delta$'s;
and, of course, possible differences between the phase shifts and real data.

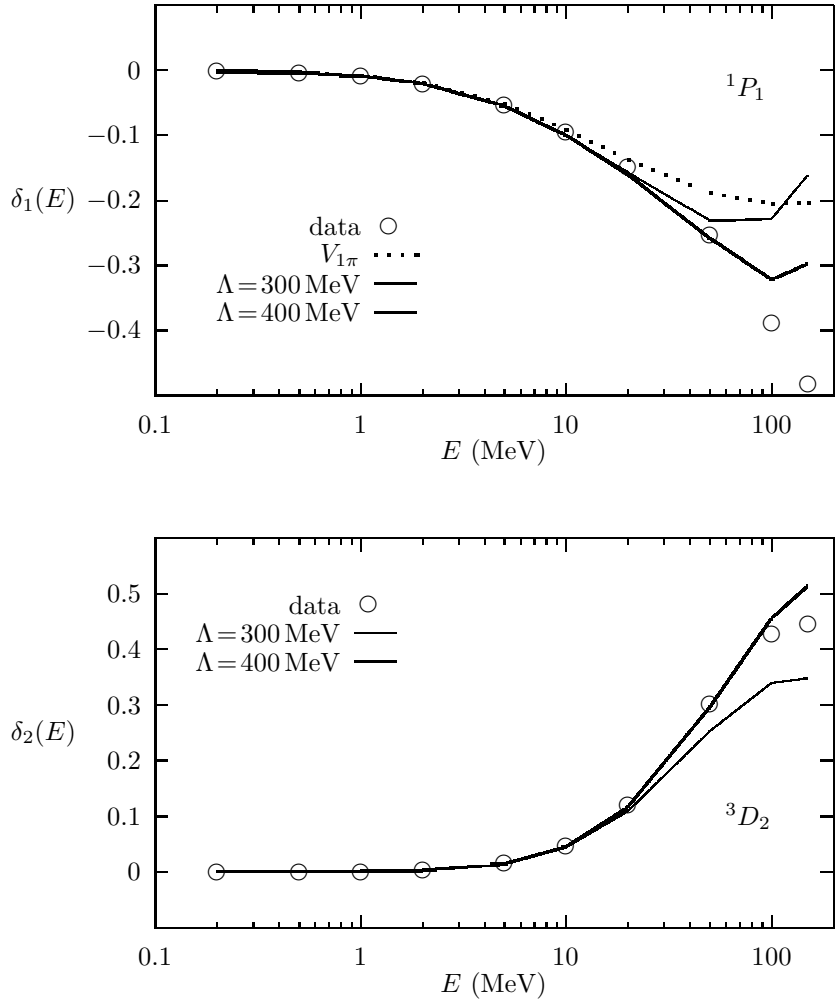
\begin{figure}
\begin{center}
% GNUPLOT: LaTeX picture
\setlength{\unitlength}{0.240900pt}
\ifx\plotpoint\undefined\newsavebox{\plotpoint}\fi
\sbox{\plotpoint}{\rule[-0.200pt]{0.400pt}{0.400pt}}%
\begin{picture}(1349,749)(0,0)
\font\gnuplot=cmr10 at 10pt
\gnuplot
\sbox{\plotpoint}{\rule[-0.200pt]{0.400pt}{0.400pt}}%
\put(220.0,215.0){\rule[-0.200pt]{4.818pt}{0.400pt}}
\put(198,215){\makebox(0,0)[r]{$-0.4$}}
\put(1265.0,215.0){\rule[-0.200pt]{4.818pt}{0.400pt}}
\put(220.0,317.0){\rule[-0.200pt]{4.818pt}{0.400pt}}
\put(198,317){\makebox(0,0)[r]{$-0.3$}}
\put(1265.0,317.0){\rule[-0.200pt]{4.818pt}{0.400pt}}
\put(220.0,419.0){\rule[-0.200pt]{4.818pt}{0.400pt}}
\put(198,419){\makebox(0,0)[r]{$-0.2$}}
\put(1265.0,419.0){\rule[-0.200pt]{4.818pt}{0.400pt}}
\put(220.0,522.0){\rule[-0.200pt]{4.818pt}{0.400pt}}
\put(198,522){\makebox(0,0)[r]{$-0.1$}}
\put(1265.0,522.0){\rule[-0.200pt]{4.818pt}{0.400pt}}
\put(220.0,624.0){\rule[-0.200pt]{4.818pt}{0.400pt}}
\put(198,624){\makebox(0,0)[r]{$0$}}
\put(1265.0,624.0){\rule[-0.200pt]{4.818pt}{0.400pt}}
\put(220.0,113.0){\rule[-0.200pt]{0.400pt}{4.818pt}}
\put(220,68){\makebox(0,0){$0.1$}}
\put(220.0,706.0){\rule[-0.200pt]{0.400pt}{4.818pt}}
\put(317.0,113.0){\rule[-0.200pt]{0.400pt}{2.409pt}}
\put(317.0,716.0){\rule[-0.200pt]{0.400pt}{2.409pt}}
\put(374.0,113.0){\rule[-0.200pt]{0.400pt}{2.409pt}}
\put(374.0,716.0){\rule[-0.200pt]{0.400pt}{2.409pt}}
\put(414.0,113.0){\rule[-0.200pt]{0.400pt}{2.409pt}}
\put(414.0,716.0){\rule[-0.200pt]{0.400pt}{2.409pt}}
\put(446.0,113.0){\rule[-0.200pt]{0.400pt}{2.409pt}}
\put(446.0,716.0){\rule[-0.200pt]{0.400pt}{2.409pt}}
\put(471.0,113.0){\rule[-0.200pt]{0.400pt}{2.409pt}}
\put(471.0,716.0){\rule[-0.200pt]{0.400pt}{2.409pt}}
\put(493.0,113.0){\rule[-0.200pt]{0.400pt}{2.409pt}}
\put(493.0,716.0){\rule[-0.200pt]{0.400pt}{2.409pt}}
\put(511.0,113.0){\rule[-0.200pt]{0.400pt}{2.409pt}}
\put(511.0,716.0){\rule[-0.200pt]{0.400pt}{2.409pt}}
\put(528.0,113.0){\rule[-0.200pt]{0.400pt}{2.409pt}}
\put(528.0,716.0){\rule[-0.200pt]{0.400pt}{2.409pt}}
\put(543.0,113.0){\rule[-0.200pt]{0.400pt}{4.818pt}}
\put(543,68){\makebox(0,0){$1$}}
\put(543.0,706.0){\rule[-0.200pt]{0.400pt}{4.818pt}}
\put(640.0,113.0){\rule[-0.200pt]{0.400pt}{2.409pt}}
\put(640.0,716.0){\rule[-0.200pt]{0.400pt}{2.409pt}}
\put(697.0,113.0){\rule[-0.200pt]{0.400pt}{2.409pt}}
\put(697.0,716.0){\rule[-0.200pt]{0.400pt}{2.409pt}}
\put(737.0,113.0){\rule[-0.200pt]{0.400pt}{2.409pt}}
\put(737.0,716.0){\rule[-0.200pt]{0.400pt}{2.409pt}}
\put(768.0,113.0){\rule[-0.200pt]{0.400pt}{2.409pt}}
\put(768.0,716.0){\rule[-0.200pt]{0.400pt}{2.409pt}}
\put(794.0,113.0){\rule[-0.200pt]{0.400pt}{2.409pt}}
\put(794.0,716.0){\rule[-0.200pt]{0.400pt}{2.409pt}}
\put(815.0,113.0){\rule[-0.200pt]{0.400pt}{2.409pt}}
\put(815.0,716.0){\rule[-0.200pt]{0.400pt}{2.409pt}}
\put(834.0,113.0){\rule[-0.200pt]{0.400pt}{2.409pt}}
\put(834.0,716.0){\rule[-0.200pt]{0.400pt}{2.409pt}}
\put(850.0,113.0){\rule[-0.200pt]{0.400pt}{2.409pt}}
\put(850.0,716.0){\rule[-0.200pt]{0.400pt}{2.409pt}}
\put(865.0,113.0){\rule[-0.200pt]{0.400pt}{4.818pt}}
\put(865,68){\makebox(0,0){$10$}}
\put(865.0,706.0){\rule[-0.200pt]{0.400pt}{4.818pt}}
\put(962.0,113.0){\rule[-0.200pt]{0.400pt}{2.409pt}}
\put(962.0,716.0){\rule[-0.200pt]{0.400pt}{2.409pt}}
\put(1019.0,113.0){\rule[-0.200pt]{0.400pt}{2.409pt}}
\put(1019.0,716.0){\rule[-0.200pt]{0.400pt}{2.409pt}}
\put(1059.0,113.0){\rule[-0.200pt]{0.400pt}{2.409pt}}
\put(1059.0,716.0){\rule[-0.200pt]{0.400pt}{2.409pt}}
\put(1091.0,113.0){\rule[-0.200pt]{0.400pt}{2.409pt}}
\put(1091.0,716.0){\rule[-0.200pt]{0.400pt}{2.409pt}}
\put(1116.0,113.0){\rule[-0.200pt]{0.400pt}{2.409pt}}
\put(1116.0,716.0){\rule[-0.200pt]{0.400pt}{2.409pt}}
\put(1138.0,113.0){\rule[-0.200pt]{0.400pt}{2.409pt}}
\put(1138.0,716.0){\rule[-0.200pt]{0.400pt}{2.409pt}}
\put(1157.0,113.0){\rule[-0.200pt]{0.400pt}{2.409pt}}
\put(1157.0,716.0){\rule[-0.200pt]{0.400pt}{2.409pt}}
\put(1173.0,113.0){\rule[-0.200pt]{0.400pt}{2.409pt}}
\put(1173.0,716.0){\rule[-0.200pt]{0.400pt}{2.409pt}}
\put(1188.0,113.0){\rule[-0.200pt]{0.400pt}{4.818pt}}
\put(1188,68){\makebox(0,0){$100$}}
\put(1188.0,706.0){\rule[-0.200pt]{0.400pt}{4.818pt}}
\put(1285.0,113.0){\rule[-0.200pt]{0.400pt}{2.409pt}}
\put(1285.0,716.0){\rule[-0.200pt]{0.400pt}{2.409pt}}
\put(220.0,113.0){\rule[-0.200pt]{256.558pt}{0.400pt}}
\put(1285.0,113.0){\rule[-0.200pt]{0.400pt}{147.672pt}}
\put(220.0,726.0){\rule[-0.200pt]{256.558pt}{0.400pt}}
\put(45,419){\makebox(0,0){$\delta_1(E)$}}
\put(752,23){\makebox(0,0){$E$ (MeV)}}
\put(1116,598){\makebox(0,0)[l]{${}^1P_1$}}
\put(220.0,113.0){\rule[-0.200pt]{0.400pt}{147.672pt}}
\sbox{\plotpoint}{\rule[-0.500pt]{1.000pt}{1.000pt}}%
\put(543,379){\makebox(0,0)[r]{data}}
\put(587,379){\circle{24}}
\put(317,623){\circle{24}}
\put(446,620){\circle{24}}
\put(543,615){\circle{24}}
\put(640,603){\circle{24}}
\put(768,570){\circle{24}}
\put(865,528){\circle{24}}
\put(962,473){\circle{24}}
\put(1091,365){\circle{24}}
\put(1188,228){\circle{24}}
\put(1245,132){\circle{24}}
\put(543,334){\makebox(0,0)[r]{$V_{1\pi}$}}
\multiput(565,334)(20.756,0.000){4}{\usebox{\plotpoint}}
\put(631,334){\usebox{\plotpoint}}
\put(317,623){\usebox{\plotpoint}}
\multiput(317,623)(20.753,-0.322){7}{\usebox{\plotpoint}}
\multiput(446,621)(20.716,-1.281){4}{\usebox{\plotpoint}}
\multiput(543,615)(20.623,-2.339){5}{\usebox{\plotpoint}}
\multiput(640,604)(20.098,-5.182){6}{\usebox{\plotpoint}}
\multiput(768,571)(19.118,-8.081){6}{\usebox{\plotpoint}}
\multiput(865,530)(18.678,-9.050){5}{\usebox{\plotpoint}}
\multiput(962,483)(19.250,-7.760){6}{\usebox{\plotpoint}}
\multiput(1091,431)(20.444,-3.583){5}{\usebox{\plotpoint}}
\multiput(1188,414)(20.743,0.728){3}{\usebox{\plotpoint}}
\put(1245,416){\usebox{\plotpoint}}
\sbox{\plotpoint}{\rule[-0.200pt]{0.400pt}{0.400pt}}%
\put(543,289){\makebox(0,0)[r]{$\Lambda\!=\!300$\,MeV}}
\put(565.0,289.0){\rule[-0.200pt]{15.899pt}{0.400pt}}
\put(317,623){\usebox{\plotpoint}}
\multiput(317.00,621.95)(28.593,-0.447){3}{\rule{17.300pt}{0.108pt}}
\multiput(317.00,622.17)(93.093,-3.000){2}{\rule{8.650pt}{0.400pt}}
\multiput(446.00,618.93)(10.729,-0.477){7}{\rule{7.860pt}{0.115pt}}
\multiput(446.00,619.17)(80.686,-5.000){2}{\rule{3.930pt}{0.400pt}}
\multiput(543.00,613.92)(4.158,-0.492){21}{\rule{3.333pt}{0.119pt}}
\multiput(543.00,614.17)(90.081,-12.000){2}{\rule{1.667pt}{0.400pt}}
\multiput(640.00,601.92)(1.841,-0.498){67}{\rule{1.563pt}{0.120pt}}
\multiput(640.00,602.17)(124.756,-35.000){2}{\rule{0.781pt}{0.400pt}}
\multiput(768.00,566.92)(1.058,-0.498){89}{\rule{0.943pt}{0.120pt}}
\multiput(768.00,567.17)(95.042,-46.000){2}{\rule{0.472pt}{0.400pt}}
\multiput(865.00,520.92)(0.838,-0.499){113}{\rule{0.769pt}{0.120pt}}
\multiput(865.00,521.17)(95.404,-58.000){2}{\rule{0.384pt}{0.400pt}}
\multiput(962.00,462.92)(0.850,-0.499){149}{\rule{0.779pt}{0.120pt}}
\multiput(962.00,463.17)(127.383,-76.000){2}{\rule{0.389pt}{0.400pt}}
\put(1091,388.17){\rule{19.500pt}{0.400pt}}
\multiput(1091.00,387.17)(56.527,2.000){2}{\rule{9.750pt}{0.400pt}}
\multiput(1188.58,390.00)(0.499,0.605){111}{\rule{0.120pt}{0.584pt}}
\multiput(1187.17,390.00)(57.000,67.787){2}{\rule{0.400pt}{0.292pt}}
\sbox{\plotpoint}{\rule[-0.400pt]{0.800pt}{0.800pt}}%
\put(543,244){\makebox(0,0)[r]{$\Lambda\!=\!400$\,MeV}}
\put(565.0,244.0){\rule[-0.400pt]{15.899pt}{0.800pt}}
\put(317,623){\usebox{\plotpoint}}
\put(317,619.84){\rule{31.076pt}{0.800pt}}
\multiput(317.00,621.34)(64.500,-3.000){2}{\rule{15.538pt}{0.800pt}}
\multiput(446.00,618.06)(15.872,-0.560){3}{\rule{15.720pt}{0.135pt}}
\multiput(446.00,618.34)(64.372,-5.000){2}{\rule{7.860pt}{0.800pt}}
\multiput(543.00,613.08)(4.333,-0.511){17}{\rule{6.667pt}{0.123pt}}
\multiput(543.00,613.34)(83.163,-12.000){2}{\rule{3.333pt}{0.800pt}}
\multiput(640.00,601.09)(1.855,-0.503){63}{\rule{3.126pt}{0.121pt}}
\multiput(640.00,601.34)(121.512,-35.000){2}{\rule{1.563pt}{0.800pt}}
\multiput(768.00,566.09)(1.062,-0.502){85}{\rule{1.887pt}{0.121pt}}
\multiput(768.00,566.34)(93.084,-46.000){2}{\rule{0.943pt}{0.800pt}}
\multiput(865.00,520.09)(0.784,-0.502){117}{\rule{1.452pt}{0.121pt}}
\multiput(865.00,520.34)(93.987,-62.000){2}{\rule{0.726pt}{0.800pt}}
\multiput(962.00,458.09)(0.645,-0.501){193}{\rule{1.232pt}{0.121pt}}
\multiput(962.00,458.34)(126.443,-100.000){2}{\rule{0.616pt}{0.800pt}}
\multiput(1091.00,358.09)(0.748,-0.501){123}{\rule{1.394pt}{0.121pt}}
\multiput(1091.00,358.34)(94.107,-65.000){2}{\rule{0.697pt}{0.800pt}}
\multiput(1188.00,296.41)(1.157,0.504){43}{\rule{2.024pt}{0.121pt}}
\multiput(1188.00,293.34)(52.799,25.000){2}{\rule{1.012pt}{0.800pt}}
\end{picture}

\vspace{2em}

% GNUPLOT: LaTeX picture
\setlength{\unitlength}{0.240900pt}
\ifx\plotpoint\undefined\newsavebox{\plotpoint}\fi
\sbox{\plotpoint}{\rule[-0.200pt]{0.400pt}{0.400pt}}%
\begin{picture}(1349,749)(0,0)
\font\gnuplot=cmr10 at 10pt
\gnuplot
\sbox{\plotpoint}{\rule[-0.200pt]{0.400pt}{0.400pt}}%
\put(220.0,201.0){\rule[-0.200pt]{4.818pt}{0.400pt}}
\put(198,201){\makebox(0,0)[r]{$0$}}
\put(1265.0,201.0){\rule[-0.200pt]{4.818pt}{0.400pt}}
\put(220.0,288.0){\rule[-0.200pt]{4.818pt}{0.400pt}}
\put(198,288){\makebox(0,0)[r]{$0.1$}}
\put(1265.0,288.0){\rule[-0.200pt]{4.818pt}{0.400pt}}
\put(220.0,376.0){\rule[-0.200pt]{4.818pt}{0.400pt}}
\put(198,376){\makebox(0,0)[r]{$0.2$}}
\put(1265.0,376.0){\rule[-0.200pt]{4.818pt}{0.400pt}}
\put(220.0,463.0){\rule[-0.200pt]{4.818pt}{0.400pt}}
\put(198,463){\makebox(0,0)[r]{$0.3$}}
\put(1265.0,463.0){\rule[-0.200pt]{4.818pt}{0.400pt}}
\put(220.0,551.0){\rule[-0.200pt]{4.818pt}{0.400pt}}
\put(198,551){\makebox(0,0)[r]{$0.4$}}
\put(1265.0,551.0){\rule[-0.200pt]{4.818pt}{0.400pt}}
\put(220.0,638.0){\rule[-0.200pt]{4.818pt}{0.400pt}}
\put(198,638){\makebox(0,0)[r]{$0.5$}}
\put(1265.0,638.0){\rule[-0.200pt]{4.818pt}{0.400pt}}
\put(220.0,113.0){\rule[-0.200pt]{0.400pt}{4.818pt}}
\put(220,68){\makebox(0,0){$0.1$}}
\put(220.0,706.0){\rule[-0.200pt]{0.400pt}{4.818pt}}
\put(317.0,113.0){\rule[-0.200pt]{0.400pt}{2.409pt}}
\put(317.0,716.0){\rule[-0.200pt]{0.400pt}{2.409pt}}
\put(374.0,113.0){\rule[-0.200pt]{0.400pt}{2.409pt}}
\put(374.0,716.0){\rule[-0.200pt]{0.400pt}{2.409pt}}
\put(414.0,113.0){\rule[-0.200pt]{0.400pt}{2.409pt}}
\put(414.0,716.0){\rule[-0.200pt]{0.400pt}{2.409pt}}
\put(446.0,113.0){\rule[-0.200pt]{0.400pt}{2.409pt}}
\put(446.0,716.0){\rule[-0.200pt]{0.400pt}{2.409pt}}
\put(471.0,113.0){\rule[-0.200pt]{0.400pt}{2.409pt}}
\put(471.0,716.0){\rule[-0.200pt]{0.400pt}{2.409pt}}
\put(493.0,113.0){\rule[-0.200pt]{0.400pt}{2.409pt}}
\put(493.0,716.0){\rule[-0.200pt]{0.400pt}{2.409pt}}
\put(511.0,113.0){\rule[-0.200pt]{0.400pt}{2.409pt}}
\put(511.0,716.0){\rule[-0.200pt]{0.400pt}{2.409pt}}
\put(528.0,113.0){\rule[-0.200pt]{0.400pt}{2.409pt}}
\put(528.0,716.0){\rule[-0.200pt]{0.400pt}{2.409pt}}
\put(543.0,113.0){\rule[-0.200pt]{0.400pt}{4.818pt}}
\put(543,68){\makebox(0,0){$1$}}
\put(543.0,706.0){\rule[-0.200pt]{0.400pt}{4.818pt}}
\put(640.0,113.0){\rule[-0.200pt]{0.400pt}{2.409pt}}
\put(640.0,716.0){\rule[-0.200pt]{0.400pt}{2.409pt}}
\put(697.0,113.0){\rule[-0.200pt]{0.400pt}{2.409pt}}
\put(697.0,716.0){\rule[-0.200pt]{0.400pt}{2.409pt}}
\put(737.0,113.0){\rule[-0.200pt]{0.400pt}{2.409pt}}
\put(737.0,716.0){\rule[-0.200pt]{0.400pt}{2.409pt}}
\put(768.0,113.0){\rule[-0.200pt]{0.400pt}{2.409pt}}
\put(768.0,716.0){\rule[-0.200pt]{0.400pt}{2.409pt}}
\put(794.0,113.0){\rule[-0.200pt]{0.400pt}{2.409pt}}
\put(794.0,716.0){\rule[-0.200pt]{0.400pt}{2.409pt}}
\put(815.0,113.0){\rule[-0.200pt]{0.400pt}{2.409pt}}
\put(815.0,716.0){\rule[-0.200pt]{0.400pt}{2.409pt}}
\put(834.0,113.0){\rule[-0.200pt]{0.400pt}{2.409pt}}
\put(834.0,716.0){\rule[-0.200pt]{0.400pt}{2.409pt}}
\put(850.0,113.0){\rule[-0.200pt]{0.400pt}{2.409pt}}
\put(850.0,716.0){\rule[-0.200pt]{0.400pt}{2.409pt}}
\put(865.0,113.0){\rule[-0.200pt]{0.400pt}{4.818pt}}
\put(865,68){\makebox(0,0){$10$}}
\put(865.0,706.0){\rule[-0.200pt]{0.400pt}{4.818pt}}
\put(962.0,113.0){\rule[-0.200pt]{0.400pt}{2.409pt}}
\put(962.0,716.0){\rule[-0.200pt]{0.400pt}{2.409pt}}
\put(1019.0,113.0){\rule[-0.200pt]{0.400pt}{2.409pt}}
\put(1019.0,716.0){\rule[-0.200pt]{0.400pt}{2.409pt}}
\put(1059.0,113.0){\rule[-0.200pt]{0.400pt}{2.409pt}}
\put(1059.0,716.0){\rule[-0.200pt]{0.400pt}{2.409pt}}
\put(1091.0,113.0){\rule[-0.200pt]{0.400pt}{2.409pt}}
\put(1091.0,716.0){\rule[-0.200pt]{0.400pt}{2.409pt}}
\put(1116.0,113.0){\rule[-0.200pt]{0.400pt}{2.409pt}}
\put(1116.0,716.0){\rule[-0.200pt]{0.400pt}{2.409pt}}
\put(1138.0,113.0){\rule[-0.200pt]{0.400pt}{2.409pt}}
\put(1138.0,716.0){\rule[-0.200pt]{0.400pt}{2.409pt}}
\put(1157.0,113.0){\rule[-0.200pt]{0.400pt}{2.409pt}}
\put(1157.0,716.0){\rule[-0.200pt]{0.400pt}{2.409pt}}
\put(1173.0,113.0){\rule[-0.200pt]{0.400pt}{2.409pt}}
\put(1173.0,716.0){\rule[-0.200pt]{0.400pt}{2.409pt}}
\put(1188.0,113.0){\rule[-0.200pt]{0.400pt}{4.818pt}}
\put(1188,68){\makebox(0,0){$100$}}
\put(1188.0,706.0){\rule[-0.200pt]{0.400pt}{4.818pt}}
\put(1285.0,113.0){\rule[-0.200pt]{0.400pt}{2.409pt}}
\put(1285.0,716.0){\rule[-0.200pt]{0.400pt}{2.409pt}}
\put(220.0,113.0){\rule[-0.200pt]{256.558pt}{0.400pt}}
\put(1285.0,113.0){\rule[-0.200pt]{0.400pt}{147.672pt}}
\put(220.0,726.0){\rule[-0.200pt]{256.558pt}{0.400pt}}
\put(45,419){\makebox(0,0){$\delta_2(E)$}}
\put(752,23){\makebox(0,0){$E$ (MeV)}}
\put(1116,288){\makebox(0,0)[l]{${}^3D_2$}}
\put(220.0,113.0){\rule[-0.200pt]{0.400pt}{147.672pt}}
\sbox{\plotpoint}{\rule[-0.500pt]{1.000pt}{1.000pt}}%
\put(511,621){\makebox(0,0)[r]{data}}
\put(555,621){\circle{24}}
\put(317,201){\circle{24}}
\put(446,201){\circle{24}}
\put(543,201){\circle{24}}
\put(640,203){\circle{24}}
\put(768,214){\circle{24}}
\put(865,241){\circle{24}}
\put(962,306){\circle{24}}
\put(1091,465){\circle{24}}
\put(1188,575){\circle{24}}
\put(1245,590){\circle{24}}
\sbox{\plotpoint}{\rule[-0.200pt]{0.400pt}{0.400pt}}%
\put(511,576){\makebox(0,0)[r]{$\Lambda\!=\!300$\,MeV}}
\put(533.0,576.0){\rule[-0.200pt]{15.899pt}{0.400pt}}
\put(317,201){\usebox{\plotpoint}}
\put(543,201.17){\rule{19.500pt}{0.400pt}}
\multiput(543.00,200.17)(56.527,2.000){2}{\rule{9.750pt}{0.400pt}}
\multiput(640.00,203.58)(6.646,0.491){17}{\rule{5.220pt}{0.118pt}}
\multiput(640.00,202.17)(117.166,10.000){2}{\rule{2.610pt}{0.400pt}}
\multiput(768.00,213.58)(1.883,0.497){49}{\rule{1.592pt}{0.120pt}}
\multiput(768.00,212.17)(93.695,26.000){2}{\rule{0.796pt}{0.400pt}}
\multiput(865.00,239.58)(0.852,0.499){111}{\rule{0.781pt}{0.120pt}}
\multiput(865.00,238.17)(95.380,57.000){2}{\rule{0.390pt}{0.400pt}}
\multiput(962.00,296.58)(0.512,0.499){249}{\rule{0.510pt}{0.120pt}}
\multiput(962.00,295.17)(127.942,126.000){2}{\rule{0.255pt}{0.400pt}}
\multiput(1091.00,422.58)(0.638,0.499){149}{\rule{0.611pt}{0.120pt}}
\multiput(1091.00,421.17)(95.733,76.000){2}{\rule{0.305pt}{0.400pt}}
\multiput(1188.00,498.59)(4.306,0.485){11}{\rule{3.357pt}{0.117pt}}
\multiput(1188.00,497.17)(50.032,7.000){2}{\rule{1.679pt}{0.400pt}}
\put(317.0,201.0){\rule[-0.200pt]{54.443pt}{0.400pt}}
\sbox{\plotpoint}{\rule[-0.400pt]{0.800pt}{0.800pt}}%
\put(511,531){\makebox(0,0)[r]{$\Lambda\!=\!400$\,MeV}}
\put(533.0,531.0){\rule[-0.400pt]{15.899pt}{0.800pt}}
\put(317,201){\usebox{\plotpoint}}
\put(543,200.34){\rule{23.367pt}{0.800pt}}
\multiput(543.00,199.34)(48.500,2.000){2}{\rule{11.684pt}{0.800pt}}
\multiput(640.00,204.40)(7.055,0.514){13}{\rule{10.440pt}{0.124pt}}
\multiput(640.00,201.34)(106.331,10.000){2}{\rule{5.220pt}{0.800pt}}
\multiput(768.00,214.41)(1.831,0.504){47}{\rule{3.074pt}{0.121pt}}
\multiput(768.00,211.34)(90.620,27.000){2}{\rule{1.537pt}{0.800pt}}
\multiput(865.00,241.41)(0.784,0.502){117}{\rule{1.452pt}{0.121pt}}
\multiput(865.00,238.34)(93.987,62.000){2}{\rule{0.726pt}{0.800pt}}
\multiput(963.41,302.00)(0.501,0.609){251}{\rule{0.121pt}{1.174pt}}
\multiput(960.34,302.00)(129.000,154.564){2}{\rule{0.800pt}{0.587pt}}
\multiput(1092.41,459.00)(0.501,0.722){187}{\rule{0.121pt}{1.355pt}}
\multiput(1089.34,459.00)(97.000,137.188){2}{\rule{0.800pt}{0.677pt}}
\multiput(1188.00,600.41)(0.558,0.502){95}{\rule{1.094pt}{0.121pt}}
\multiput(1188.00,597.34)(54.729,51.000){2}{\rule{0.547pt}{0.800pt}}
\put(317.0,201.0){\rule[-0.400pt]{54.443pt}{0.800pt}}
\end{picture}
\end{center}
\caption{Phase shifts (in radians) for different channels from the
effective theory. $P$-wave results are for the uncorrected
($V_{1\pi}$) theory, and for the full effective theory with different
values of $\Lambda$. $D$-wave results are for the uncorrected theory
with different values of $\Lambda$; there are no $\Lambda^{-2}$ or
$\Lambda^{-4}$ corrections for this $D$-wave channel.}
\label{many_phase-fig}
\end{figure}

In Figure~\ref{many_phase-fig} I show results for the ${}^1P_1$ and
${}^3D_2$ phase shifts.  The effective potentials for these two cases
are, through order~$\Lambda^{-4}$,
\be
V(^1P_1) = 3\,\alpha_\pi\,v_\Lambda(r) +
           c(^1P_1)\,\frac{\nabla^2\delta_{1/\Lambda}^3(\rv)}{\Lambda^4},
\ee
and
\be
V(^3D_2) = -\alpha_\pi\left[\,v_\Lambda(r) + 2\,v_T(r)\right].
\ee
The data confirm that the long-range potential plays a
much more important role for $P$-waves, which have no
$\Lambda^{-2}$ contact term, than for $S$-waves. 
And the $D$-wave analysis has no contact terms at all until
order~$\Lambda^{-6}$. In both these cases the phase shifts are fairly
linear in the couplings, unlike the low-energy $S$-wave phase shifts.

\begin{figure}
\begin{center}
% GNUPLOT: LaTeX picture
\setlength{\unitlength}{0.240900pt}
\ifx\plotpoint\undefined\newsavebox{\plotpoint}\fi
\sbox{\plotpoint}{\rule[-0.200pt]{0.400pt}{0.400pt}}%
\begin{picture}(1349,749)(0,0)
\font\gnuplot=cmr10 at 10pt
\gnuplot
\sbox{\plotpoint}{\rule[-0.200pt]{0.400pt}{0.400pt}}%
\put(220.0,171.0){\rule[-0.200pt]{4.818pt}{0.400pt}}
\put(198,171){\makebox(0,0)[r]{$10^{-6}$}}
\put(1265.0,171.0){\rule[-0.200pt]{4.818pt}{0.400pt}}
\put(220.0,337.0){\rule[-0.200pt]{4.818pt}{0.400pt}}
\put(198,337){\makebox(0,0)[r]{$10^{-4}$}}
\put(1265.0,337.0){\rule[-0.200pt]{4.818pt}{0.400pt}}
\put(220.0,502.0){\rule[-0.200pt]{4.818pt}{0.400pt}}
\put(198,502){\makebox(0,0)[r]{$10^{-2}$}}
\put(1265.0,502.0){\rule[-0.200pt]{4.818pt}{0.400pt}}
\put(220.0,668.0){\rule[-0.200pt]{4.818pt}{0.400pt}}
\put(198,668){\makebox(0,0)[r]{$1$}}
\put(1265.0,668.0){\rule[-0.200pt]{4.818pt}{0.400pt}}
\put(220.0,113.0){\rule[-0.200pt]{0.400pt}{4.818pt}}
\put(220,68){\makebox(0,0){0.1}}
\put(220.0,706.0){\rule[-0.200pt]{0.400pt}{4.818pt}}
\put(317.0,113.0){\rule[-0.200pt]{0.400pt}{2.409pt}}
\put(317.0,716.0){\rule[-0.200pt]{0.400pt}{2.409pt}}
\put(374.0,113.0){\rule[-0.200pt]{0.400pt}{2.409pt}}
\put(374.0,716.0){\rule[-0.200pt]{0.400pt}{2.409pt}}
\put(414.0,113.0){\rule[-0.200pt]{0.400pt}{2.409pt}}
\put(414.0,716.0){\rule[-0.200pt]{0.400pt}{2.409pt}}
\put(446.0,113.0){\rule[-0.200pt]{0.400pt}{2.409pt}}
\put(446.0,716.0){\rule[-0.200pt]{0.400pt}{2.409pt}}
\put(471.0,113.0){\rule[-0.200pt]{0.400pt}{2.409pt}}
\put(471.0,716.0){\rule[-0.200pt]{0.400pt}{2.409pt}}
\put(493.0,113.0){\rule[-0.200pt]{0.400pt}{2.409pt}}
\put(493.0,716.0){\rule[-0.200pt]{0.400pt}{2.409pt}}
\put(511.0,113.0){\rule[-0.200pt]{0.400pt}{2.409pt}}
\put(511.0,716.0){\rule[-0.200pt]{0.400pt}{2.409pt}}
\put(528.0,113.0){\rule[-0.200pt]{0.400pt}{2.409pt}}
\put(528.0,716.0){\rule[-0.200pt]{0.400pt}{2.409pt}}
\put(543.0,113.0){\rule[-0.200pt]{0.400pt}{4.818pt}}
\put(543,68){\makebox(0,0){1}}
\put(543.0,706.0){\rule[-0.200pt]{0.400pt}{4.818pt}}
\put(640.0,113.0){\rule[-0.200pt]{0.400pt}{2.409pt}}
\put(640.0,716.0){\rule[-0.200pt]{0.400pt}{2.409pt}}
\put(697.0,113.0){\rule[-0.200pt]{0.400pt}{2.409pt}}
\put(697.0,716.0){\rule[-0.200pt]{0.400pt}{2.409pt}}
\put(737.0,113.0){\rule[-0.200pt]{0.400pt}{2.409pt}}
\put(737.0,716.0){\rule[-0.200pt]{0.400pt}{2.409pt}}
\put(768.0,113.0){\rule[-0.200pt]{0.400pt}{2.409pt}}
\put(768.0,716.0){\rule[-0.200pt]{0.400pt}{2.409pt}}
\put(794.0,113.0){\rule[-0.200pt]{0.400pt}{2.409pt}}
\put(794.0,716.0){\rule[-0.200pt]{0.400pt}{2.409pt}}
\put(815.0,113.0){\rule[-0.200pt]{0.400pt}{2.409pt}}
\put(815.0,716.0){\rule[-0.200pt]{0.400pt}{2.409pt}}
\put(834.0,113.0){\rule[-0.200pt]{0.400pt}{2.409pt}}
\put(834.0,716.0){\rule[-0.200pt]{0.400pt}{2.409pt}}
\put(850.0,113.0){\rule[-0.200pt]{0.400pt}{2.409pt}}
\put(850.0,716.0){\rule[-0.200pt]{0.400pt}{2.409pt}}
\put(865.0,113.0){\rule[-0.200pt]{0.400pt}{4.818pt}}
\put(865,68){\makebox(0,0){10}}
\put(865.0,706.0){\rule[-0.200pt]{0.400pt}{4.818pt}}
\put(962.0,113.0){\rule[-0.200pt]{0.400pt}{2.409pt}}
\put(962.0,716.0){\rule[-0.200pt]{0.400pt}{2.409pt}}
\put(1019.0,113.0){\rule[-0.200pt]{0.400pt}{2.409pt}}
\put(1019.0,716.0){\rule[-0.200pt]{0.400pt}{2.409pt}}
\put(1059.0,113.0){\rule[-0.200pt]{0.400pt}{2.409pt}}
\put(1059.0,716.0){\rule[-0.200pt]{0.400pt}{2.409pt}}
\put(1091.0,113.0){\rule[-0.200pt]{0.400pt}{2.409pt}}
\put(1091.0,716.0){\rule[-0.200pt]{0.400pt}{2.409pt}}
\put(1116.0,113.0){\rule[-0.200pt]{0.400pt}{2.409pt}}
\put(1116.0,716.0){\rule[-0.200pt]{0.400pt}{2.409pt}}
\put(1138.0,113.0){\rule[-0.200pt]{0.400pt}{2.409pt}}
\put(1138.0,716.0){\rule[-0.200pt]{0.400pt}{2.409pt}}
\put(1157.0,113.0){\rule[-0.200pt]{0.400pt}{2.409pt}}
\put(1157.0,716.0){\rule[-0.200pt]{0.400pt}{2.409pt}}
\put(1173.0,113.0){\rule[-0.200pt]{0.400pt}{2.409pt}}
\put(1173.0,716.0){\rule[-0.200pt]{0.400pt}{2.409pt}}
\put(1188.0,113.0){\rule[-0.200pt]{0.400pt}{4.818pt}}
\put(1188,68){\makebox(0,0){100}}
\put(1188.0,706.0){\rule[-0.200pt]{0.400pt}{4.818pt}}
\put(1285.0,113.0){\rule[-0.200pt]{0.400pt}{2.409pt}}
\put(1285.0,716.0){\rule[-0.200pt]{0.400pt}{2.409pt}}
\put(220.0,113.0){\rule[-0.200pt]{256.558pt}{0.400pt}}
\put(1285.0,113.0){\rule[-0.200pt]{0.400pt}{147.672pt}}
\put(220.0,726.0){\rule[-0.200pt]{256.558pt}{0.400pt}}
\put(45,419){\makebox(0,0){$|\Delta \delta_l(E)|$}}
\put(752,23){\makebox(0,0){$E$ (MeV)}}
\put(220.0,113.0){\rule[-0.200pt]{0.400pt}{147.672pt}}
\put(374,643){\makebox(0,0)[r]{${}^1S_0$}}
\put(396.0,643.0){\rule[-0.200pt]{15.899pt}{0.400pt}}
\put(317,199){\usebox{\plotpoint}}
\multiput(317.00,199.58)(0.850,0.499){149}{\rule{0.779pt}{0.120pt}}
\multiput(317.00,198.17)(127.383,76.000){2}{\rule{0.389pt}{0.400pt}}
\multiput(446.00,275.58)(2.340,0.496){39}{\rule{1.948pt}{0.119pt}}
\multiput(446.00,274.17)(92.958,21.000){2}{\rule{0.974pt}{0.400pt}}
\multiput(543.00,296.59)(6.371,0.488){13}{\rule{4.950pt}{0.117pt}}
\multiput(543.00,295.17)(86.726,8.000){2}{\rule{2.475pt}{0.400pt}}
\multiput(640.00,304.58)(0.529,0.499){239}{\rule{0.523pt}{0.120pt}}
\multiput(640.00,303.17)(126.914,121.000){2}{\rule{0.262pt}{0.400pt}}
\multiput(768.00,425.58)(0.884,0.499){107}{\rule{0.805pt}{0.120pt}}
\multiput(768.00,424.17)(95.328,55.000){2}{\rule{0.403pt}{0.400pt}}
\multiput(865.00,480.58)(1.081,0.498){87}{\rule{0.962pt}{0.120pt}}
\multiput(865.00,479.17)(95.003,45.000){2}{\rule{0.481pt}{0.400pt}}
\multiput(962.00,525.58)(1.378,0.498){91}{\rule{1.198pt}{0.120pt}}
\multiput(962.00,524.17)(126.514,47.000){2}{\rule{0.599pt}{0.400pt}}
\multiput(1091.00,572.58)(1.883,0.497){49}{\rule{1.592pt}{0.120pt}}
\multiput(1091.00,571.17)(93.695,26.000){2}{\rule{0.796pt}{0.400pt}}
\multiput(1188.00,598.59)(3.730,0.488){13}{\rule{2.950pt}{0.117pt}}
\multiput(1188.00,597.17)(50.877,8.000){2}{\rule{1.475pt}{0.400pt}}
\sbox{\plotpoint}{\rule[-0.400pt]{0.800pt}{0.800pt}}%
\put(374,598){\makebox(0,0)[r]{${}^1P_1$}}
\put(396.0,598.0){\rule[-0.400pt]{15.899pt}{0.800pt}}
\put(317,188){\usebox{\plotpoint}}
\multiput(317.00,189.41)(0.779,0.501){159}{\rule{1.443pt}{0.121pt}}
\multiput(317.00,186.34)(126.004,83.000){2}{\rule{0.722pt}{0.800pt}}
\multiput(446.00,272.41)(0.825,0.502){111}{\rule{1.515pt}{0.121pt}}
\multiput(446.00,269.34)(93.855,59.000){2}{\rule{0.758pt}{0.800pt}}
\multiput(543.00,331.41)(0.919,0.502){99}{\rule{1.664pt}{0.121pt}}
\multiput(543.00,328.34)(93.546,53.000){2}{\rule{0.832pt}{0.800pt}}
\multiput(640.00,384.41)(1.037,0.502){117}{\rule{1.852pt}{0.121pt}}
\multiput(640.00,381.34)(124.157,62.000){2}{\rule{0.926pt}{0.800pt}}
\multiput(768.00,446.41)(1.290,0.503){69}{\rule{2.242pt}{0.121pt}}
\multiput(768.00,443.34)(92.346,38.000){2}{\rule{1.121pt}{0.800pt}}
\multiput(865.00,484.41)(1.983,0.504){43}{\rule{3.304pt}{0.121pt}}
\multiput(865.00,481.34)(90.142,25.000){2}{\rule{1.652pt}{0.800pt}}
\multiput(962.00,506.09)(1.926,-0.503){61}{\rule{3.235pt}{0.121pt}}
\multiput(962.00,506.34)(122.285,-34.000){2}{\rule{1.618pt}{0.800pt}}
\multiput(1091.00,475.41)(0.504,0.501){185}{\rule{1.008pt}{0.121pt}}
\multiput(1091.00,472.34)(94.907,96.000){2}{\rule{0.504pt}{0.800pt}}
\multiput(1188.00,571.41)(0.774,0.503){67}{\rule{1.432pt}{0.121pt}}
\multiput(1188.00,568.34)(54.027,37.000){2}{\rule{0.716pt}{0.800pt}}
\sbox{\plotpoint}{\rule[-0.600pt]{1.200pt}{1.200pt}}%
\put(374,553){\makebox(0,0)[r]{${}^3D_2$}}
\put(396.0,553.0){\rule[-0.600pt]{15.899pt}{1.200pt}}
\put(317,160){\usebox{\plotpoint}}
\multiput(317.00,162.24)(0.814,0.500){148}{\rule{2.259pt}{0.120pt}}
\multiput(317.00,157.51)(124.310,79.000){2}{\rule{1.130pt}{1.200pt}}
\multiput(446.00,241.24)(0.847,0.500){104}{\rule{2.342pt}{0.120pt}}
\multiput(446.00,236.51)(92.139,57.000){2}{\rule{1.171pt}{1.200pt}}
\multiput(543.00,298.24)(0.930,0.500){94}{\rule{2.538pt}{0.120pt}}
\multiput(543.00,293.51)(91.731,52.000){2}{\rule{1.269pt}{1.200pt}}
\multiput(640.00,350.24)(1.065,0.500){110}{\rule{2.860pt}{0.120pt}}
\multiput(640.00,345.51)(122.064,60.000){2}{\rule{1.430pt}{1.200pt}}
\multiput(768.00,410.24)(1.244,0.500){68}{\rule{3.285pt}{0.121pt}}
\multiput(768.00,405.51)(90.183,39.000){2}{\rule{1.642pt}{1.200pt}}
\multiput(865.00,449.24)(1.522,0.500){54}{\rule{3.938pt}{0.121pt}}
\multiput(865.00,444.51)(88.828,32.000){2}{\rule{1.969pt}{1.200pt}}
\multiput(962.00,481.24)(20.656,0.509){2}{\rule{26.100pt}{0.123pt}}
\multiput(962.00,476.51)(74.828,6.000){2}{\rule{13.050pt}{1.200pt}}
\multiput(1091.00,487.24)(0.912,0.500){96}{\rule{2.496pt}{0.120pt}}
\multiput(1091.00,482.51)(91.819,53.000){2}{\rule{1.248pt}{1.200pt}}
\multiput(1188.00,540.24)(0.832,0.500){58}{\rule{2.312pt}{0.121pt}}
\multiput(1188.00,535.51)(52.202,34.000){2}{\rule{1.156pt}{1.200pt}}
\end{picture}
\end{center}
\caption{Errors in the phase shifts (in radians) for different
channels from the effective theory corrected through order~$\Lambda^{-4}$.}
\label{all_dphase-fig}
\end{figure}
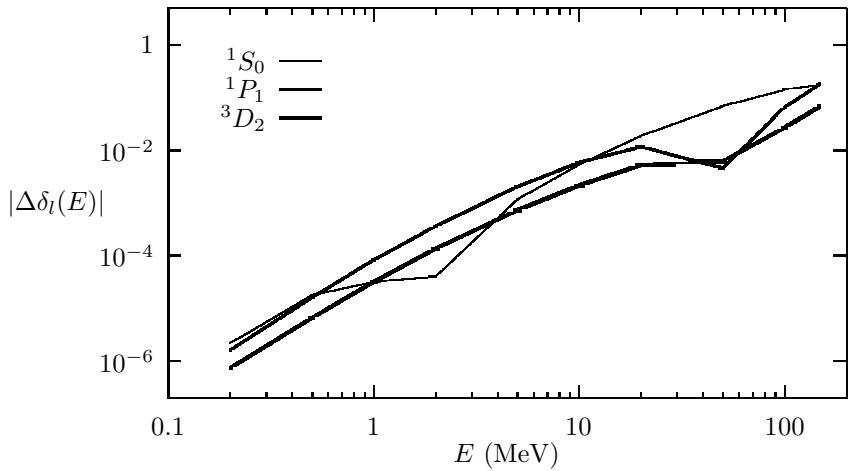

The relative errors are much smaller for our $S$-wave phase shifts
than for the others. The absolute magnitude of the errors
in each channel, however, 
is roughly the same when the effective theories for each
channel are corrected to the same order in~$1/\Lambda$. Also the
energy dependence of these errors is the same. This is illustrated in
Figure~\ref{all_dphase-fig}. It is exactly what is
expected based upon the effective theory.

\subsection{Multinucleon Systems}
Our nucleon-nucleon analysis is readily extended to systems of three or more
nucleons. The potentials tuned using nucleon-nucleon scattering can be
carried over without change. In addition new contact terms involving three
or more nucleons can be added, but these tend to be very high order
in~$\Lambda^{-1}$, and so can often be neglected. For example, the leading
three-body contact potential is order~$\Lambda^{-5}$. Of course there
are in addition
various long-range three-body potentials predicted by the chiral theory.
This subject is discussed in detail in the literature.\cite{vankolck94}

\subsection{The Problem with Dimensional Regularization}

I have emphasized in this paper that the exact details of the cutoff are
unimportant; just about any choice should give results equivalent to
those presented here. Dimensional regularization, however, is an
exception.\footnote{Much
of the discussion in this section is based on a series of
useful conversations I had with Martin Savage and David Kaplan at
the INT in Seattle.} 
Dimensional regularization is a popular scheme developed for
perturbative analyses of gauge theories.
A recent paper has uncovered a pathology when dimensional regularization is
applied to the nucleon-nucleon problem.\cite{kaplan96} The authors find
that the dimensionally regulated version of the ${}^1S_0$~theory we describe
above fails
completely by energies of~10--20\,MeV, well below the energy at
which the cutoff theory fails.
They conclude that the rules for
applying effective field theories must be modified. In this I believe they
are incorrect, especially given the obvious 
success of the cutoff version of the
theory. They have uncovered a problem, but the problem is with
dimensional regularization\,---\,or, more precisely, with minimally-subtracted
dimensional regularization\,---\,and not with effective field theory. 

The authors trace the problem in the dimensionally regulated theory to
the unusually long scattering length~$a_s$ in the ${}^1S_0$~channel.
As we have seen,  an effective field theory generally
has residual errors that are some power of 
the typical momentum~$q$ divided by the ultraviolet
cutoff~$\Lambda$. When the ${}^1S_0$~theory is dimensionally regulated,
the effective~$\Lambda$ in the $q/\Lambda$~errors is
tied to the scattering length, and vanishes as
$a_s\!\to\!\infty$. Therefore this version of the 
theory is accurate only for very small momenta when $a_s$~is large.

Large scattering lengths arise when
there is a bound state just below threshold ($a_s$
positive) or a resonance just above threshold ($a_s$ negative).
We can see how $\Lambda$ is related to the scattering length by analyzing
a simplified problem: Consider a system with a barely bound $S$~state
caused by a short-range interaction that we model with a single
contact term:
\be\label{modelV}
V(r) = \frac{4\pi}{\Lambda^2}\,\delta^3(\rv)
\ee
We analyze this problem three ways. First we use first-order perturbation
theory. In this approximation, the scattering amplitude is simply
\be
T = \frac{4\pi}{\Lambda^2},
\ee
and therefore the scattering length is given by
\be
\frac{1}{a_s} \equiv \lim_{p\!\to\!0} \frac{4\pi}{m}\,{\rm Re}\frac{1}{T}
= \frac{\Lambda^2}{m}.
\ee
Thus $\Lambda\!=\!\sqrt{m/a_s}$, and 
large scattering lengths imply small~$\Lambda$'s. Generally
bound states do not appear in first-order perturbation theory, since a
bound state is characterized by particles interacting again and again,
forever  (not just once). Consequently, the presence of a large positive
scattering length  in this analysis does not indicate a bound state
near threshold. This is confirmed by the
fact that the potential here must be repulsive, not attractive, if
$a_s$~is to be positive.

Next we analyze the simplified problem nonperturbatively, using
minimally-subtracted dimensional regularization. The contact potential,
being separable, is easily iterated to all orders, and gives the amplitude
\be \label{Tdimreg}
T = \frac{T_1}{1-T_2/T_1}
\ee
where $T_1=4\pi/\Lambda^2$ is the amplitude in first order, and 
\be
T_2 = -(T_1)^2\,\frac{{\rm i} m p}{4\pi}
\ee
is the dimensionally-regulated second-order amplitude. Minimal subtraction
implies that the second-order amplitude is purely imaginary. Consequently
the only effect of iterating the potential to all orders is to unitarize the
amplitude. The scattering length obtained from~\eq{Tdimreg} is identical
to that obtained from first-order perturbation theory, and again
$\Lambda\!=\!\sqrt{m/a_s}$ vanishes as $a_s$~becomes large. 
When the analysis is
extended to included higher-order contact terms, the small~$\Lambda$
restricts the model's utility to very low energies.\cite{kaplan96}

Finally we redo the nonperturbative calculation using a cutoff on the
loop-integral in $T_2$, which we implement in such a way that the
interaction is still separable. While the potential in the previous two
analyses was necessarily repulsive, here we flip the sign of~$V$,
making it attractive, in order to get a positive scattering length.
Then
\be
T_1 = - \frac{4\pi}{\Lambda^2},
\ee
and 
\be
T_2 = -(T_1)^2\,\left[ \frac{{\rm i} m p}{4\pi} + \frac{m\Lambda}{2\pi^2}
+ \order(p^2m/\Lambda)\right].
\ee
Substituting into the formula for the nonperturbative amplitude,
\eq{Tdimreg}, we find that
\be\label{as-cutoff}
\frac{1}{a_s} = \frac{2\Lambda}{\pi} - \frac{\Lambda^2}{m}.
\ee
Thus large~$a_s$  can coexist with large $\Lambda$ if there is a delicate
cancellation between the two terms on the right-hand side. In this case
$\Lambda$ goes to $2m/\pi$, which is large, as $a_s$~goes to infinity.

Minimally-subtracted dimensional regularization is well known to drop
power-law divergent terms that are physically significant, and that is
what happens here. The dimensionally regulated calculation
drops the linear divergence in $T_2$, and as a result captures only
one of the two terms in~\eq{as-cutoff}. In the cutoff theory, the small
binding energy results from a nearly exact cancellation between the
kinetic and potential energies, each of which is separately quite
large\,---\,of order~$\Lambda/a_s m$. The dimensionally regulated
calculation, however, cannot model this cancellation because, in effect, it
drops the power-divergent kinetic-energy term.

The cutoff theory behaves similarly to the dimensionally regulated
theory when cutoff~$\Lambda$ is small compared with~$1/a_s$. 
To see this, we multiply the potential
by a coupling constant~$c$, thereby decoupling the range and strength of the
potential. Then $m$ is replaced by
$cm$ in~\eq{as-cutoff} for the scattering length, and we can solve for
the value of~$c$ corresponding to a particular~$a_s$:
\be
c =
\frac{\Lambda^2}{m}\left(\frac{2\Lambda}{\pi}-\frac{1}{a_s}\right)^{-1}
\ee
for any~$\Lambda$. 
When $\Lambda\!\ll\!1/a_s$ the coupling constant becomes negative,
that is the potential becomes repulsive, 
as in the dimensionally regulated analysis. The
linear divergence in the loops is negligible for such small~$\Lambda$'s,
and therefore first-order perturbation theory gives the correct scattering
length, again as in the dimensionally regulated analysis. And the
theory fails at relatively low energies since~$\Lambda\!\ll\!1/a_s$.

Large scattering lengths can arise in a range of situations that lie
between two different physical extremes. One extreme
is realized by adjusting the
magnitude of the potential so that the first-order Born approximation gives
the correct scattering length, while at the same time extending the
potential's range so that 
corrections from second order and beyond are negligible
(other than the imaginary parts always required by unitarity). 
The other extreme is realized by tuning
a  short-range potential to create a bound state
or resonance near threshold; the presence of a bound state implies that
contributions from all orders in the the potential are important.
Minimal subtraction with dimensional regularization is well suited
only to the first limit. Unfortunately the physics of low-energy
nucleon-nucleon scattering corresponds to the second limit.

The problem encountered in the dimensionally regulated calculation
can be avoided in perturbative analyses since
high-order terms that cause trouble can be dropped by hand.
In nonperturbative calculations, like that needed for $S$-wave
nucleon-nucleon scattering, the problem can be easily repaired by
modifying minimal subtraction; other,
more baroque variations are also possible.\cite{kaplan2}  
But the problem can be avoided altogether by
not using dimensional regularization. Unlike in perturbative
QCD, for example, there is no compelling reason for using dimensional
regularization in the nucleon-nucleon problem. Simple cutoffs of the sort
used in this paper are effective, robust,  easy to use, and more physical.

\subsection{Evaluation}

Chiral perturbation theory, and, in particular, effective
theories of the sort I have described  
provide a systematic framework for studying low-energy nuclear
physics. This framework 
casts traditional low-energy nuclear theory as the
rigorous low-energy limit of QCD. Where it works, the effective theory
is far more efficient to use than QCD, and is preferred.
Of course, the effective theory is only useful at momenta below
500-1000\,MeV (or perhaps 300\,MeV). 
Renormalization theory indicates that quarks and QCD are essential to a
systematic and rigorous study of processes at higher
momenta. One can, of course, add a parameter to
a nucleon-nucleon potential model in order to fix a particular
piece of high-energy data; but usually that extra parameter will not
improve anything else. By contrast, when we added the
$\Lambda^{-4}$~correction to our effective theory we improved the fit at
0.1\,MeV, where we tuned, but also at all higher energies up to
100--150\,MeV. And we improved the fit in a predictable manner.

We have seen that the effective  theory is quantitatively
very successful at explaining low-energy
nucleon-nucleon scattering. This is particularly impressive
given that the tuned
parameters were not derived from a global fit, but rather from the
lowest-energy data, to
the left of all the curves plotted. What impresses me more
than the size of the errors are their systematics: the predictable
dependence on the energy and on the number of contact terms in the
potential, the relationships between different channels, and so on. 
Thus plots of the errors, $\Delta\delta_l(E)$, are more interesting
than plots of the phase shifts, $\delta_l(E)$, themselves. 
The systematics can only be observed if the theory is tuned to the
lowest-energy data, or if energy-dependent truncation errors are
included in the weights assigned to data in a global fit.

Using an effective
potential model rather than a Bethe-Salpeter approach in the chiral
field theory is a small technical advance. It
simplifies the treatment of retardation effects, resulting in simpler
potentials. 

The effective theory is successful, but is it successful enough? A
major application of nuclear theory in our energy region is to
problems in nuclear
structure. These typically involve nucleon energies of order
the Fermi energy, roughly~50\,MeV. Our analysis
gives results that are accurate to within 5--15\%, depending upon the
value of~$\Lambda$, in this region. This suggests that the
$1/\Lambda^2$~expansion converges for these energies, and that a
systematic analysis of nuclear matter 
is possible without recourse to QCD. Precision will
improve substantially if two-pion exchange doubles the threshold
cutoff. And it will also improve substantially when
$\Lambda^{-6}$~terms are added. 
In the not too distant future,
lattice QCD will begin to generate numbers for the various coupling
constants in the effective theory, and the connection to the
fundamental theory will be complete. In the meantime we can continue
to extract the couplings experimentally from the abundance of accurate
nucleon-nucleon data.

\section*{Acknowledgements}
I thank Ron Horgan, with whom I first started thinking about
chiral models and nuclear physics almost ten years ago. I also thank
my former student Lisa~Angelos, who provided several key insights in
the early stages of this work.  I  thank the
Institute for Nuclear Theory, in Seattle, where the first version of
these lectures was created. I owe much to the summer's worth of
conversations I had with Martin Savage, David Kaplan, and others at
the INT.   I also benefited from several conversations with Eric
Braaten. Finally 
I am very grateful to the organizers of the {\em VIII Jorge Andre
Swieca Summer School\/} for inviting me to speak on this subject.


\section*{References}
\begin{thebibliography}{99}
\bibitem{tasitalk} This introduction is a severe condensation of an
earlier summer school lecture: G.P. Lepage, {\em What is Renormalization?},
in {\em From Actions to Answers,} edited by
T. DeGrand and D. Toussaint, World Scientific Press
(Singapore, 1990). It contains references to the general literature
on renormalization theory; see especially K.G. Wilson's Nobel lecture:
K.G.\ Wilson, \Journal{ Rev.\ Mod.\ Phys.}{55}{583}{1983}.


\bibitem{angelos}This point is emphasized in E.~Angelos, Cornell University
Ph.D. Thesis, 1996.

\bibitem{savage} I thank Martin Savage explaining the multiple
solutions to me.

\bibitem{lepage-brodsky} See, for example, G.P. Lepage and
S.J. Brodsky, \Journal{\PRD}{22}{2157}{1976}.

\bibitem{labelle} For recent examples see P. Labelle, G.P. Lepage and
U. Magnea, \Journal{\PRL}{72}{2006}{1994}; and 
M. Nio and T. Kinoshita, \Journal{\PRD}{55}{7267}{1997}.

\bibitem{bbl} G.T. Bodwin, E. Braaten and G.P. Lepage, 
\Journal{\PRD}{51}{1125}{1995}.

\bibitem{ordonez96} A particularly useful reference 
is C. Ordonez, L. Ray and U. van Klock, 
\Journal{\PRD}{53}{2086}{1995}. They reference much of the older
material as well.

\bibitem{chi-pth} Ordonez et al\,\cite{ordonez96} summarize the
essential ingredients needed for nucleon-nucleon scattering.

\bibitem{nijmegen} Partial wave analysis of the Nijmegen University
group, obtained from the World Wide Web page
http:://nn-online.sci.kun.nl/.

\bibitem{vankolck94} See, for example, U. van Klock,
\Journal{\PRC}{49}{2932}{1994}. 

\bibitem{kaplan96} D.B. Kaplan, M.J. Savage and M.B. Wise,
\Journal{\NPB}{478}{629}{1996}; 
M. Luke and A.V. Manohar, \Journal{\PRD}{55}{4129}{1997}.

\bibitem{kaplan2} D.B. Kaplan, University of Washington preprint
(1996), e-print nucl-th/9610052.

\end{thebibliography}
\end{document}